\documentstyle[11pt,amssymb,epsf]{article}
\begin{document}
%
%
%
%
\newcommand{\bbbone}{{\mathchoice {\rm 1\mskip-4mu l} {\rm 1\mskip-4mu l}    {\rm 1\mskip-4.5mu l} {\rm 1\mskip-5mu l}}}
\newcommand{\bA}{{\Bbb A}}
\newcommand{\bB}{{\Bbb B}}
\newcommand{\bC}{{\Bbb C}}
\newcommand{\bD}{{\Bbb D}}
\newcommand{\bE}{{\Bbb E}}
\newcommand{\bF}{{\Bbb F}}
\newcommand{\bG}{{\Bbb G}}
\newcommand{\bH}{{\Bbb H}}
\newcommand{\bI}{{\Bbb I}}
\newcommand{\bJ}{{\Bbb J}}
\newcommand{\bK}{{\Bbb K}}
\newcommand{\bL}{{\Bbb L}}
\newcommand{\bM}{{\Bbb M}}
\newcommand{\bN}{{\Bbb N}}
\newcommand{\bO}{{\Bbb O}}
\newcommand{\bP}{{\Bbb P}}
\newcommand{\bQ}{{\Bbb Q}}
\newcommand{\bR}{{\Bbb R}}
\newcommand{\bS}{{\Bbb S}}
\newcommand{\bT}{{\Bbb T}}
\newcommand{\bU}{{\Bbb U}}
\newcommand{\bV}{{\Bbb V}}
\newcommand{\bW}{{\Bbb W}}
\newcommand{\bX}{{\Bbb X}}
\newcommand{\bY}{{\Bbb Y}}
\newcommand{\bZ}{{\Bbb Z}}
%
%
\newcommand{\al}{\alpha}
\newcommand{\be}{\beta}
\newcommand{\ga}{\gamma}
\newcommand{\de}{\delta}
\newcommand{\ep}{\epsilon}
\newcommand{\et}{\eta}
\newcommand{\ze}{\zeta}
\newcommand{\th}{\theta}
\newcommand{\io}{\iota}
\newcommand{\ka}{\kappa}
\newcommand{\la}{\lambda}
\newcommand{\rh}{\rho}
\newcommand{\si}{\sigma}
\newcommand{\ta}{\tau}
\newcommand{\up}{\upsilon}
\newcommand{\ph}{\phi}
\newcommand{\ch}{\chi}
\newcommand{\ps}{\psi}
\newcommand{\om}{\omega}
\newcommand{\vphi}{\varphi}
\newcommand{\vth}{\vartheta}
\newcommand{\veps}{\varepsilon}
%
%
\newcommand{\Ga}{\Gamma}
\newcommand{\De}{\Delta}
\newcommand{\Ep}{\Epsilon}
\newcommand{\Th}{\Theta}
\newcommand{\La}{\Lambda}
\newcommand{\Si}{\Sigma}
\newcommand{\Up}{\Upsilon}
\newcommand{\Ph}{\Phi}
\newcommand{\Ps}{\Psi}
\newcommand{\Om}{\Omega}
%
%
\newcommand{\bGa}{{{\rm I}\kern-.16em \Gamma}}
%
%
\newcommand{\alq}{{\bar\alpha}}
\newcommand{\beq}{{\bar\beta}}
\newcommand{\gaq}{{\bar\gamma}}
\newcommand{\deq}{{\bar\delta}}
\newcommand{\epq}{{\bar\epsilon}}
\newcommand{\etq}{{\bar\eta}}
\newcommand{\zeq}{{\bar\zeta}}
\newcommand{\thq}{{\bar\theta}}
\newcommand{\ioq}{{\bar\iota}}
\newcommand{\kaq}{{\bar\kappa}}
\newcommand{\laq}{{\bar\lambda}}
\newcommand{\rhq}{{\bar\rho}}
\newcommand{\siq}{{\bar\sigma}}
\newcommand{\taq}{{\bar\tau}}
\newcommand{\upq}{{\bar\upsilon}}
\newcommand{\phq}{{\bar\phi}}
\newcommand{\chq}{{\bar\chi}}
\newcommand{\psq}{{\bar\psi}}
\newcommand{\omq}{{\bar\omega}}
\newcommand{\vepsq}{{\bar\varepsilon}}
\newcommand{\vthq}{{\bar\vth}}
\newcommand{\vphq}{{\bar\vph}}
\newcommand{\Psq}{{\bar\Ps}}
%
%
\newcommand{\cA}{{\cal A}}
\newcommand{\cB}{{\cal B}}
\newcommand{\cC}{{\cal C}}
\newcommand{\cD}{{\cal D}}
\newcommand{\cE}{{\cal E}}
\newcommand{\cF}{{\cal F}}
\newcommand{\cG}{{\cal G}}
\newcommand{\cH}{{\cal H}}
\newcommand{\cI}{{\cal I}}
\newcommand{\cJ}{{\cal J}}
\newcommand{\cK}{{\cal K}}
\newcommand{\cL}{{\cal L}}
\newcommand{\cM}{{\cal M}}
\newcommand{\cN}{{\cal N}}
\newcommand{\cO}{{\cal O}}
\newcommand{\cP}{{\cal P}}
\newcommand{\cQ}{{\cal Q}}
\newcommand{\cR}{{\cal R}}
\newcommand{\cS}{{\cal S}}
\newcommand{\cT}{{\cal T}}
\newcommand{\cU}{{\cal U}}
\newcommand{\cV}{{\cal V}}
\newcommand{\cW}{{\cal W}}
\newcommand{\cX}{{\cal X}}
\newcommand{\cY}{{\cal Y}}
\newcommand{\cZ}{{\cal Z}}
%
%
%
\newcommand{\tG}{{\tilde G}}
\newcommand{\prP}{{\bf P}}
\newcommand{\db}{{\mkern2mu\mathchar'26\mkern-2mu\mkern-9mud}}
\newcommand{\tr}{\mbox{ tr }}
\newcommand{\del}{\partial}
\newcommand{\gtoas}[1]{{\quad\mathop{\longrightarrow}\limits_{#1}\quad}}
\newcommand{\ve}[1]{{\bf #1}}
\newcommand{\edp}[1]{e^{\prime\prime}(\ve{#1})}
\newcommand{\nat}[1]{\{ 1,\ldots,#1 \} }
\newcommand{\natz}[1]{\{ 0,\ldots,#1 \} }
\newcommand{\abs}[1]{{\left\vert #1 \right\vert}}
\newcommand{\norm}[1]{{\left\Vert #1 \right\Vert}}
\newcommand{\oo}[1]{{\mathaccent'27 #1}}
\newcommand{\Pol}[1]{\hbox{ pol}(#1)}
\newcommand{\openkrnl}[1]{\mathop{#1}\limits^{\;_\zer}}
\newcommand{\Perm}[1]{{\cal S}_{#1}}
\newcommand{\Ref}[1]{$(\ref{#1})$}
%
%
\newcommand{\True}[1]{\; \bbbone\left( #1 \right) \;}
\newcommand{\sfrac}[2]{{\textstyle \frac{#1}{#2}}}
\newcommand{\ssfrac}[2]{{\scriptstyle \frac{#1}{#2}}}
\newcommand{\vol}{{\hbox{ vol }}}
\newcommand{\dbar}{{\mkern2mu\mathchar'26\mkern-2mu\mkern-9mud}}
\newcommand{\Imp}{\Longrightarrow}
\newcommand{\Equiv}{\Longleftrightarrow}
\newcommand{\dotcup}{\mathop{\mathop{\cup}\limits^\cdot}}
\newcommand{\const}{\hbox{ \rm const }}
\newcommand{\supp}{\hbox{ \rm supp }}
\newcommand{\half}{\frac{1}{2}}
\newcommand{\ili}{\int\limits}
\newcommand{\sli}{\sum\limits}
\newcommand{\pli}{\prod\limits}
\newcommand{\lli}{\lim\limits}
\newcommand{\Laplace}{\Delta}
\newcommand{\LC}{{\Laplace_C}}
\newif\ifintremark
\newcommand{\intremark}[1]{\ifintremark\par\medskip {\bf Internal Remark:}
#1 \hfill$\clubsuit$\par\medskip\else\fi}
\newcommand{\noin}{\noindent}
\newcommand{\nonu}{\nonumber}
\newif\ifspr
\sprfalse
\ifspr 
\renewcommand{\abstractname}{\hglue 15pt Abstract}
\else
\newcommand{\E}{{\rm e}}
\newcommand{\I}{{\rm i}}
\fi
\newcommand{\Gast}{{\Ga^*}}
\newcommand{\Gainf}{{\Ga^*_\infty}}
\newcommand{\Last}{{\La^*}}
\newcommand{\Sk}{{\bf k}}
\newcommand{\Sl}{{\bf l}}
\newcommand{\Sp}{{\bf p}}
\newcommand{\Sq}{{\bf q}}
\newcommand{\Sr}{{\bf r}}
\newcommand{\Sx}{{\bf x}}
\newcommand{\Sy}{{\bf y}}
\newcommand{\Sz}{{\bf z}}
\newcommand{\SQ}{{\bf Q}}
\newcommand{\SLa}{\bL}
\newcommand{\Nta}{n_\ta}
\newcommand{\Xk}{K}
\newcommand{\Xp}{P}
\newcommand{\Xq}{Q}
\newcommand{\Xr}{R}
\newcommand{\Xv}{V}
\newcommand{\Xw}{W}
\newcommand{\Xx}{X}
\newcommand{\Xy}{Y}
\newcommand{\Xz}{Z}
\newcommand{\psorpsq}{{\mathop{\ps}\limits^{\scriptscriptstyle{(-)}}}}
\newcommand{\psorpsqh}{\hat {\mathop{\ps}\limits^{\scriptscriptstyle{(-)}}}}
\renewcommand{\tr}{\mbox{ tr}}
\newcommand{\Geff}[2]{\cG (#1,#2)}
\newcommand{\Lp}[1]{\De_{#1}}
\newcommand{\Lpshow}[1]{\half (\frac{\de}{\de \ps}, #1 \frac{\de}{\de \ps})}
\newcommand{\Gr}[4]{G_{#1#2}(#3 \mid #4)}
\newcommand{\Ir}[4]{I_{#1#2}(#3 \mid #4)}
\newcommand{\QQ}[4]{Q_{#1#2}(#3 \mid #4)}
\newcommand{\Qrmt}[1]{\QQ{m}{r}{t}{#1}}
\newcommand{\QQrm}[1]{Q_{mr}(#1)}
\newcommand{\Grmt}[1]{\Gr{m}{r}{t}{#1}}
\newcommand{\Grmz}[1]{\Gr{m}{r}{0}{#1}}
\newcommand{\GGrm}[1]{G_{mr}(#1)}
\newcommand{\Grmto}[1]{\Gr{m_1}{r_1}{t}{#1}}
\newcommand{\GGrmo}[1]{G_{m_1r_1}(#1)}
\newcommand{\Grmtt}[1]{\Gr{m_2}{r_2}{t}{#1}}
\newcommand{\GGrmt}[1]{G_{m_2r_2}(#1)}
\newcommand{\Irmt}[1]{\Ir{m}{r}{t}{#1}}
\newcommand{\Irmto}[1]{\Ir{m_1}{r_1}{t}{#1}}
\newcommand{\Irmtt}[1]{\Ir{m_2}{r_2}{t}{#1}}
\newcommand{\Irmnl}{I_{mr}^{(\Nta,L)}}
\newcommand{\Irm}{{I_{mr}}}
\newcommand{\CW}[1]{D_{#1}} 
\newcommand{\CWb}[1]{\bar D_{#1}} 
\newcommand{\CWt}[1]{{\tilde D_{#1}}} 
\newcommand{\CWtd}[1]{\dot{\tilde D_{#1}}} 
\newcommand{\cDt}[1]{\cD_t^{(#1)}}
\newcommand{\bdtd}{\dot\bD_t}
\newcommand{\ul}[1]{\underline{#1}}
\newcommand{\WO}[1]{\Om_{#1}}
\newcommand{\Mats}[1]{\bM_{#1}}
\newcommand{\Hat}[1]{\widehat{#1}} 
\newcommand{\dlinks}[1]{\frac{\del_L}{\del #1}}
\newcommand{\Norm}[1]{|\!|\!| #1 |\!|\!|}
\newcommand{\vv}{(-1)}
\newcommand{\PP}{P}
\newcommand{\zil}{i}
\newcommand{\FS}{\cS}
\newcommand{\ept}{\ep_t}
\newcommand{\epz}{\ep_0}
\newcommand{\omb}{\om_\be}
\newcommand{\kK}{K}
\newcommand{\Mo}{M_1}
\newcommand{\Kzmr}{{K_{mr}^{(0)}}}
\newcommand{\Ko}{{K^{(1)}}}
\newcommand{\Kdrr}{{L_{3,r}}}
\newcommand{\Kvr}{{L_{4,r}}}
\newcommand{\Kfmr}{{L_{5,m,r}}}
\newcommand{\kaS}{\tilde\ka}
\newcommand{\kaSN}{\tilde\ka^{(N)}}
\newcommand{\IS}{\tilde I}
\newcommand{\ISN}{\tilde I^{(N)}}
\newcommand{\QS}{\tilde Q}
\newcommand{\QSN}{\tilde Q^{(N)}}
\newcommand{\BL}{B^{(L)}}
\newcommand{\UNR}{U^{(N)}}
\newcommand{\kz}{{k_0}}
\newcommand{\MM}{M}
\newcommand{\MP}{\Pi}
\newcommand{\Mm}{m}
\newcommand{\dd}{\de}
\newcommand{\ff}{f_t}
\newcommand{\fff}{\tilde f}
\newcommand{\nn}{n}
\newcommand{\nnn}{\cN}
\newcommand{\Vo}{{\cV_1}}
\newcommand{\coco}{{M_0}}
\newcommand{\coce}{{M_1}}
\newcommand{\cocz}{{M_2}}
\newcommand{\Sisk}{{\Si_{\rm sk}}}
\newcommand{\Spi}{\pi\mkern-9.7mu\pi}
\renewcommand{\Ref}[1]{Eq.\ $(\ref{#1})$}
\newcommand{\nonRPA}{non-ladder}
\newcommand{\Seps}{\veps}
\newcommand{\Teps}{\veps_\ta}
\newcommand{\Sign}{\veps}
\newcommand{\Ez}{E_0}
%
%
\ifspr
\else
\newtheorem{theorem}{Theorem}
\newtheorem{proposition}{Proposition}
\newtheorem{definition}{Definition}
\newtheorem{lemma}{Lemma}
\newtheorem{remark}{Remark}
\newtheorem{corollary}{Corollary}
\newenvironment{proof}{\par\noin {\it Proof:} \hspace{7pt}}%
{\hfill\hbox{\vrule width 7pt depth 0pt height 7pt} \par\vspace{10pt}}
\fi

\begin{titlepage}

\ifspr
\journalname{Communications in Mathematical Physics}
\title{Continuous renormalization for fermions
and Fermi liquid theory}
\bigskip
\author{Manfred Salmhofer}
\institute{Mathematik, ETH-Zentrum, 8092 Z\" urich, Switzerland,
manfred@math.ethz.ch} 
\else
\title{Continuous renormalization for fermions \\
and Fermi liquid theory}
\bigskip
\author{Manfred Salmhofer\\ \medskip
Mathematik, ETH-Zentrum,8092 Z\" urich, Switzerland\\
manfred@math.ethz.ch} 
\bigskip

\date{May 1997}
\bigskip
\fi

\maketitle

\begin{abstract}
\noin
I derive a Wick ordered 
continuous renormalization group equation for fermion systems
and show that a determinant bound applies directly to this equation.
This removes factorials in the recursive equation for the 
Green functions, and thus improves the combinatorial behaviour.
The form of the equation is also ideal for the investigation
of many-fermion systems, where the propagator is singular on a surface. 
For these systems, I define a criterion for Fermi liquid behaviour
which applies at positive temperatures.
As a first step towards establishing such behaviour in $d \ge 2$, 
I prove basic regularity properties of the interacting Fermi surface 
to all orders in a skeleton expansion.
The proof is a considerable simplification of previous ones.
\end{abstract}

\end{titlepage}

\section{Introduction}\label{sect1}

\noin
In this paper, I begin a study of fermionic quantum field theory
by a continuous Wick ordered renormalization group equation (RGE).
As an example, I take the standard 
many-fermion system of solid state quantum field theory,
but the method applies to general fermionic models
with short-range interactions.
I show that a determinant of propagators appears in the RGE
and I use a determinant bound to prove that a factorial 
which would appear in bosonic theories is removed from 
the recursion for the fermionic Green functions. This 
may lead to convergence of perturbation theory in the absence 
of relevant couplings, but I do not address the convergence problem,
which is related to the solution of a particular combinatorial recursion,
in this paper.
A short account of this work has appeared in \cite{S2}.

Continuous RGEs were invented by Wegner \cite{Weg} and 
Wilson \cite{Wil}. Polchinski \cite{Pol} found a beautiful way
to use them for a proof of perturbative renormalizability of
$\ph ^4$ theory. His method was simplified in \cite{KKS}, 
and extended to composite operator renormalization 
and to gauge theories by Keller and Kopper \cite{KK}. 
Keller \cite{K} also proved local Borel summability.
While equivalent to the Gallavotti-Nicol\` o \cite{Gal,GN,FHRW} method,
the continuous RGE is much simpler technically.
An application of continuous RG methods to nonperturbative bosonic
problems \cite{BY,BDH} requires many new ideas and a combination
with cluster expansion techniques, to control the combinatorics. 
It is one of the points of this paper
that for fermions, the straightforward adaption of the method
yields a determinant bound which improves the combinatorics
of fermionic theories as compared to bosonic ones, and may lead
to nonperturbative bounds.
The determinant structure is not visible in the form of the flow equation
used in \cite{Pol,KKS,KK}, because the flow equation in that 
form is a one-loop equation which has too little structure.
A key ingredient for the present analysis is Wick ordering, 
which was first used in the context of continuous 
RGEs for scalar field theories by Wieczerkowski \cite{Wie}.
I show that for fermions, the Wick ordered RGE contains a determinant
of propagators to which a Gram inequality applies directly.
A closer look at the way the Feynman graph expansion is generated
by the RGE shows that the sign cancellations
bring the combinatorial factors for fermions nearer to that 
of a planar field theory.  
This reduction does, however, not lead to 
a planar field theory in the strict sense  
because of certain binomial factors in the recursion. 

The direct application of the determinant bound shown here
requires the interaction between the fermions to be 
short-range. This prevents a straightforward application 
to systems with abelian gauge fields by simply integrating
over the gauge fields. One model to which the method applies
directly is the Gross-Neveu model (which has been constructed 
rigorously  \cite{GK,FMRS}). I show here only the most basic
power counting bounds by leaving out all relevant and marginal 
couplings, but it is possible to take them into account by 
renormalization.  

A class of physically realistic models with a short-range interaction
is that of nonrelativistic many-fermion models. 
In these models, there is a significant complication 
of the analysis because the singularity of the fermion propagator
in momentum space is not at a point, but instead on the 
Fermi surface, which is a $(d-1)$-dimensional subset of 
momentum space. Only in one dimension,
the singularity is pointlike -- the `surface' becomes a point. 

The interest in these models has resurged recently because of
the discovery of high-temperature superconductivity.
Before that 
it was taken for granted that Fermi liquid (FL) behaviour holds in all 
dimensions $d \ge 2$, and Luttinger liquid behaviour
in one dimension (the latter has been proven \cite{BG,BGPS,BM}). 
At certain doping values, however,
strong deviations from FL behaviour are seen in the high-$T_c$ 
materials. The discussion following these discoveries 
revealed that the former arguments for FL behaviour
contained logical gaps.
Like \cite{FT,FMRT,FST1,FST2,FST3,S,FKLT}, 
the present work is not aimed at an understanding of these 
deviations, but at the more modest goal of determining first
under which conditions FL behaviour occurs. 

Before doing so, it is necessary to give a definition of 
what would constitute FL behaviour. 
At zero temperature, the noninteracting Fermi gas has a 
discontinuity in the occupation number density; 
this is also a property one would require of 
a zero temperature FL. This discontinuity  
is absent for Luttinger liquids.
It is, however, not sufficient for FL behaviour because 
there is a one-dimensional model which has both such a discontinuity
and some Luttinger liquid features in the spectral density
\cite{Mp}. Moreover, in the standard models of many-fermion systems
such a step never occurs because superconductivity sets in
below a critical temperature, and it 
smoothes out the step in the zero-temperature Fermi distribution.
It is thus desirable to give a definition of FL behaviour
at temperatures above the critical temperature for superconductivity.
This is not at all straightforward because there
is no clean characterization of FL behaviour {\em at a fixed 
temperature}. I propose to look at a whole range of temperatures and 
values of the coupling constant to bring out the characteristic features
of a FL.

I define an equilibrium Fermi liquid as a system in which the perturbation 
expansion in the coupling constant $\la$ converges for the 
{\em skeleton Green functions} in the region
$|\la| \log \be $ small enough (here $\be$ is the inverse temperature)
and where the self-energy fulfils certain regularity conditions.
The skeleton Green functions are defined in detail in Section \ref{sect6};
in them, self-energy insertions are left out, so that the Fermi surface 
stays fixed. I discuss in Section \ref{sect7} how they are related
to the exact Green functions. 
The logarithmic dependence of the radius of convergence on $\be$ comes from
the Cooper instability; the difference between Fermi liquids and 
Luttinger liquids is in the regularity properties of the self-energy.
This is discussed in detail in Section \ref{thepurpose}.

The goal is to show that the standard many-fermion systems 
are Fermi liquids in that sense. A proof of this requires 
a combination of the  
regularity techniques of \cite{FST1,FST2,FST3} for renormalization
with the sector technique of \cite{FMRT} in the determinant 
bound (see Section \ref{thepurpose} for further discussion). 
I do not give a complete proof in this paper but only 
a part of it, by showing 
a determinant bound and some of the required regularity properties of the 
selfenergy in perturbation theory.
The hope is that the determinant bound will lead to convergence,
so that the method developed here, which is somewhat simpler than
e.g.\ the one in \cite{FMRT}, will work nonperturbatively. 
A different representation for fermionic 
Green functions that provides a simplification 
and leads to nonperturbative bounds 
is given in \cite{FKT}.

In Section \ref{sect2}, I review the Grassmann integral for many-fermion
systems briefly, 
to give a self-contained motivation for the study of such systems,
and to fix notation. 
The Fermi liquid criterion is formulated in Section \ref{thepurpose}.
Section \ref{sect3} contains the general renormalization group 
equation and the determinant formula \Ref{Qrmtdef}. 
Section \ref{sect4} contains the determinant bound
and an application to systems where the propagator 
has point singularities.
In Section \ref{sect5}, I show the existence of the thermodynamic
limit for the many-fermion system in perturbation theory.
In Section \ref{sect6}, 
I prove bounds on the skeleton self-energy 
that are needed to renormalize the full theory
in perturbation theory. Again, the RGE in the form 
of \cite{Pol,KKS,KK} would not be very convenient for this
because it is a one-loop equation, whereas the crucial 
effects for regularity all start at two loops. 
They can be seen in a simple way in the Wick ordered RGE.
Details about Wick ordering and the derivation 
of the determinant formula \Ref{Qrmtdef} are deferred to the Appendix. 

\medskip\noin
{\bf Acknowledgement: }
I thank Volker Bach, Walter Metzner, Erhard Seiler, 
and Christian Wieczerkowski for discussions.
I also thank Christian Lang for his hospitality at a very pleasant
visit to the University of Graz, where this work was started.

\section{Many-fermion systems}\label{sect2}
The model is defined on a spatial lattice with spacing $\Seps$. 
Continuum models are obtained in the limit $\Seps\to 0$;
lattice models, such as the Hubbard model, are obtained by
fixing $\Seps$. 

Let $d \ge 2$ be the spatial dimension, $\Seps > 0$, $L \in \bR$
be such that $\frac{L}{2\Seps} \in \bN$, and let $\bG$ be any lattice
of maximal rank in $\bR^d$, e.g., $\bG = \bZ^d$. Let $\SLa$ be the 
torus $\SLa = \Seps \bG / L \bG$. 
The number of points of this lattice is
$|\SLa| = (\frac{L}{\Seps})^d$.
Let $\int_\SLa d\Sx\; F(\Sx) = \Seps^d \sum_{\Sx \in \SLa} F(\Sx)$
and $\de_\SLa (\Sx,\Sx') = \Seps^{-d} \de_{\Sx,\Sx'}$. 

Let $\cF_\SLa$ be the Fock space generated by the spin one half
fermion operators satisfying the canonical anticommutation relations \cite{BR},
i.e.\ for all $\Sx,\Sx' \in \SLa$,
\begin{equation}
c_\al (\Sx) c^+_{\al'} (\Sx') +  c^+_{\al'} (\Sx')c_\al (\Sx) 
= \de_{\al\al'} \; \de_\SLa (\Sx,\Sx') 
.\end{equation}
Here $\al \in \{-1,1\}$ is the spin of the fermion
in units of $\frac{\hbar}{2}$.
The free part of the Hamiltonian $H_\SLa(c,c^+)=H_0+\la V$ is 
\begin{equation}
H_0 = - \sli_{\al \in \{-1,1\}}
\ili_{\SLa} d\Sx \; \ili_{\SLa} d\Sy \; 
T(\Sx,\Sy)\; c^+_\al(\Sx) c_\al (\Sy)
.\end{equation}
For a one-band model on a lattice with fixed spacing $\Seps$,
$T(\Sx,\Sy) = t_{\Sx-\Sy}=t_{\Sy-\Sx}$
describes hopping from a site $\Sy$ to another site $\Sx$ with 
an amplitude $t_{\Sx-\Sy}=t_{\Sy-\Sx}$. 

The interaction is multiplied by a small coupling constant $\la$;
I assume it to be a normal ordered density-density interaction
\begin{equation}
V(c,c^+) = - \ili_{\SLa} d\Sx \; \ili_{\SLa} d\Sy \; 
\sli_{\al,\si\in \{-1,1\}}
v(\Sx-\Sy) c^+_\al (\Sx) c^+_\si (\Sy) c_\al (\Sx) c_\si (\Sy)
\end{equation}
In other words, it is a special type of a four-fermion interaction. 

For instance, the simplest 
Hubbard model is given by $\la = \frac{U}{2}$ where $U$ is 
the usual Hubbard-$U$, and by $v(\Sx-\Sy)=\de_\SLa (\Sx, \Sy)$, and the 
hopping term is $t_{\Sx-\Sy}=t$ if $|{\Sx-\Sy}|=1$ and zero 
otherwise, where $t$ is the hopping parameter. 

At temperature $T$ and chemical potential $\mu$, the grand canonical 
partition function is given by 
$Z_\SLa = \tr_{\cF_\SLa}\left( \E^{-\be (H_\SLa-\mu N_\SLa )}\right)$
with $N_\SLa  = \int_\SLa d\Sx \; n(\Sx)$, 
and $\be=\frac{1}{k_B T}$. 
Observables are given by expectation values of functions, 
mainly polynomials, of the $c$ and $c^+$, 
\begin{equation}\label{Hamexp}
\langle \cO \rangle_\SLa  = \frac{1}{Z_\SLa}
\tr \left( \E^{-\be (H_\SLa -\mu N_\SLa )} \cO (c,c^+)\right)
\end{equation}
A basic question is whether the expected values of observables
have a finite thermodynamic limit and whether an expansion in $\la$ can
be used to get their behaviour at small or zero temperature $T$. 
For instance, one would like to expand the two-point function 
$\langle c^+(\Sx) c(\Sy)\rangle_\SLa  = 
\sum_{r=0}^\infty \la^r G_{2,r}^{L,\Seps} (\Sx,\Sy)$.
It is by now well-known that the result of a naive expansion $\E^{-\la V}$ 
in powers of $\la$ is that at $T=0$, $\lim_{L \to \infty}
G_{2,r}^{L,\Seps} = \infty$ for all $r \ge 3$ (see, e.g., \cite{FT,FST1}).
At positive temperature $T$, this {\em unrenormalized} expansion 
converges for $|\la| \le \const T^d$; see Section 
\ref{sect4}. To get a better $T$-dependence of the radius of convergence,
one has to renormalize. Because of the BCS instability, the best
one can hope for in general is a bound $|\la| \log \sfrac{1}{T} < \const$
for the region of convergence. This is part of the Fermi liquid
criterion formulated below.

\subsection{Grassmann integral representation}
\label{Grarep}
The standard Grassmann integral representation is obtained by 
applying the Lie product formula 
\begin{equation}\label{Trotto}
\E^{-\be(H_\SLa - \mu N_\SLa)} = 
\lli_{\Nta \to \infty} \left(
\E^{-\Teps (H_0-\mu N_\SLa)} \; \E^{-\Teps \la V}\right)^{\Nta}
\end{equation}
to the trace for $Z_\La$ and $\langle \cO \rangle$. 
The spacing in imaginary-time direction is 
$\Teps = \frac{\be}{\Nta}$.
The limit exists in operator norm because on the finite lattice $\SLa$, 
all operators are just finite-dimensional matrices. 

Inserting the orthonormal basis of $\cF_\SLa$ between the factors
in \Ref{Trotto} and rearranging,
I get $Z_\SLa = \lim_{\Nta \to \infty} Z_{\SLa,\Nta}$,
where $Z_{\SLa,\Nta}$ is given by a finite-dimensional 
Grassmann integral, as follows. 
Let $\Nta$ be even and $\bT = \{\ta=n\Teps:\; n \in \bZ,\; -\frac{\Nta}{2} \le n < \frac{\Nta}{2}\}$,
let $\La = \bT \times \SLa$, and $\cA$ be the 
Grassmann algebra generated by $\ps_\si(x)$, $\psq_\si(x)$, 
with $\si \in \{1,-1\}$ and $x = (\ta,\Sx) \in \La$. Fix some 
ordering on $\La$ and denote the usual Grassmann measure \cite{Ber,BrM} by 
$D_\La \ps D_\La \psq = \prod_{x,\si}d\ps_\si(x)\; d\psq_\si(x)$.
Then \Ref{Trotto} implies
$Z_{\SLa,\Nta} = \cN_\La \int D_\La \ps D_\La \psq \;\E^{- S_\La(\ps,\psq)}$
where $\cN_\La $ is a normalization factor that depends on 
$\Seps, L$, and $\Nta$, and where 
\begin{equation}
S_\La (\ps,\psq) =  \int_\bT d\ta
\left(\sli_\si \ili_\SLa d\Sx\; \psq_\si(\ta,\Sx)
\del_\ta \ps_\si (\ta,\Sx)
\; - \; H_\La(\ps(\ta),\psq(\ta)) \right)
\end{equation}
Here I have used the notations $\ps(\ta)(\Sx) = \ps(\ta,\Sx)$,
$\int_\bT d\ta F(\ta) = \Teps \sum_{\ta \in \bT} F(\ta)$,
and $\del_\ta \ps (\ta) = \Teps^{-1}(\ps_\si(\ta+\Teps)-\ps_\si(\ta))$,
and the sum over $\ta$ runs over $\bT$, 
with antiperiodic boundary conditions \cite{Lue}.
For $\Nta<\infty$ and $L < \infty$, this is a finite-dimensional
Grassmann integral. The limit $\Nta \to \infty$, and afterwards
$L \to \infty$, will be taken only for the effective action. 
No infinite-dimensional Grassmann integration will be required.

To do the Fourier transformation, it will be convenient to deal 
with periodic functions defined on an interval of double length in $\ta$, 
and to impose the antiperiodicity as an antisymmetry condition: 
let $\bT_2 = \Teps \bZ / 2 \be \bZ$,
in other words, $\bT_2 = \{ \ta \in \Teps \bZ: -\be \le \ta < \be\}$
with periodic boundary conditions. Thus the fields $\ps$ and $\psq$ are 
periodic  with respect to translations of 
$\ta $ by $2\be$, and antiperiodicity 
with respect to translations by $\be$ is imposed by setting 
\begin{equation}\label{psanti}
\psorpsq(\ta+\be,\Sx)  = -\psorpsq (\ta,\Sx)
\end{equation}
for all $x\in \La$. With the further notation  
$\int_{\La} dx\; F(x) = \int_{\bT} d\ta \int_\SLa d\Sx\; F(\ta,\Sx)$,
and $\de_\La (x,x') = 
{\Teps}^{-1} \Seps^{-d} \de_{\ta,\ta'} \de_{\Sx,\Sx'}$,
the action is
$S(\ps,\psq) = S_2(\ps,\psq) + \la S_4(\ps,\psq)$
where 
\begin{equation}
S_2(\ps,\psq) = \sli_{\si,\si'} \ili_\La dx\; \ili_\La dx'\;
\psq_\si(x) \;  a(x,\si,x',\si') \ps_{\si'} (x')
\end{equation}
with 
\begin{equation}\label{aadef}
a(x,\si,x',\si') = \de_{\si\si'} \left(
(\del_\ta + \mu)\de_\La (x,x') - T(\Sx,\Sx')  \de_{\bT_2} (\ta,\ta')\right)
\end{equation}
and 
\begin{equation}\label{S4def}
S_4(\ps,\psq) = \sli_{\si,\si'} \ili_\La dx\; \ili_\La dx'\;
\psq_\si(x) \ps_\si(x) 
v(\ta,\Sx,\ta',\Sx')
\psq_{\si'}(x') \ps_{\si'}(x')
\end{equation}
with $v(\ta,\Sx,\ta',\Sx')=\de_{\bT_2} (\ta,\ta') v(\Sx-\Sx') $. 
For the present work, the interaction does not have to be 
instantaneous. Retardation effects, like from phonons, 
are allowed. That is, $v(\ta,\Sx,\ta',\Sx')$ may have a dependence
on $\ta$ and need not be local in $\ta$.  

The operator $a$ appearing in $S_2$
is invertible because the antiperiodicity condition
removes the zero modes of the discretized time derivative.
In other words, the Matsubara frequencies for fermions are 
nonzero at positive temperature (this will become explicit 
in the next section).

\subsection{The propagator in Fourier space}
\label{ss22}
Fourier transformation with the 
antiperiodicity conditions \Ref{psanti} is described in Appendix
\ref{fourapp}. The Fourier transforms of $\ps$ and $\psq$ are
\begin{equation}
\hat {\mathop{\ps_\si}\limits^{_{(-)}}}(p) 
= \ili_\La dx \;\E^{-\I px}\; {\mathop{\ps_\si}\limits^{_{(-)}}}(x)
\end{equation}
where, for $p=(\om,\Sp)$ and $x=(\ta,\Sx)$, $px=\om\ta+\Sp\Sx$. 
If $\SLa^*$ is the dual lattice to $\SLa$, the momentum $p$ is
in $\Last = \bM_{\Nta} \times \SLa^*$, where 
\begin{equation}\label{Matsdef}
\Mats{\Nta}=\{ \om_n = \frac{\pi}{\be}(2n+1): \; n \in \bZ, \;
-\frac{\Nta}{2} \le n < \frac{\Nta}{2}\}
\end{equation}
is the set of Matsubara frequencies $\om_n$. With the notation
$\int_{\Last} dp \; F(p) = 
\frac{1}{\be}\sum_{\om \in\Mats{\Nta}} \;
\int_{\SLa^*} d\Sp \; F(\om,\Sp)$,
where $\int_{\SLa^*} d\Sp = L^{-d} \sum_{\Sp \in \SLa^*}$, the inverse
Fourier transform is 
$\ps_\si(x) = \int_{\Last} dx\; \E^{\I px}\;\hat \ps_\si (p)$.
The Fourier transform of the hopping term 
is $\hat T(\Sp,\Sq) = \de_{\SLa^*} (\Sp+\Sq,0) \tilde T (\Sq)$, 
with $\tilde T(\Sq) = \ili_{\SLa} d\Sz\; \E^{\I \Sq\Sz}t_\Sz$.
Denoting 
\begin{equation}
E(\Sp) = \tilde T(\Sp) -\mu
,\end{equation}
where $\mu$ is the chemical potential,
\begin{equation}\label{Hatomdef}
\Hat{\om} = \frac{1}{\I \Teps} \left( \E^{\I \Teps\om} -1\right)
,\end{equation}
and $\de_{\Last} (p+p',0) = \de_{\SLa} (\Sp+\Sp',0) \;
{\Teps}^{-1} \de_{-\om,\om'}$,
the Fourier transform of the operator $a$ in the quadratic part of the 
action is 
\begin{equation}
\hat a (p,\si,p',\si') = 
\de_\La (p+p',0) \de_{\si\si'} \; 
\left(\I\Hat{\om'}-E(\Sp')\right)
\end{equation}
In other words, the matrix with entries $\hat a (p,\si,-p',\si')$ is diagonal,
and for temperature $T = \sfrac{1}{\be} > 0$, 
all diagonal entries are nonzero because 
\begin{equation}\label{Homnonz}
\abs{\mbox{ Re } \Hat{\om}} =
\abs{\om \; \frac{\sin(\Teps\om)}{\Teps\om}}
\ge \frac{1}{2\be}
.\end{equation}
Thus the inverse of $a$, the propagator $c=a^{-1}$, exists; 
it has the Fourier transform
\begin{equation}\label{mfprop1}
\hat c (p,\si,p',\si') = 
\de_\Last (p+p',0) \de_{\si\si'} \; 
\frac{1}{\I \Hat{\om'}-E(\Sp')}
\end{equation}
In the formal continuum limit $\Teps \to 0$,
$\Hat{\om} \to \om$, so one gets the usual formula $(i\om-E(\Sp))^{-1}$.
The partition function of the system of independent fermions ($\la=0$) is 
\begin{equation}\label{freeZ}
\int D_\La\ps\; D_\La \psq \; \E^{(\psq,A \ps)} =
\det A = \pli_{p} \left(\I\Hat{\om(p)}-E(\Sp)\right)
\end{equation}
which is nonzero by \Ref{Homnonz}.

\subsection{The class of models}
\label{hyposuse}
Denote the dual to $\bG$ by $\cB$, the first Brillouin zone
of the infinite lattice. For instance, for $\bG = \Seps\bZ^d$,
$\cB=\bR^d/\sfrac{2\pi}{\Seps}\bZ^d$. The assumptions for the 
class of models are: there is $\kz \ge 2$ such that 
the { dispersion relation} 
$E\in C^\kz(\cB,\bR)$, and for all $\Sp\in \cB$, $E(-\Sp)=E(\Sp)$ holds. 
The interaction $\hat v$ is a $C^\kz$ function from $\bR \times \cB$ to $\bR$,
all its derivatives up to order $\kz$ are bounded functions on
$\cB \times \bR$, $\hat v(-p_0,\Sp) = \overline{\hat v (p_0,\Sp)}$,
and the limit $p_0\to \infty$ of $\hat v$ exists and is $C^\kz$ in 
$\Sp$. There is $g_0 > 0$ such that for all $\Sp$ on the Fermi surface 
$\FS = \{ \Sp: E(\Sp) =0\}$,
$\abs{\nabla E(\Sp )} \ge g_0$ holds. 
The Fermi surface is a subset of an $\Seps$--independent bounded
region of momentum space (hence compact), it is strictly convex and has 
positive curvature everywhere. In particular, there is $\Vo>0$ 
such that for all $L$ and $\Seps$
\begin{equation}\label{Vobou}
\ili_{\SLa^*} d\Sk \True{|E(\Sk)| \le 2} \le \Vo
.\end{equation}
The constant
\begin{equation}\label{EMAX}
E_{\rm max} = \sup\limits_{\Sp \in \cB} \abs{E(\Sp)}
\end{equation}
is independent of $\Seps$.

Under these hypotheses, there is
$\epz > 0$ and a $C^2$-diffeomorphism $\Spi$ from 
$(-2\epz,2\epz) \times S^{d-1}$ to an open neighbourhood of
the Fermi surface $\FS $ in $\cB$,
$(\rh,\th) \mapsto \Spi(\rh,\th)$, such that
\begin{equation}\label{stgz}
E(\Spi(\rh,\th))=\rh \quad \mbox{ and } \quad 
\abs{\del_\rh\Spi(\rh,\th)} \le  \frac{2}{g_0}
\end{equation}
(see \cite{FST1}, Lemma 2.1, and \cite{FST2}, Section 2.2; 
$\epz$ was called $r_0$ there). 
Let $J(\rh,\th) = \det \Spi'(\rh,\th)$ and denote 
\begin{equation}
J_0 = \sup\limits_{|\rh| \le \epz \atop \th \in S^{d-1}} \abs{J(\rh,\th)}
\quad \mbox{ and }\quad
J_1= \sup\limits_{|\rh| \le \epz} \ili_{S^{d-1}} d\th \;\abs{J(\rh,\th)}
.\end{equation}
I assume that $\epz \le 1$ and (for convenience in stating some 
bounds) that
$\be\epz=\frac{\epz}{k_B T}\ge 6$. With the units chosen 
in a natural way, i.e., with typical bandwidths of electron volts, 
this corresponds to temperatures $T$ up to $1000$ Kelvin
if $\epz$ is of order one, which 
seems a sufficient temperature range to study conduction in crystals.
Note, however, that $\epz$ depends on the Fermi surface and thus on the 
filling factor. A typical example is the discretized Laplacian
\begin{equation}
E(\Sk) = B \frac{1}{\Seps^2} \sli_{\nu=1}^d (1-\cos(\Seps k_\nu)) - \mu
.\end{equation}
For $\Seps \to 0$ and $B=\sfrac{1}{m}$, 
$E(\Sk) \to \Sk^2/2m-\mu$, the Jellium dispersion 
relation, which satisfies the above hypotheses if large $|\Sk|$
are cut off. For $\Seps=1$ and $B={\rm t}$, one gets the 
tight-binding dispersion relation with hopping parameter ${\rm t}/2$,
which satisfies the above hypotheses if $\mu \ne{\rm t}d$ (half-filling).
In the limit $\mu \to {\rm t}d$, $\epz \to 0$ 
in the Hubbard model. This implies that to have bounds uniform in the 
filling, one has to stay away from half-filling. The energy $\epz$ sets the 
scale where the low-energy behaviour sets in. The effective four-point interaction at that scale can differ substantially from the original 
interaction. For a discussion, see Section \ref{sect7} and \cite{S}.

\subsection{Nambu formalism}

It will be useful for deriving the component form of the RGE 
to rename the Grassmann variables
such that the distinction between $\ps$ and $\psq$ is in another
index. This is a variant of the usual `Nambu formalism'; 
see, e.g., \cite{BW}. Let
\begin{equation}\label{Gadef}
\Ga = \La \times \{-1,1\} \times \{1,2\} 
,\end{equation}
and denote $\Xx=(x,\si,i) \in \Ga$. For $x \in \La$ and 
$\si \in \{-1,1\}$, the fields are defined as 
$\ps(x,\si,1) = \psq_\si(x)$ and $\ps(x,\si,2) = \ps_\si(x)$.
The antiperiodicity condition reads $\ps(x+\be e_\ta,\si,i ) = -\ps(x,\si,i)$
with $e_\ta$ the unit vector in $\ta$-direction. 
The Grassmann algebra generated by the $(\ps(\Xx))_{\Xx \in \Ga}$
is denoted by $\cA_\Ga[\ps]$. Given another set of Grassmann variables
$(\et (\Xx))_{\Xx \in \Ga}$, the Grassmann algebra generated by 
the $\ps$ and $\et$ is denoted by $\cA_\Ga[\ps,\et]$. Furthermore, denote
$\int_\Ga d\Xx\; F(\Xx) = \sum_{i=1}^2 \; \sum_{\si\in \{-1,1\}}\; 
\int_\La dx\; F(x,\si,i)$ and $\de_\Ga ((x,\si,i),(x',\si',i')) = \de_{ii'}\;\de_{\si\si'}\;\de_\La (x,x')$,
and define a bilinear form on $\cA_\Ga[\ps,\et]$ by 
\begin{equation}
\left( \psi, \eta\right)_\Ga = \ili_\Ga d\Xx \; \ps(\Xx)\; \et(\Xx )
= -\left(\eta,\psi\right)_\Ga
.\end{equation}
Then $S_2 = \half \left( \ps, \; A \; \ps \right)_\Ga$
where, for $\Xx=(x,\si,i)$ and $\Xx'=(x',\si',i')$,
\begin{equation}
 A(\Xx,\Xx') = \cases{ 0 & if $i=i'$ \cr
a(x,\si,x',\si') & if $i=1$ and $i'=2$ \cr
-a(x',\si',x,\si) & if $i=2$ and $i'=1$}
\end{equation}
with $a$ given by \Ref{aadef}. In other words, when written as 
a matrix in the index $i$, $A$ takes the form 
\begin{equation}
\left(\matrix{ 0 & a \cr  - a^T & 0 \cr}
\right)
\end{equation}
with $(a^T)(x,\si,x',\si') = a(x',\si',x,\si)$ denoting the 
transpose of $a$. Since $a$ is invertible, $A$ is invertible as well.

With this, $Z_{\SLa,\Nta}= \cN_\La \det a\; \tilde Z_{\La}$
where $\tilde Z_{\La}= \int d\mu_C (\ps) \E^{-\la S_4(\ps)}$,
where $C=A^{-1}$, and  $d\mu_C$ is the linear functional 
(`Grassmann Gaussian measure') defined by 
$d\mu_C (\ps) = \left(\det a\right)^{-1}\; 
D_\Ga \ps \; \E^{\half (\ps,A\ps)_\Ga}$.
The constant 
$\cN_\La \det a $ drops out of all correlation functions
and can therefore be omitted. The `measure' $d\mu_C$ is normalized, 
$\int d\mu_C (\ps) =1$,
and its characteristic function is 
\begin{equation}\label{charfu}
\int d\mu_C (\ps) \; \E^{(\et,\ps)_\Ga} = 
\E^{\half (\et,\; C\; \et)_\Ga}
.\end{equation}
All moments of $d\mu_C$ can be obtained by differentiating
\Ref{charfu} with respect to $\eta $ and setting $\eta=0$;
see also the next subsection. 

\subsection{The connected Green functions}
In the correspondence between the system, as defined by 
the Hamiltonian $H_\SLa$ and $\cF_\SLa$, to the Grassmann integral, 
I have so far only discussed the partition function itself. 
In the path integral representation of \Ref{Hamexp}, 
with a  polynomial observable $\cO(c,c^+)$, 
one simply gets a factor $\cO (\ps(0),\psq(0))$ in the Grassmann 
integral. 
The {\em $m$-point Green functions}
of the system determined by $C$ and $V$ are
\begin{equation}\label{Graexp}
\left\langle \pli_{k=1}^m \ps(\Xx_k) \right\rangle = 
\frac{1}{Z_\La} \int d\mu_C (\ps) \; 
\E^{-\la V(\ps)} \pli_{k=1}^m \ps(\Xx_k)
.\end{equation}
They determine the expected values of all polynomials
by linearity.

On a finite lattice, the limit $\Nta \to \infty$ of $\tilde Z_\La$ 
exists by the Lie product formula \Ref{Trotto}, and for 
$\la $ small enough (depending on $L$, $\Seps$, $\be$, and $\mu$), 
it is nonzero since the trace of the matrix $\E^{-\be(H_\SLa - \mu N_\SLa)}$
over the finite-dimensional space $\cF_\SLa$ is a continuous function 
of $\la$, which is nonzero at $\la =0$ by \Ref{Homnonz} and \Ref{freeZ}.
A similar argument applies to the numerator of \Ref{Hamexp}.
Thus the limits of numerator and denominator in \Ref{Hamexp}
as $\Nta\to \infty$ exist separately. Therefore one can take 
this limit in numerator and denominator through the same sequence,
i.e., take $\Nta$ to be the same in numerator and denominator.
It follows that with the special choice $\cO (\ps(0),\psq(0))$,
all expectation values in the Hamiltonian picture can be 
expressed as the limit $\Nta \to \infty$ of \Ref{Graexp},
with a special choice of the polynomial in the fields.
Thus the correlation functions given by \Ref{Graexp}
include as a special case the expectation values 
of polynomials in the creation and annihilation operators. 

Let $(\et(\Xx))_{\Xx \in \Ga}$ be a family of Grassmann generators. 
The partition function with source terms is 
$Z_\Ga(\et) = \int d\mu_C (\ps) \; 
\E^{-\la V(\ps) \; + \; (\et,\ps)_\Ga}$.
Let $\frac{\de}{\de \et (\Xx)} = {\Teps}^{-1} \; 
\Seps^{-d} \; \frac{\del}{\del \et(\Xx)}$,
then 
\begin{equation}
\left\langle \pli_{k=1}^n \ps(\Xx_k) \right\rangle = 
\frac{1}{Z_\Ga(0)}
\left\lbrack \pli_{k=1}^n  \frac{\de}{\de \et(\Xx_k)}\;
Z_\Ga (\et) \right\rbrack_{\et=0}
.\end{equation}
Thus, if one knows $Z_\Ga(\et)$ one can derive all  
correlation functions. 

It is convenient to study the connected correlation functions, 
defined as 
\begin{equation}
\left\langle \pli_{k=1}^n \ps(\Xx_k) \right\rangle_c = 
\left\lbrack \pli_{k=1}^n  \frac{\de}{\de \et(\Xx_k)}\;
\log Z_\Ga (\et) \right\rbrack_{\et=0}
\end{equation}
instead. Since $Z_\Ga (\et)$ is the exponential of $\log Z_\Ga (\et)$,
one can reconstruct all correlation functions from the connected ones. 

It is even more convenient to transform the sources $\et$, to get 
the amputated connected Green functions. They are generated by 
\begin{equation}
\cG_{\rm eff} (\ch) = \log \int d\mu_C (\ps) \; \E^{-\la V(\ps+\ch) }
\end{equation}
A shift in the measure shows that 
$\cG_{\rm eff} (\ch) = \half \left( \ch, C^{-1} \ch\right)_\Ga +
\log Z_\Ga \left( C^{-1} \ch\right)$.
so that the study of $\cG_{\rm eff} $ is equivalent to 
that of $\log Z_\Ga$.

The {\em selfenergy} $\Si(p)$ 
is defined as the one-particle irreducible part
of the two-point function. In terms of the connected amputated
two-point Green function $G_2$, which is the coefficient of the
quadratic part
(in $\ps$) of the effective action  $\cG_{\rm eff}(\ps)$, 
it is $\Si (p) = G_2(p) (1-C G_2 (p))^{-1}$.

\subsection{Criteria for Fermi liquid behaviour}
\label{thepurpose}
In the following I give a definition of Fermi liquid behaviour
which is linked to the question of convergence of the expansion in the 
coupling constant $\la$, and I discuss in some detail 
the physical motivation for this definition, 
the results that have been proven in this direction,
and its relation to other notions of FL behaviour. 

In most many-fermion 
models, one cannot expect the expansion in $\la$ to converge uniformly 
in the temperature, not even after renormalization.
In particular, the Cooper instability produces a superconducting
ground state, and thus a nonanalyticity in $\la$,
if the temperature is low enough. 
This happens even if the initial interaction is repulsive
\cite{KL,FKST}. Nesting instabilities can produce other types of
symmetry breaking, such as antiferromagnetic ordering, which 
may compete or coexist with superconductivity, but the 
conditions I posed, in particular the curvature of the Fermi surface,
remove these instabilities at low temperatures 
(which temperatures are `low' depends on the scale $\epz$). 

Let the {\em skeleton Green functions} be defined as the 
connected amputated $m$-point correlation functions 
where self-energy insertions are left out. 
These functions are the solution of a natural truncation of
the renormalization group equation; they are defined precisely 
in Section \ref{sect6}. In particular, the skeleton selfenergy 
$\Sisk$ is the second Legendre transform of the two-point function.

\begin{definition}\label{FLDef}
The $d$-dimensional many-fermion system with dispersion relation $E$
and interaction $V$ shows (equilibrium) Fermi liquid behaviour
if the thermodynamic limit of the Green functions exists
for $|\la| < \la_0(\be)$, and 
if there are constants $\coco,\coce,\cocz>0$ (independent of $\be$
and $\la$), such that the following holds. The perturbation 
expansion for the skeleton Green functions converges for all 
$(\la,\be)$ with $\abs{\la} \log \be < \coco$, and 
for all $(\la,\be)$ with $\abs{\la} \log \be \le \sfrac{\coco}{2}$,
the skeleton self-energy $\Sisk :\bR \times \cB \to \bC$ 
satisfies the regularity conditions 
\begin{enumerate}
\item $\Sisk$ is twice differentiable in $p$ and 
\begin{equation}\label{zwiff}
\max\limits_{|\al|=2}\norm{\del^\al\Sisk}_\infty \le \coce
\end{equation}
\item the restriction to the Fermi surface
$\Sisk\vert_{\{0\}\times \FS}\in C^\kz(\FS,\bR)$,
and 
\begin{equation}
\max\limits_{|\al|=\kz}\norm{\del^\al\Sisk}_\infty \le \cocz
.\end{equation}
Here $\kz>d$ is the 
degree of differentiability of the dispersion relation $E$
(given in Section \ref{hyposuse}).
\end{enumerate} 
\end{definition}
Nothing is special about the factor $\sfrac{1}{2}$ in 
the condition $\abs{\la} \log \be \le \sfrac{\coco}{2}$.
One could instead also have taken any fixed compact subset of 
$\{z: |z|<\coco\}$. The derivatives mean, when taken in 
$p_0$, a difference $\sfrac{\be}{2\pi} (\Sisk (p_0+\sfrac{2\pi}{\be},\Sp)-
\Sisk(p_0,\Sp))$. The maximum runs over all multiindices $\al\in \bN_0^{d+1}$.

This definition only concerns equilibrium properties of
Fermi liquid behaviour; it does not touch phenomena like
zero sound, which require an analysis of the response to 
perturbations that depend on real time. 
It is natural in that it defines
a Fermi liquid above the critical temperature for superconductance: 
at a given $\la$, the value of $T$ for which the convergence 
breaks down is $T_c \propto \E^{-\coco/|\la|}$, which is 
the usual BCS formula. Convergence of perturbation 
theory above $T_c$ implies that 
the usual Fermi liquid formulas are valid there.
Convergence is stated only for skeleton quantities
because that is all one can show. This convergence
and the regularity properties of the self-energy, imply that 
the exact Green functions (no restriction to skeletons) 
are continuous in $\la$, that the exact selfenergy $\Si$ is
$C^2$ in $\la$ and $p$, and that $\Si$ obeys a bound similar to 
\Ref{zwiff}. The Green functions are not analytic in $\la$
because otherwise already the unrenormalized expansion,
which diverges termwise, would converge.
The regularity properties $(1)$ and $(2)$
ensure that the exact 
Green functions can be reconstructed from the skeleton Green functions
by renormalization.
The usual skeleton expansion argument \cite{skelmist}, 
where finiteness only of the skeleton
self-energy, but not of its derivatives, is shown, is insufficient
to do that; one has to prove regularity properties $(1)$ and $(2)$. 
This was discussed in detail in \cite{FST2}, see also 
Section \ref{sect7}.

The condition $\kz>d$ is necessary to make the regularized propagator
summable in position space. It is required in the proof of 
Lemma \ref{tdlemma} (and in the proofs in \cite{FMRT},
only that there the dispersion relation was taken $C^\infty$). 
In the absence of level crossing, the free
dispersion relation $E(\Sk)$ is usually even real analytic in $\Sk$.
However, when reconstructing the exact Green functions from the 
skeleton Green functions, one needs regularity of the dispersion relation
of the interacting system, and thus regularity property $(2)$, 
which is rather hard to verify even in perturbation theory. 
Thus it is desirable to take the smallest possible $\kz>d$. 

Because $\Si$ obeys a bound similar to \Ref{zwiff}, 
one can do the usual first-order Taylor expansion in the momenta
to get 
\begin{equation}
\Si(p) = p_0 (\del_0\Si)(0,{\bf P}(\Sp)) + 
(\Sp - {\bf P}(\Sp)) \cdot \nabla \Si(0,{\bf P}(\Sp))
+ \tilde\Si(p)
\end{equation}
from which one obtains a finite wave function renormalization
\begin{equation}
Z(\Sp) = 1+i (\del_0\Si)(0,{\bf P}(\Sp))
\end{equation}
and a finite correction to the Fermi velocity, and the Taylor remainder
vanishes quadratically in the distance 
of the momentum $(p_0,\Sp)$ to its projection $(0,{\bf P}(\Sp))$
to the Fermi surface. 

This property distinguishes
Fermi liquids from other possible states of the many-fermion system,
such as Luttinger liquids: In one dimension
(where `Luttinger liquid behaviour' has been proven \cite{BG,BGPS,BM}),
the second derivative of even the second order skeleton
 selfenergy grows like 
$\be$ for large $\be$ and thus violates the condition that 
the second derivative should be bounded independently of $\be$
for $|\la|\log\be \le \sfrac{\coco}{2}$.
Note that this distinction can only be made if $\be$ is allowed to vary;
at fixed $\be$, the requirement that something is bounded independently 
of $\be$ is trivial. This is the reason why a whole range of 
values of $\be$ and $\la$ is included in Definition \ref{FLDef}. 

A full proof that the models obeying the hypotheses stated in 
\ref{hyposuse} are Fermi liquids in the sense of Definition 
\ref{FLDef} is not within the scope of this paper,
but several ingredients for such a proof are 
already in place.

The convergence of the skeleton expansion follows for $d=2$ 
spatial dimensions from a modification of the method in 
\cite{FMRT}. The required modification is to put in the 
four-point functions. This is not difficult because 
at positive temperature, these functions have no singularities,
but are bounded (in momentum space) by $\const \log\be$.
If the recursion \Ref{untrecu} has an exponentially bounded
solution, the method developed here can also be extended 
to prove this convergence. 

Regularity property $(1)$ is proven in Section \ref{sect6}.
Regularity property $(2)$ was proven in two dimensions in 
perturbation theory for $\kz=2+h$, $h< \half$
\cite{FST1,FST2,FST3}. The methods 
developed in Section \ref{sect6} provide a simplification
of most of these proofs, and can also be extended to give a full 
proof of $(2)$. 

There is one case in which $(2)$ becomes trivial:
for the Jellium dispersion relation $\Sk^2/2m-\mu$, 
$\Si\vert_\FS$ is a constant, hence $C^\infty$. 
In that case, it is possible to establish 
Fermi liquid behaviour
by a combination of the determinant bound of Section \ref{sect4},
the sector decomposition of \cite{FMRT}, and the overlapping loop 
method of \cite{FST1}, without the major complications of the 
regularity proofs.  

A natural question is  
if there are criteria independent of temperature that
can be applied also in the zero temperature limit for `Fermi liquid 
behaviour'. For the class of models specified in Section \ref{hyposuse},
the simplest criterion is that the \nonRPA\  skeleton 
Green functions are analytic in the coupling constant $\la$, 
that the \nonRPA\  skeleton selfenergy $\Si^{(N)}$ is $C^1$,
and that $\Si^{(N)}\vert_\FS$ is $C^\kz$, with bounds uniform in $\be$. 
The \nonRPA\  skeleton Green functions are obtained by removing
all ladder contributions (defined in Section 
\ref{sect6}) to the Green functions.
The regularity implies that the Fermi velocity and the wave function
renormalization are finite uniformly in $\be$, and 
even at zero temperature, for $d\ge 2$
(which is the usual criterion for Fermi liquids),
whereas they still diverge 
in one dimension as $\be \to \infty$ \cite{BG,BGPS,BM}. 

The Fermi liquid criterion given in Definition \ref{FLDef} 
is more natural than the one using the \nonRPA\  Green functions:
if the regularity conditions $(1)$ and $(2)$ of Definition \ref{FLDef}
hold, the transition from 
the exact Green functions to the skeleton Green functions is a matter
of convenience, but the replacement of the skeleton Green functions
by the \nonRPA\  skeleton Green functions changes the model 
drastically because it removes superconductivity.
Evidently, a definition referring to a modified model is not
as natural. 
Morover, as mentioned above,
Fermi liquid behaviour is observed only above the critical
temperature for superconductance anyway. 
In Section \ref{sect6}, I define the \nonRPA\  skeleton functions precisely
and prove the above statements about $\Si^{(N)}$
uniformly in the temperature in perturbation theory. 
A nonperturbative proof requires the same extensions of the analysis
as mentioned in the previous paragraph.

In \cite{FKLT}, an 
{\em asymmetric} model, in which the symmetry $E(\Sk) = E(-\Sk)$
does not hold, was introduced, and a proof was outlined 
that such models are Fermi liquids down to zero temperature. 
The asymmetry of the Fermi surface removes the Cooper instability 
at zero relative momentum $q$ of the Cooper pair, i.e., 
it implies that the four-point function has no singularity at
relative momentum $q=0$ (which is where the usual Cooper pairing comes from).
The regularity conditions on the self-energy, which are
crucial for Fermi liquid behaviour, were not verified in \cite{FKLT}.
Doing this is quite a bit harder than 
in the $(\Sk \to -\Sk)$-symmetric case \cite{FST2,FST3}.
At zero temperature, regularity property $(1)$ has to be replaced by 
$(1'):\quad\Si \in C^{2-\ep}$
because the selfenergy is not $C^2$ at zero temperature
(the second derivative grows as a power of $\log \be$; 
for a detailed discussion
of these problems see \cite{FST2}). This modified regularity property $(1')$
and $(2)$ were proven in perturbation theory for a general class 
of two-dimensional models with a strictly convex Fermi surface,
which includes the non-$(\Sk \to -\Sk)$-symmetric Fermi surfaces, 
in \cite{FST1}--\cite{FST3}. More precise conditions on the dispersion
relation that imply absence of the Cooper instability also
at nonzero relative momentum $q$  were also formulated 
in \cite{FST2}. It is not sufficient just to have a $\Sk \to -\Sk$
nonsymmetric surface to achieve that; 
one also needs that the curvature at a point on
the Fermi surface and at its antipode differ except at finitely
many points (for details, see Hypothesis $(H4')$ of \cite{FST2}
and the geometrical discussion in Appendix C of \cite{FST2}).

\section{The renormalization group equation}\label{sect3}
In this section I derive the continuous RGE 
for fermionic models. I first derive it for the generating 
function, and then turn to the component form which is obtained by 
expanding the effective action in Wick ordered monomials
of the fields. 
Let $\Ga$ be a finite set, for a function $\Xx \mapsto F(\Xx)$ 
from $\Ga$ to any linear space let
$\int_\Ga d\Xx\; F(\Xx) = \veps_{\Ga} \sum_{\Xx\in \Ga} F(\Xx)$, where 
$\veps_\Ga>0$ is a constant, let $\de(\Xx,\Xx')=\veps_\Ga^{-1} \de_{\Xx\Xx'}$,
and define the bilinear form
$(f,g) = \int_\Ga d\Xx\; f(\Xx) g(\Xx)$.
Let $\cA$ be the finite-dimensional Grassmann algebra generated by 
the generators $(\ps(\Xx),\ch(\Xx),\et(\Xx))_{\Xx \in \Ga}$ and
let $\frac{\de}{\de \ps(\Xx)}$
be the fermionic derivative normalized such that 
$\frac{\de}{\de \ps(\Xx)} \ps(\Xx') = \de(\Xx,\Xx')$; 
recall that the fermionic derivatives anticommute.

\subsection{The RGE for the generating function}
\label{gRGEsuse}
For $t \ge 0$ let $C_t$ be an invertible, antisymmetric linear 
operator acting on functions defined on $\Ga$, i.e.\ 
$(C_t f)(\Xx) = \int_\Ga d\Xx C_t (\Xx,\Xx') f(\Xx')$
with 
\begin{equation}
C_t (\Xx',\Xx) = - C_t (\Xx,\Xx')
.\end{equation}
Let $C_t$ be continuously differentiable in $t$; denote 
$\frac{\del C_t}{\del t} = \dot C_t$. Let $d\mu_{C_t}$ be the 
linear functional (Grassmann Gaussian measure) with characteristic
function 
\begin{equation}\label{charfu2}
\int d\mu_{C_t} (\ps) \; \E^{(\et,\ps)_\Ga} = 
\E^{\half (\et,\; C_t\; \et)_\Ga}
.\end{equation}
The integrals of arbitrary monomials are obtained from 
this formula by taking derivatives with respect to $\et$. 
The measure is normalized: $\int d\mu_{C_t} (\psi) =1$.

Let $V(\ps)\in \cA$ have no constant part, $\la \in \bC$, and
$\Geff{0}{\ps}=\la V(\ps) $. 
The {\em effective action} at $t>0$ is 
\begin{equation}\label{Geffdef}
\Geff{t}{\ps} = \log \int d\mu_{C_t} (\ch) \; 
\E^{\Geff{0}{\ch+\ps}}
.\end{equation}
Because the measure is normalized, 
$\cG(t,\ps)$ is a well-defined formal power series in $\la$. 
By the nilpotency of the Grassmann variables,
$\E^{\cG(t,\ps)}$ is a polynomial in $\la$
(the degree of which grows with $\la$).
Thus $\cG(t,\ps) $ is analytic in $\la$ for $|\la| < \la_0(\Ga)$.

\begin{proposition}\label{funRGE}
Let 
\begin{equation}\label{LpCtdef}
\Lp{C_t} =\half \left(\frac{\de}{\de \ps}, C_t \frac{\de}{\de \ps}\right)_\Ga
=\half \ili_\Ga d\Xx \ili_\Ga d\Xx' \; \frac{\de}{\de \ps (\Xx)}\;
C_t(\Xx,\Xx') \frac{\de}{\de \ps(\Xx')}\; 
.\end{equation}
Then
\begin{equation}\label{Geffdif}
\frac{\del}{\del t}\; \E^{\Geff{t}{\ps}}=
\dot \Lp{C_t} \E^{\Geff{t}{\ps}}
\end{equation}
and 
\begin{equation}\label{Geff3}
\E^{\Geff{t}{\ps}}  = \E^{\Lp{C_t}} \; \E^{\Geff{0}{\ps}} 
.\end{equation}
If $\Geff{0}{\ps}$ is an element of the even subalgebra,
then for all $t > 0$, $\Geff{t}{\ps}$ is an element of the even subalgebra,
and it satisfies the {\em renormalization group equation}
\begin{equation}\label{RGE}
\frac{\del}{\del t}\Geff{t}{\ps} = 
\dot \Lp{C_t} \Geff{t}{\ps} + 
\half\left( \frac{\de \Geff{t}{\ps}}{\de \ps}, 
\dot C_t \frac{\de\Geff{t}{\ps}}{\de \ps}\right)
.\end{equation}
\end{proposition}

\begin{proof}
For any $F(\ps) \in \cA$, define $F(\frac{\de}{\de \et}) $
by replacing every factor $\ps(\Xx)$ by $\frac{\de}{\de \et(\Xx)}$
in the polynomial expression for $F$. Then 
$F(\ps) = \lbrack F(\sfrac{\de}{\de \et})\;  \E^{(\et,\ps)_\Ga}
\rbrack_{\et =0}$
(the derivatives $\sfrac{\de}{\de \et}$ also generate a finite-dimensional
Grassmann algebra, so the expansion for $F$ terminates at some 
power). Since Grassmann integration is a continuous operation,
and by \Ref{charfu2},
\begin{eqnarray}\label{Geff2}
\E^{\Geff{t}{\ps}} &=&  \left\lbrack
\E^{\Geff{0}{\frac{\de}{\de \et}}} \;
\int d\mu_{C_t} (\ch) \; \E^{(\et,\ch+\ps)_\Ga}
\right\rbrack_{\et =0}
\nonu \\
& = & \left\lbrack
\E^{\Geff{0}{\frac{\de}{\de \et}}} \;
\E^{\half(\et, C_t\et)_\Ga} \; \E^{(\et,\ps)_\Ga}
\right\rbrack_{\et =0}
.\end{eqnarray}
 
For any formal power series
$f(z) = \sum f_k x^k$, 
\begin{equation}
f(\Lp{C_t} ) \; \E^{(\et,\ps)_\Ga} = f\left(\frac{1}{2} 
(\et,\; C_t\; \et)_\Ga\right)
\; \E^{(\et,\ps)_\Ga}
,\end{equation}
so $\E^{\frac{1}{2} (\et,\; C_t\; \et)_\Ga + (\et,\ps)_\Ga}=
\E^{\Lp{C_t}} \; \E^{(\et,\ps)_\Ga}$.
Since $\Lp{C_t} $ is bilinear in the derivatives, it commutes with
all factors that depend only on $\et$ and can be taken out in front
in \Ref{Geff2}. This implies \Ref{Geff3}. Since $\Lp{C_t} $
also commutes with $\dot\Lp{C_t} $, \Ref{Geffdif} follows.

If $\Geff{0}{\ps}$ is an element of the even subalgebra,
the same holds for $\Geff{t}{\ps}$ by \Ref{Geff3}, 
since every application of $\Lp{C_t}$ removes two fields. 
Thus performing the derivatives with respect to $\ps$
gives \Ref{RGE}.
\end{proof}

\subsection{The component RGE in position space}

Let $V(\ps)$ be an element of the even subalgebra. 
The effective action has the expansion
\begin{equation}
\Geff{t}{\ps} = \sli_{r=1}^\infty \la^r \cG_r(t,\ps)
\end{equation}
with $\cG_r(t,\ps)$ polynomials in the Grassmann algebra. 
As explained in Section \ref{gRGEsuse}, this expansion 
converges for $\Ga$ finite, 
but the radius of convergence $\la_0$ depends on $\Ga$
and $C_t$.
For the models discussed in Section \ref{sect2}, 
this means that $\la_0$ goes to zero in the limit 
$L\to \infty$ and $\be \to \infty$.

Assume that $C = \lim_{t\to \infty} C_t$ exists and let 
$\CW{t} = C- C_t$ 
so that $\dot C_t = -\dot \CW{t}$. The application 
will be that $C$ is the covariance of the model and $C_t$ is 
part of it, so that in $\Geff{t}{\ps}$, part of the fields
have been integrated over. $\CW{t}$ is then the covariance of 
the unintegrated fields. 

I expand the polynomial $\cG_r(t,\ps) \in \cA$ in the 
basis for the Grassmann algebra given by the 
{\em Wick ordered monomials} $\WO{\CW{t}} (\ps(\Xx_1)\ldots \ps(\Xx_p))$,
\begin{equation}\label{expaWO}
\cG_r(t,\ps) = \sli_{m=0}^{\bar m(r)}\; \ili_{\Ga^m} d\ul{\Xx} \;
\Grmt{\ul{\Xx} } \;
\WO{\CW{t}} \left(\pli_{k=1}^m \ps(\Xx_k)\right)
,\end{equation}
where $\Grmt{\Xx_1, \ldots ,\Xx_m} $ is the {\em connected, amputated
$m$--point Green function} and $\ul{\Xx} = (\Xx_1, \ldots ,\Xx_m)$. 
Details about Wick ordering
are provided in Appendix \ref{appWick}.
A short formula is  
\begin{equation}\label{WiLa}
\WO{\CW{t}} (\ps(\Xx_1)\ldots \ps(\Xx_p)) =
\E^{-\Lp{\CW{t}}} \; \ps(\Xx_1)\ldots \ps(\Xx_p)
.\end{equation}
I use the symbol $\WO{\CW{t}} (\ps(\Xx_1)\ldots \ps(\Xx_p))$ rather than 
$:\ps(\Xx_1)\ldots \ps(\Xx_p):$ to indicate clearly with respect 
to which covariance Wick ordering is done, because this will be 
important.  

The $\Grmt{\Xx_1, \ldots ,\Xx_m}$ are assumed to be 
totally antisymmetric, that is, for all $\pi \in \Perm{m}$,
$\Grmt{\Xx_{\pi(1)}, \ldots ,\Xx_{\pi(m)}} = 
\Sign(\pi) \Grmt{\Xx_1, \ldots ,\Xx_m}$,
because any part of $G$ that is not antisymmetric would cancel
in \Ref{expaWO}. 

Application of $\frac{\del}{\del t}$ to \Ref{expaWO} gives a sum of
two terms since two factors depend on $t$. By \Ref{WiLa},
\begin{eqnarray}
\frac{\del}{\del t}
\WO{\CW{t}} (\ps(\Xx_1)\ldots \ps(\Xx_p)) &=&
-\dot \Lp{\CW{t}}\WO{\CW{t}} (\ps(\Xx_1)\ldots \ps(\Xx_p)) 
\nonu\\
&=& \dot \Lp{C_t}\WO{\CW{t}} (\ps(\Xx_1)\ldots \ps(\Xx_p))
.\end{eqnarray}
When multiplied by $\Grmt{\Xx_1, \ldots ,\Xx_m}$
and integrated over $\Xx_1, \ldots ,\Xx_m$, this gives 
$\dot \Lp{C_t} \Geff{t}{\ps}$. 
Thus the term linear in $\cG$ drops out of \Ref{RGE} by Wick ordering 
with respect to $\CW{t}$, and \Ref{RGE} now reads
\begin{equation}
\sli_{m=0}^{\bar m(r)} \;\ili_{\Ga^m} d\ul{\Xx} \;
\WO{\CW{t}} \left(\pli_{k=1}^m \ps(\Xx_k)\right)\;
\frac{\del}{\del t} \Grmt{\ul{\Xx} } =
\half \cQ_r(t,\ps)
\end{equation}
where $\cQ_r(t,\ps)$ is defined by 
\begin{equation}\label{toWm}
\left( \frac{\de \Geff{t}{\ps}}{\de \ps}, 
C_t \frac{\de\Geff{t}{\ps}}{\de \ps}\right)
=  \sli_{\la=1}^\infty \la^r \cQ_r(t,\ps)
\end{equation}
Being an element of the Grassmann algebra, $\cQ_r(t,\ps)$ 
has the representation
\begin{equation}\label{beingQ}
\cQ_r(t,\ps) = \sli_{m=0}^{\bar m(r)} \;\ili_{\Ga^m} d\ul{\Xx} \;
\Qrmt{\ul{\Xx} } \;
\WO{\CW{t}} \left(\pli_{k=1}^m \ps(\Xx_k)\right)
.\end{equation}
To obtain the $\Qrmt{\ul{\Xx} } $, one has to 
rewrite the product of the two Wick monomials in \Ref{toWm}.
This is done in Appendix \ref{ReWick}. The result is

\begin{proposition}\label{Qmrtvors}
$\QQ{m}{1}{t}{\ul{\Xx}} =0$, and for $r\ge 2$,
\begin{eqnarray}\label{Qrmtdef}
\Qrmt{\ul{\Xx}} & = & \int d\ka_{mr} 
\;  \ili_{\Ga^\zil} d\ul{\Xv} \;\ili_{\Ga^\zil} d\ul{\Xw} \; 
\left( -\frac{\del}{\del t} \det \cDt{\zil}(\ul{\Xv},\ul{\Xw})\right)
\nonu \\
& &  \Grmto{\ul{\Xx_1},\ul{\Xv} } \;
\Grmtt{\ul{\tilde{\Xw}},\ul{\Xx_2} } 
,\end{eqnarray}
where $\int d\ka_{mr}$ stands for the sum
\begin{eqnarray}
\int d\ka_{mr}&& \kern -26pt%
(r_1,m_1,r_2,m_2,\zil) \; F(r_1,m_1,r_2,m_2,\zil) 
\nonu \\
&=& 
\sli_{r_1,r_2\ge 1 \atop r_1+r_2=r} \; 
\sli_{(m_1,m_2,\zil) \in \cM_{r_1r_2m}}
\ka_{m_1m_2\zil} \;F(r_1,m_1,r_2,m_2,\zil)
\end{eqnarray}
with positive weights
$\ka_{m_1m_2\zil} = {m_1 \choose \zil} \; {m_2 \choose \zil}$,
so that $d\ka_{mr}$ is a positive measure. $\cM_{r_1r_2m}$
is the set of $(m_1,m_2,\zil)$ such that $\zil \ge 1$, 
$1 \le m_1 \le \bar m(r_1)$, $ 1 \le m_2 \le \bar m(r_2)$, 
$m_1+m_2 = m+ 2\zil$, and $m_1$ and $m_2$ are even.
$\ul{\Xx_1}= (\Xx_1,\ldots,\Xx_{m_1-\zil})$,
$\ul{\Xx_2}= (\Xx_{m_1-\zil+1},\ldots,\Xx_{m})$,
$\ul{\Xv} =  (\Xv_1,\ldots, \Xv_\zil)$,
$\ul{\Xw} =  (\Xw_1,\ldots, \Xw_\zil)$, and
$\ul{\tilde \Xw}=(\Xw_\zil, \ldots, \Xw_1 )$,
and $\cDt{\zil}(\ul{\Xv} ,\ul{\Xw})$ is the $\zil\times\zil$ matrix
$(\cDt{\zil}(\ul{\Xv} ,\ul{\Xw}))_{kl} =
\CW{t} (\Xv_k,\Xw_l)$. 
\end{proposition}
Comparison of the coefficients gives the {\em component form of the 
RGE} 
\begin{equation}\label{comRGE}
\frac{\del}{\del t} \Grmt{\Xx_1, \ldots ,\Xx_m} =
\half \bA_m \Qrmt{\Xx_1, \ldots ,\Xx_m}
\end{equation}
where $\bA_m$ is the antisymmetrization operator
\begin{equation}
(\bA_m f) (\Xx_1, \ldots ,\Xx_m) = 
\frac{1}{m!} \sli_{\pi \in \Perm{m}} \Sign(\pi)
f(\Xx_{\pi(1)}, \ldots ,\Xx_{\pi(m)})
.\end{equation}
The important feature of \Ref{Qrmtdef} is that
the determinant of the propagators appears in this equation. 
The Gram bound for this determinant 
improves the combinatorics by a factorial.
I now discuss the graphical interpretation
of the equation and the determinant, 
to motivate why this improvement can be regarded as a `planarization'
of the graphs.

\subsection{The graphical interpretation}
\label{plansuse}

\begin{figure}
\epsfxsize=5in
\centerline{\epsffile{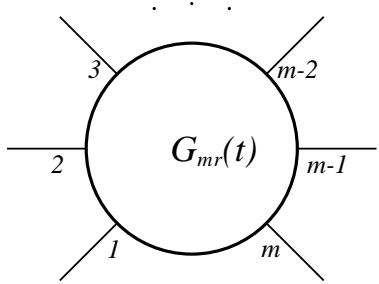}}
\caption{The vertex corresponding to $\GGrm{t}$}
\label{fig1}
\end{figure}

\begin{figure}
\epsfxsize=4in
\centerline{\epsffile{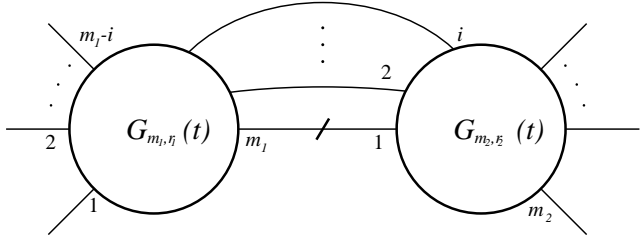}}
\caption{A graph contributing to the right hand side of the RGE}
\label{fig2}
\end{figure}

The component form of the RGE has a straighforward 
graphical interpretation. If one associates the vertex drawn in 
Figure \ref{fig1} to $\Grmt{\Xx_1,\ldots,\Xx_m}$, adopts the 
convention that the variables occurring on the internal lines
of a graph are integrated, and writes out the determinant as
\begin{equation}
\det \cDt{\zil}(\ul{\Xv} ,\ul{\Xw}) =
\sli_{\pi \in \Perm{\zil}} \Sign(\pi) 
\pli_{k=1}^\zil \CW{t} (\Xv_k,\Xw_{\pi(k)})
\end{equation}
the right hand side of \Ref{Qrmtdef} appears as the signed sum 
over graphs with two vertices $\GGrmo{t}$ and $\GGrmt{t}$,
obtained by joining leg number $m_1-\zil+k$ of vertex $1$ with leg 
number $\zil-\pi(k)+1$ of vertex 2, for all $k \in \nat{i}$. 
The graph for $\pi = \mbox{id}$ is drawn in Figure \ref{fig2}. 
The expansion in terms of Feynman graphs is generated 
by iteration of the equivalent integral equation
\begin{equation}\label{icomRGE}
\Grmt{\ul{\Xx}} = \Grmz{\ul{\Xx}} + 
\half \bA_m \ili_0^t dt \; \Qrmt{\ul{\Xx}}
\end{equation}
with the initial condition $\Grmz{\ul{\Xx}}=-\de_{r1}V_m(\ul{\Xx})$, 
where $V_m(\ul{\Xx})$ is the coefficient of $\WO{C} (\ps(\Xx_1)\ldots
\ps(\Xx_m))$ in the original interaction. It is evident from 
Figure \ref{fig2} that only connected graphs contribute to this
sum. 

Note that the only planar graphs appearing in this sum are from 
$\pi(j)=j+k$ mod $\zil$ for $k \in \natz{\zil-1}$, and that for 
$k \ne 0$, these permutations produce planar graphs only if 
$\zil=m_1$ or $\zil=m_2$.  
For all other permutations, the graphs arising are nonplanar. 
For bosons, the determinant is replaced by a permanent, and
one can permute the integration variables 
so that the derivative of the permanent 
(and hence the sum over permutations)
gets replaced by 
\begin{equation}\label{plapla}
\zil\;\zil ! \; \pli_{k=2}^\zil D_t (\Xv_k,\Xw_k)\; 
\frac{\del}{\del t} D_t (\Xv_1,\Xw_1)
.\end{equation}
so that the planar graph drawn in Figure \ref{fig2}
is the only one contributing to the right hand side.
Thus this factor $\zil!$ distinguishes between the combinatorics
of the exact theory and a `planarized' theory, 
in which $\zil!$ is replaced by $\zil^2$
(the second factor $\zil$ comes from doing the derivative 
in the determinant, see Section \ref{sect4}). 
The `planarized' theory does contain more than the sum 
over all planar graphs because of the binomial factors
(the antisymmetrization operation $\bA_m$ does not change the combinatorics
because it contains an explicit factor $1/m!$). 
In the next section, I  
bound the determinant by $\const^\zil$ and thereby reduce the 
combinatorics of the fermionic theory to that of the
planarized theory. 

\section{Fermionic sign cancellations}\label{sect4}

\subsection{The determinant bound}\label{detbouss}

\Ref{Qrmtdef} already suggests that a determinant bound
similar to the one used in Lemma 1 of \cite{FMRT} can
be applied to the RGE. Before applying this bound, the derivative 
with respect to $t$ has to be performed, and some factors need
to be arranged to avoid the factor $\abs{\La}$ that appeared in 
\cite{FMRT}. The reason it does not appear here is that 
only connected graphs contribute to the effective action
(whereas the partition function itself was bounded in 
Lemma 1 of \cite{FMRT}).
In the RGE, this is very easily seen without 
a reference to graphs.

Since the determinant is multilinear in the columns of the matrix,
the derivative with respect to $t$ produces a sum of terms where
every column gets differentiated. Expanding along each differentiated 
column gives 
\begin{equation}
\frac{\del}{\del t} \det \cDt{\zil}(\ul{\Xv},\ul{\Xw})
=
\sli_{l=1}^\zil \sli_{l'=1}^\zil (-1)^{l+l'}
\dot\CW{t} (V_l,W_{l'})
\det \cDt{\zil-1}(\ul{\Xv}^{(l)},\ul{\Xw}^{(l')})
\end{equation}
with 
$\ul{\Xv}^{(l)} = (\Xv_1, \ldots, \Xv_{l-1},\Xv_{l+1},\ldots,\Xv_\zil)$
and a similar expression for $\ul{\Xw}^{(l')}$.
The sign is cancelled by rearranging 
\begin{eqnarray}
\Grmto{\ul{\Xx}_1,\ul{\Xv}} &=& 
(-1)^{i-l} \Grmto{\ul{\Xx}_1,\ul{\Xv}^{(l)},V_l} \nonu \\ 
\Grmtt{\ul{\tilde\Xw},\ul{\Xx}_2} &=&
(-1)^{\zil-l'} \Grmtt{\Xw_{l'},\ul{\tilde\Xw}^{(l')},\ul{\Xx}_2}
.\end{eqnarray}
Upon renaming of the integration variables, the summand becomes 
independent of $l$ and $l'$, so the sum gives a factor $\zil^2$. 
Thus 
\begin{eqnarray}
\Qrmt{\ul{\Xx}} &=& \int d\ka_{mr} \; \zil^2 \int d\Xv\; d\Xw\; 
\dot \CW{t} (\Xv,\Xw) \int d\ul{\Xy} \int d\ul{\Xz} \\ 
&& \det \cDt{\zil-1}(\ul{\Xy},\ul{\Xz}) 
\Grmto{\ul{\Xx}_1,\ul{\Xy},\Xv}
\Grmtt{\Xw,\ul{\tilde\Xz},\ul{\Xx}_2} \nonu
.\end{eqnarray}
Let $\Norm{\,\cdot\,}$ be the norm \cite{GK}
\begin{equation}
\Norm{F_m} = \max\limits_{p \in \nat{m}} \;
\sup\limits_{\Xx_p}
\int \pli_{q=1\atop q\ne p}^m d\Xx_q \;
\abs{F_m(\Xx_1,\ldots,\Xx_m)}.
\end{equation}

\begin{lemma} \label{unidet} 
Assume that 
\begin{equation}\label{ati}
\sup\limits_{\ul{\Xy},\ul{\Xz}} \abs{
\det \cDt{\zil-1}(\ul{\Xy},\ul{\Xz})} \le A_{\zil-1} (t)
.\end{equation}
Then
\begin{equation}
\Norm{\QQrm{t}} \le \int d\ka_{mr} \zil^2 A_{\zil-1}(t) \;
\Norm{\dot\CW{t}} \;\Norm{\GGrmo{t}}\; \Norm{\GGrmt{t}}
.\end{equation}
\end{lemma}

\begin{proof}
Let $p \in \nat{m}$. Without loss of generality, let
$p \le m_1-\zil$, so that $\Xx_p$ is a component of $\ul{\Xx}_1$
(the other case is similar by the symmetry of the sum for $\QQrm{t}$ in 
$m_1$ and $m_2$). By \Ref{ati},
\begin{equation}
\Norm{\QQrm{t}} \le \int d\ka_{mr} \zil^2 A_{\zil-1}(t) \;
\sup\limits_{\Xx_p}
\int \pli_{q=1\atop q\ne p}^{m_1-\zil} d\Xx_q 
\int d\Xv \; \ph(V,\ul{X}_1) \; \ch(V) 
\end{equation}
with 
\begin{equation}
\ph(V,\ul{X}_1) =
\int d\ul{\Xy} \; \abs{\Grmto{\ul{\Xx}_1,\ul{\Xy},V}}
\end{equation}
and 
\begin{equation}
\ch(V) = \int d\Xw\; d\ul{\Xz}\; d\ul{\Xx}_2\;
\abs{\dot\CW{t}(\Xv,\Xw)} \;
\abs{\Grmtt{\Xw,\ul{\tilde\Xz},\ul{\Xx}_2}}
.\end{equation}
The bound 
\begin{equation}
\ch(V) \le \sup\limits_{\Xv} 
\int d\Xw \abs{\dot\CW{t}(\Xv,\Xw)} 
\sup\limits_{\Xw}
\int d\ul{\Xz}\; d\ul{\Xx}_2\;\abs{\Grmtt{\Xw,\ul{\tilde\Xz},\ul{\Xx}_2}}
\end{equation}
gives the result.
\end{proof}

From now on I assume that $\La$ is a discrete torus and that
$\Last $ is its dual. Let $\nnn$ be a set with $|\nnn|=\nn$ elements,
let $\Ga=\La \times \nnn\times\{1,2\}$ and 
$\Ga^* = \Last\times \nnn\times\{1,2\}$.
$\nnn$ can be thought of as the index set containing spin and 
colour indices. The last factor $\{1,2\}$ distinguishes between
the usual $\psi$ and $\psq$, as discussed in Section \ref{sect2}. 

\begin{lemma} \label{gramdet}
Let $\Xx=(x,\si,j)\in \Ga$, and for every $k \in \Last$, let
$\MM (k)$ be a symmetric matrix in $ \cM(\nn,\bR)$ 
with eigenvalues $\Mm_\rh(k)$ satisfying $|\Mm_\rh(k)| \le 1$.
Let $\ff: \Last \to \bC$, and let 
\begin{equation}\label{Dform}
\CW{t} (\Xx,\Xx') = \de_{j',3-j} \ili_{\Last} dk \; \E^{ik(x-x')}\;
(-1)^j \;\ff((-1)^j k) \MM_{\si,\si'} \left((-1)^j k\right)\; 
.\end{equation}
Then $\CW{t} (\Xx',\Xx) = -\CW{t} (\Xx,\Xx') $, and 
\begin{equation}\label{grambound}
\abs{\det \cDt{\zil-1} (\ul{\Xy},\ul{\Xz})} \le
\left(\; \ili_\Last dk\; \abs{\ff(k)} \right)^{\zil-1}
.\end{equation}
\end{lemma}

\begin{proof} Since $\Last = - \Last$ and $(-1)^j=-(-1)^{j'}$, 
a change of variables $ k \to (-1)^j k$ implies
\begin{equation}
\CW{t} (\Xx,\Xx') = \de_{j+j',3} \; (-1)^j \;\ili_{\Last} dk \; 
\ff(k)\;\MM_{\si,\si'} (k)\; \E^{ik((-1)^jx+(-1)^{j'}x')}
.\end{equation}
The antisymmetry of $\CW{t}$ now follows from the symmetry of $\MM$.
Let $\MP_\rh(k)$ be the spectral projection to the eigenspace of 
$\Mm_\rh(k)$,
so that $\MM(k)=\sum_\rh \Mm_\rh (k) \MP_\rh (k)$, and denote the 
scalar product on the spin space by $[\cdot,\cdot]$ so that 
$\MM(k)_{\si\si'} = [e_\si, \MM(k) \; e_{\si'}]$, where the
$e_\si$ are orthonormal. Then 
\begin{eqnarray}
\CW{t} (\Xx,\Xx') &=& \ili_{\Last_1} dk\; 
\ili_0^{2\pi} \frac{d\vphi}{2\pi}\;
\sli_\rh \big[a_t(\Xx)(k,\vphi,\rh), \; \MP_\rh (k)\;
b_t(\Xx')(k,\vphi,\rh)\big] 
\nonu \\
&=& \big\langle a_t(\Xx)\;,\;b_t(\Xx')\big\rangle
\end{eqnarray}
with $\Last_1 = \{ k \in \Last: \ff (k) \ne 0\}$, and 
\begin{eqnarray}
a_t(\Xx)(k,\vphi,\rh) &=& (-1)^j \; 
\E^{-ik(-1)^j x} \; \E^{-i\vphi(j-3)} |\ff(k)|^{1\over 2} e_\si
\nonu\\
b_t(\Xx')(k,\vphi,\rh) &=& 
\E^{ik(-1)^{j'} x'} \; \E^{i\vphi j'} 
\; \ff(k) \Mm_\rh(k)\; |\ff(k)|^{-{1\over 2}} e_{\si'}
.\end{eqnarray}
Gram's bound \cite{BeBe}, 
\begin{equation}
\abs{\det \cDt{\zil-1} (\ul{\Xy},\ul{\Xz})} \le
\pli_{k=1}^{\zil-1} 
\Big(
\langle a_t(\Xy_k),a_t(\Xy_k)\rangle \; 
\langle b_t(\Xz_k),b_t(\Xz_k)\rangle \Big)^{1\over 2}
,\end{equation}
and 
\begin{equation}
\sli_\rh |\Mm_\rh (k)|^2 \langle e_\si,\MP_\rh (k) e_\si\rangle
\le \sli_\rh \langle e_\si,\MP_\rh (k) e_\si\rangle \le 1
\end{equation}
imply \Ref{grambound}.
\end{proof}

\begin{corollary} \label{mfcor}
If $\ff(k) = \CWt{t}(k)$ and $\MM_{\si\si'} (k) = \de_{\si\si'}$ (as in the above many-fermion systems) then 
\begin{equation}
|\det \cDt{\zil-1} (\ul{\Xy},\ul{\Xz})| \le
\left(\int_\Last dk\; \abs{\CWt{t}(k)}\right)^{\zil-1}
.\end{equation}
\end{corollary}

\subsection{Power counting for point singularities}
In this section, I show some basic power counting bounds
for the Green functions obtained from the truncation that
all marginal or relevant couplings are left out. 

\begin{theorem}\label{opoco}
Assume that $\CW{t}$ is of the form \Ref{Dform}, 
that 
\begin{equation}\label{einnorm}
\ili_\Last dk\; \abs{\CWt{t}(k)} \le \De_1 \E^{-t}
,\end{equation}
and that 
\begin{equation}\label{sumnorm}
\Norm{\dot\CW{t}} \le \De_2 \E^{t}
.\end{equation}
Let $\int d\tilde\ka_{mr}$ denote the measure obtained by
 restricting the sum in $\int d\ka_{mr}$ to $m_1\ge 4$ and
$m_2\ge 4$ and replacing $G_{4,r_k}$ by $v \de_{r_k,1}$ whenever
it appears in the sum on the right hand side of \Ref{Qrmtdef}. 
Then the solution $\tilde G_{mr}(t)$ 
of \Ref{comRGE} with initial condition
$G_{mr} (0) = G_{mr}^{(0)}$ satisfies
\begin{equation}\label{nppoco}
\Norm{\tilde G_{mr}(t)} \le 
\cases{ \ga_{mr} \E^{t({m\over 2} -2)} & if $m \ge 6$\cr
\ga_{4r} (1+t) & if $m=4$\cr
\ga_{2r} & if $m=2$}
\end{equation}
with $\ga_{mr}$ defined recursively as
\begin{equation}\label{Recu}
\ga_{mr} = \Norm{G_{mr}^{(0)}} +
3 \De_2 \frac{1}{m} \int d\tilde\ka_{mr} \zil^2 \De_1^{\zil-1}
\ga_{m_1r_1} \ga_{m_2r_2}
.\end{equation}
\end{theorem}

\begin{proof}
Induction in $r$, with \Ref{nppoco} and \Ref{Recu} as the 
inductive hypothesis. The case $r=1$ is trivial. 
Let $r\ge 2$, and the statement hold for all $r'<r$.
By Lemma \ref{gramdet}, \Ref{ati} holds with 
$A_\zil (t) = \De_1^{\zil-1} \E^{-t(i-1)}$. Thus
\begin{equation}
\Norm{\QQrm{t}} \le \De_2
\int d\tilde\ka_{mr} \zil^2 \De_1^{\zil-1} \E^{-t(i-2)}
\Norm{\tilde\GGrmo{t}} \;\Norm{\tilde\GGrmt{t}}
.\end{equation}
By the inductive hypothesis and 
$m_1+m_2-2\zil = m$, $\Norm{\,\cdot\,}$ of
the right hand side of \Ref{icomRGE}
is bounded by 
\begin{equation}
\Norm{\GGrm{0}} + \int d\tilde\ka_{mr} \zil^2 \De_1^{\zil-1}
\De_2 \ga_{m_1r_1}\ga_{m_2r_2} \;
\half \ili_0^t ds \; \E^{s({m\over 2}-2)}
\end{equation}
To complete the induction step, this has to be bounded
by the right  hand side of \Ref{nppoco}. If $m \ge 6$, 
$\frac{m}{2}-2 \ge \frac{m}{6} \ge 1$, so 
\begin{equation}
\half \ili_0^t ds\; \E^{s({m\over 2} -2)} \le 
\frac{1}{m-4} \E^{t({m\over 2} -2)} \le \frac{3}{m}
\E^{t({m\over 2} -2)}
.\end{equation}
If $m=4$, $\half\int_0^t ds \le \frac{2}{m}(1+t)$. 
If $m=2$, $\frac{m}{2}-2 =-1$, so 
$\half \int_0^t ds\; \E^{s({m\over 2} -2)} \le 
\sfrac{1}{2} = \sfrac{1}{m}$.
This implies \Ref{nppoco}, with $\ga_{mr}$ given by 
\Ref{Recu}. 
\end{proof}

\begin{remark} In graphical language, the truncation 
removes all two-legged and all nontrivial fourlegged insertions
(`nontrivial' means that the four-legged vertices are still there).
\end{remark}

\begin{remark} \label{untrplan}
The solution to \Ref{Recu} is bounded by the solution to the
untruncated recursion $\ga_{m1}=v \de_{m4}$,
and for $r \ge 2$,
\begin{equation}\label{untrecu}
\ga_{mr} = \ga_{mr}^{(0)} + A \frac{1}{m} 
\int d\ka_{mr} \zil^2 B^{\zil-1} \ga_{m_1r_1}\ga_{m_2r_2}
\end{equation}
with $A=3\De_2$ and $B=\De_1$. 
The constants $A$ and $B$ can be scaled out of the 
recursion. If the initial interaction is a 
four-fermion interaction, $g_{m,r}=m\ga_{mr} A^{-1}
B^{1-{m\over 2}}$ satisfies the recursion
$g_{m,1}=w \de_{m4}$, with $w=4AB$, and for $r \ge 2$,
\begin{equation}\label{grecu}
g_{m,r} = \sum\limits_{s=1}^{r-1} \sum\limits_{k \ge 0}
\sum\limits_{\mu=2\atop\mu \;\rm even}^{2(s+1)}
{\mu-1 \choose k}{m+2k+1-\mu \choose k}
g_{\mu, s} \; g_{m+2k+2-\mu,r-s}
.\end{equation}
I do not provide bounds for the solution $g_{m,r}$ here;
if $g_{m,r} \le \const^r$, then the above bounds imply 
that $\sum_r \la^r \tilde \GGrm{t}$ is analytic in $\la$.
This behaviour is suggested by the absence of the factor $\zil!$
that would appear for bosons in the recursion; see \cite{S2}.
\end{remark}

The propagators $C_t$ and $\CW{t}$  for the renormalization group 
equation are defined using a partition of unity
$\ch_1+\ch_2=1$, $\ch_i \in C^\infty (\bR^+_0,[0,1])$ with 
\begin{equation}\label{zerleg}
\ch_1(x) = \cases{1 & if $x \le \frac{1}{4}$ \cr 0 & if $x \ge 1$}
\end{equation}
$\ch_1'(x) < 0$ for all $x \in (\frac{1}{4},1)$, and  
$\norm{\ch_1'}_\infty \le 2$. 

\begin{proposition}\label{GNtoy} 
Let $\La$ be $d$-dimensional, and 
$\cB$ be the infinite-volume limit of $\Last$. For instance, for
$\La = \Seps \bZ^d / L \bZ^d$, with $\sfrac{L}{2\Seps} \in 2\bN$,
$\cB = \bR^d/\sfrac{2\pi}{\Seps} \bZ^d$. Let $\CW{t}$ be given by
\Ref{Dform} with $\norm{\MM(k)} \le 1$, $\ff(0)=0$, and for $k \ne 0$ 
\begin{equation}
\ff(k) = \fff(k) \ch_1(\E^{2t} \fff(k)^2)
,\end{equation}
with $\fff\in C^{d+1}(\cB\setminus\{0\},\bC)$, 
satisfying for all $|\al| \le d+1$ and all $|k| \le 1$,
$|D^\al \fff (k)| \le F_d |k|^{1-d-|\al|}$.
Then \Ref{einnorm} and \Ref{sumnorm} hold. 
\end{proposition}

\begin{proof}
\Ref{einnorm} holds because $\int_{\Last\setminus \{0\}}
|k|^{1-d} \ch_1(k^2 \E^{2t}) dk$ is a Riemann sum approximation
to the convergent integral $\int_\cB |k|^{1-d} \ch_1(k^2 \E^{2t}) 
\sfrac{d^dk}{(2\pi)^d}$. The infinite-volume analogue of \Ref{sumnorm}
is usually proven by integration by parts, using repeatedly 
\begin{equation}\label{original}
(x_\nu-x_\nu') D_t(x-x') = - \ili_\cB \frac{d^dk}{(2\pi)^d}
\E^{\I (x-x')k} \frac{\del}{\del k_\nu} \hat D_t(k),
\end{equation}
which implies that $|\CW{t}(\Xx,\Xx')|$ falls off at least as 
$(1+\E^{-t}|x-x'|)^{-d-1}$ for large $|x-x'|$. On the torus at 
finite $L$, one iterates instead the summation by parts formula
\begin{equation}
D(x-x') (1-\E^{\I a(x-x')}) =
\ili_\Last dk\; \E^{\I k (x-x')} 
\left( \hat D (k) - \hat D(k-a)\right)
\end{equation}
which holds for all $a \in \Last$, decomposes $\La$ into $2^d$
parts and chooses $a$ appropriately to get an analogue of 
\Ref{original}. This works uniformly in $L$ because $\MM (0) = 0$.
A similar argument is given in more detail in the proof of Lemma
\ref{tdlemma}.
\end{proof}

\begin{remark} Let $d\ge 2$, $\La = \Seps \bZ^d / L \bZ^d$ and 
\begin{equation}
L_\Seps(k) =\frac{2}{\Seps^2} \sli_{\nu=1}^d (1-\cos(\Seps k_\nu))
.\end{equation}
The choices $\fff(k) =L_\Seps(k)^{{1-d\over 2}}$, $\MM=1$ 
(the toy model of \cite{FMRT}), and
\begin{equation}
\fff(k) = \left(\sli_{\nu=1}^d \left(\frac{1}{\Seps}\sin(k_\nu\Seps)\right)^2
+ (\Seps L_\Seps(k))^2\right)^{-1/2}
\end{equation}
and
\begin{equation}
\MM(k) = \fff(k) \left( 
\sli_{\nu=1}^d \I \ga_\nu \frac{1}{\Seps}\sin(k_\nu\Seps) 
+ \Seps L_\Seps(k)\right)
\end{equation}
(Wilson fermions) satisfy the hypotheses of 
Proposition \ref{GNtoy}. For all $t>0$, the infinite-volume limit
$L \to \infty$ and the continuum limit $\Seps \to 0$ of the 
Green functions $\tilde G_{mr}(t)$ exist and satisfy \Ref{nppoco}.
In the first case, $\fff(k) \to |k|^{1-d}$,
in the second case, 
$\MM(k) \fff(k) \to\sfrac{/\mkern-10mu p}{p^2}$ as $\Seps \to 0$.
The Euclidean Dirac matrices satisfy the Clifford algebra 
$\ga_\mu\ga_\nu+\ga_\nu\ga_\mu=\de_{\mu\nu}$, and 
can be chosen hermitian, $\ga_\mu^*=\ga_\mu$. 
It is possible to adapt the matrix structure of Lemma \ref{gramdet}
to satisfy the antisymmetry condition on $\CW{t}$ also for this case. 
\end{remark}

\begin{remark} In ultraviolet renormalizable theories,  
the signs in the exponents of 
\Ref{einnorm} and \Ref{sumnorm} are reversed. 
A power counting theorem similar to the infrared power counting 
Theorem \ref{opoco} can be proven provided that in the initial 
interaction, at most quartic interactions appear. The statement is
then that for $m \ge 6$, the Green function is bounded by 
$\const \E^{-t({m\over 2} -2)}$. 
\end{remark}

\section{The thermodynamic limit of the many-fermion system}
\label{sect5}
In this section, I apply the RGE to 
the many-fermion systems defined in Section \ref{sect2}.
For $d=1$, the singularity is pointlike,
so the power counting bound Theorem \ref{opoco} applies. 
For $d \ge 2$, the analogue of Theorem \ref{opoco} 
gives only weaker bounds, which, e.g., 
in $d=2$ would mean that the four-point function is still
relevant and the six-point function is marginal. 
This is not the actual behaviour; showing better bounds 
in $d\ge 2$ requires a refinement using 
the sector technique of \cite{FMRT} 
and is deferred to another paper. 
I also show a simple bound for the full Green functions
that takes into account the sign cancellations. If 
the coefficients $\ga_{mr}$ given by \Ref{untrecu} 
are exponentially bounded, this bound implies that the 
unrenormalized expansion converges in a region $|\la|\be^d < \const$. 
I also give a simple proof that the thermodynamic limit exists in perturbation theory.

One can also show bounds on the expansion coefficients
for finite $\Nta$ and $L$. 
Since this is mainly a tedious repetition of the 
infinite-volume proofs with integrals replaced by Riemann sums
(it also requires that $\mu$ is chosen such
that the Fermi surface contains no points of the 
finite-$L$ momentum space lattice),
I will content myself with indicating where this is necessary.

For convenience, I call the limit $\Nta \to \infty$ and  
$L \to \infty$ the {\em thermodynamic limit}, although the first limit
would more aptly be called the time-continuum limit. 
The limit $\Nta \to \infty$
has to be taken first, because I want to apply \Ref{Trotto} for 
operators on a finite-dimensional space only. However, it will turn out
that for $T>0$, the order of the two limits does not matter. 
Since I want to take the limit $\Nta \to \infty$,
I can assume that $\Nta \ge 2\be(\epz+E_{\rm max})$.
Thus Lemma \ref{LemA1} applies, with $\Ez=\ept$, for all $t \ge 0$.

\subsection{The component RGE in Fourier space}
\label{momRGE}
Let $\La=\bT\times \SLa$ and $\Last = \Mats{\Nta}\times\SLa^*$, 
as given in Sections \ref{Grarep} and \ref{ss22}.
Let $\Ga=\La \times \{-1,1\}\times\{1,2\}$
and $\Gast=\Last\times\{-1,1\}\times\{1,2\}$.
The Fourier transforms of the $\GGrm{t}$ are, 
with $(\ul{\Xp}=(\Xp_1,\ldots,\Xp_m)$,
\begin{equation}
\hat \Grmt{\ul{\Xp}} = 
\ili_{\La^m} \pli_{k=1}^m dx_k\;
\E^{-\I (p_1x_1 + \ldots + p_mx_m)}
\Grmt{\ul{\Xx}}
.\end{equation}  
For $\Xk = (k,\si,j)\in \Gast$ let $\sim \Xk = (-k,\si,3-j)$,
and let 
$\de_{\Gast} (\Xk,\Xk') = \de_{\Last} (k,k')\de_{\si\si'}\de_{jj'}$.
Assume that the Fourier transform of the propagator $\CW{t}$ is of the form
\begin{equation}\label{e93}
\hat \CW{t} (\Xk,\Xk') = \de_{\Gast} (\Xk, \sim \Xk') \; 
\CWb{t} (\Xk')
\end{equation}
where 
\begin{equation}\label{e94}
\CWb{t}(\Xk) = 
(-1)^j \CWt{t} ((-1)^j k)
.\end{equation}
This combination of signs implies that $\CW{t}(\Xx,\Xy) = -
\CW{t}(\Xy,\Xx)$. The propagator 
$\CWt{t}$ for the 
many-fermion system is given in \Ref{e112}. 
The Fourier transform of $\QQrm{t}$ is 
\begin{eqnarray}\label{Qfour1}
\hat\Qrmt{\ul{\Xp}} &=& 
\int d \ka_{mr} \; \zil! \zil \int d\Xk_1 \ldots d\Xk_\zil\;
(-\dot{\CWb{t}}(\Xk_1))\;
\pli_{j=2}^\zil \CWb{t}(\Xk_j)
\nonu\\
&&
\hat \Grmto{\ul{\Xp}^{(1)},\ul{\Xk}} \;
\hat \Grmtt{\sim\ul{\Xk},\ul{\Xp}^{(2)} }
\end{eqnarray}
with $\ul{\Xp}^{(1)}=(\Xp_1, \ldots, \Xp_{m_1-\zil})$, 
$\ul{\Xp}^{(2)}=(\Xp_{m_1-\zil+1},\ldots,\Xp_{m_1+m_2})$,
$\ul{\Xk} =(\Xk_1, \ldots, \Xk_\zil)$, and 
$\sim\ul{\Xk}=(\sim\Xk_\zil, \ldots, \sim\Xk_1)$.
Here the arguments of $\GGrmt{t}$ have been permuted
and relabelled such that
the determinant is transformed into a sum with the same sign 
for all terms (see Appendix \ref{ReWick}); 
this gives the extra factor $\zil!$. 

By translation invariance in space $\Sx$ and time $\ta$, 
\begin{equation}\label{Idelta}
\hat \Grmt{\Xp_1,\ldots,\Xp_m} =
\de_{\Last} (p_1+\ldots+p_m,0) 
\Irmt{\Xp_1,\ldots, \Xp_m}
\end{equation}
with a totally antisymmetric function 
$\Irmt{\Xp_1,\ldots, \Xp_m}$ of $(\Xp_1,\ldots, \Xp_m)\in {\Ga^*}^m$
that satisfies 
\begin{equation}\label{senkre}
\sli_{\mu=1}^m \nabla_{p_\mu} \Irmt{\Xp_1,\ldots, \Xp_m}=0
.\end{equation}
A priori, \Ref{Idelta} implies only the existence of a function 
$\hat \Irmt{\Xp_1,\ldots, \Xp_m}$, defined
only for those $\Xp_1,\ldots, \Xp_m$ for which
$p_1+\ldots +p_m=0$. However, since 
$H=\{{p_1,\ldots, p_m}: \; p_1+\ldots +p_m=0\}$
is a linear subspace of $(\Last)^m$, one can simply extend 
$\tilde I$ to a function on all space by defining 
$I=\hat I \circ \Pi_H$ with $\Pi_H$ the projection 
to the subspace $H$. Since $\Pi_H$ is symmetric in all its arguments,
$I$ is totally antisymmetric, and \Ref{senkre} holds because
$(1,\ldots,1) \perp H$ (in \Ref{senkre}, $\nabla_{p_\mu}$
is a difference operator which becomes the gradient in the 
limit where the momenta become continuous).

The product of the two $\de_\Last$ in  \Ref{Qfour1} can be combined
to cancel the $\de_\Last$ in the relation between $\hat G$ and $I$,
and to remove the integration over $k_1$. 
Thus the RGE in Fourier space is
$\sfrac{\del}{\del t} \Irmt{\ul{\Xp}} = \half \bA_m \Qrmt{\ul{\Xp}}$,
with
\begin{eqnarray}\label{momcom} 
\Qrmt{\ul{\Xp}} &=&
\int d\ka_{mr}\; \zil! \zil\;\int d\Xk_2\ldots d\Xk_\zil
\sli_{\si_1,j_1}
(-\dot{\CWb{t}} (\Xk_1)) \;
\nonu\\
&&
\pli_{s=2}^\zil \CWb{t}(\Xk_s)\;
\Irmto{\ul{\Xp}^{(1)},\ul{\Xk}} 
\; \Irmtt{\sim\ul{\Xk},\ul{\Xp}^{(2)} }
\end{eqnarray}
where $\Xk_1=(k_1,\si_1,j_1)$ and $k_1$ is fixed as
$k_1=-(k_2+\ldots + k_\zil) + p_1 + \ldots + p_{m_1-\zil+1}$.
I use the same symbol for the $Q$ in position and in momentum space
since it will be always clear from the context which one is meant.

\begin{remark}\label{recurs}
The equivalent integral equation is
\begin{equation}\label{equinteFou}
\Irmt{\ul{\Xp}} = \Ir{m}{r}{0}{\ul{\Xp}} +
\half \bA_m \;
\ili_0^t ds \; \QQ{m}{r}{s}{\ul{\Xp}}.
\end{equation}
The sum $\int d\ka_{mr}$ contains the sum over $r_1\ge 1$ and $r_2\ge 1$, 
with the restriction $r_1+r_2=r$. Thus 
only $r_1 < r$ and $r_2 <r$ occur in this sum. Therefore \Ref{equinteFou},
together with an initial condition $\Ir{m}{r}{0}{\ul{\Xp}}$, uniquely 
determines the family of functions $\Irmt{\ul{\Xp}}$. 
Iteration of \Ref{equinteFou} generates the usual perturbation expansion. 
\end{remark}

\subsection{Bounds on the finite-volume propagator}
The propagator $\hat c$ for the many-fermion system is given in 
\Ref{mfprop1}. If $\Sk$ is such that $E(\Sk)=0$, then $\hat c$ becomes
of order $\be$ for small $|\om|$.
In the temperature zero limit, $\be \to \infty$, this becomes a singularity.
This is the reason why renormalization is necessary. The renormalization group
flow is parametrized by $t\ge 0$, where 
\begin{equation}
\ept = \epz \E^{-t}
\end{equation}
is a decreasing energy scale. The fixed energy scale $\epz$ was 
specified in Section \ref{hyposuse}.  
The limit of interest is $t \to \infty$. 

The uncutoff propagator for the many-fermion system is given in 
\Ref{mfprop1}. Thus, for $k = (\om,\Sk)\in \Last$ let 
\begin{equation}\label{e112}
\CWt{t} (k) = \frac{1}{\I\Hat{\om}-E(\Sk)}\;
\ch_1 \left( {\ept}^{-2} \abs{\I\Hat{\om} -E(\Sk)}^2\right)
\end{equation}
where $\Hat{\om}$ is defined in \Ref{Hatomdef}, 
$\ch_1$ is given in \Ref{zerleg}, and define
$\tilde C_t (k)$ similarly, with $\ch_1$ replaced by
$\ch_2=1-\ch_1$.
Then $\tilde C_t(k) + \CWt{t}(k) = (\I\Hat{\om}-E(\Sk))^{-1}$
is independent of $t$. The functions
$\tilde C_t$ and $\CWt{t}$ define operators $\hat \CW{t} (\Xk,\Xk')$ and 
$\hat C_t (\Xk,\Xk')$ on the functions on $\Gast$ by \Ref{e94} and 
\Ref{e93}.
Denote $\CWtd{t} = \frac{\del}{\del t}\CWt{t}$, and let 
$\True{A} =1$ if the event $A$ is true and $0$ otherwise.

\begin{proposition}\label{elepro}
$\CWt{t}$ is a $C^\infty$ function of $t$ that
vanishes identically if $t > \log \frac{\be\epz}{2}$.
If $t \le log \frac{\be\epz}{2}$, then 
\begin{eqnarray}\label{cwtsupp}
\mbox{ \rm supp } \CWt{t} &\subset& \{ k \in \Last : \abs{\I\Hat{\om} -E(\Sk)}
\le \ept \} \nonu\\
\mbox{ \rm supp } \CWtd{t} &\subset& \{ k \in \Last : 
\frac{1}{2}\ept  \le \abs{\I\Hat{\om} -E(\Sk)}
\le \ept \}.
\end{eqnarray}
Moreover
\begin{equation}\label{cwtdbe}
\abs{\CWtd{t}(k) } \le  4 \ept^{-1}  \True{\abs{\I\Hat{\om} -E(\Sk)}
\le \ept } \nonu\\
 \le 2 \be \True{\abs{\I\Hat{\om} -E(\Sk)}
\le \ept } ,
\end{equation}
$\abs{\CWt{t}(k) } \le \frac{\be}{2} \True{\abs{\I\Hat{\om} -E(\Sk)}
\le \ept }$,
\begin{equation}\label{cwtint}
\ili_\Last dk \; \abs{\CWtd{t}(k)} \le 4\Vo
\quad \mbox{ and } \quad
\ili_\Last dk \; \abs{\CWt{t}(k)} \le \Vo \log\sfrac{\be\epz}{2}
,\end{equation}
where $\Vo$ is the constant in \Ref{Vobou}. 
\end{proposition}

\begin{proof}
$\ch_1(x)=0$ if $x \ge 1$, so $\CWt{t}(k) \ne 0$ implies 
$\abs{\I\Hat{\om} -E(\Sk)} \le \ept \le 1 $. By Lemma \ref{LemA1},
this implies $|\om| \le \sfrac{\pi}{2}\ept$. Since $|\om| \ge
\sfrac{\pi}{\be}$, $\CWt{t} \ne 0$ only for $t \le \log\sfrac{\be\epz}{2}$.
Since $|\I\Hat{\om}-E(\Sk)| \ge |\mbox{Re }\Hat{\om}| \ge \sfrac{2}{\be}$,
the stated properties of $\CWt{t}$ follow. 
The $t$-derivative gives
$|\sfrac{\del}{\del t }\CWt{t}(k)| = 2 \ept^{-2} 
|\I\Hat{\om} -E(\Sk)| \;
|\ch_1' (\ept^{-2}|\I\Hat{\om} -E(\Sk)|^2)|$.
Since $\ch_1'(x) =0$ for $x \not\in(\frac{1}{4},1)$,
$\CWtd{t}(k) \ne 0$ implies 
$\sfrac{1}{2} \ept \le \abs{\I\Hat{\om} -E(\Sk)} \le \ept$
which implies \Ref{cwtdbe} and \Ref{cwtsupp}. 
\Ref{cwtint} follows from these inequalities 
by Lemma \ref{LemA1}, by \Ref{Vobou}, and by 
$\CWt{t} = \int_t^{\log (\be\epz/2)} ds\; \CWtd{s}$.
\end{proof}

\begin{remark} The bounds \Ref{cwtint} are crude because 
the restriction $|E(\Sk)| \le \ept$ was replaced by 
$|E(\Sk)| \le 2$ when \Ref{Vobou} was applied. To get a better bound,
one has to require that no points of the finite--volume lattice
in momentum space is on the Fermi surface $\FS$. Because only then 
$\int_\SLa dk \True{|E(\Sk)| \le \ept} \le \const \ept$ holds
uniformly in $L$.
\end{remark}

\subsection{Existence of the thermodynamic limit in perturbation theory}
The proof of existence of the thermodynamic limit will proceed 
inductively in $r$, because of the recursive structure of the 
RGE mentioned in Remark \ref{recurs}. It will be an
application of the dominated convergence theorem to 
\Ref{equinteFou}. To this end, it is necessary to make the 
integration region independent of $\Nta$ and $L$. 
Although $\Irm(t)$, given by \Ref{equinteFou}, appears evaluated at $\ul{\Xp}=(\Xp_1,\ldots,\Xp_m) 
\in \Ga^m$ on the RHS of \Ref{equinteFou} only, the integral defines
the $t$-derivative of a function defined on $\Gainf^m$, where
$\Gainf=\bM(\be) \times \cB \times \{1,-1\}\times\{1,2\}$ 
with $\cB=\bG^*$ the first Brillouin zone of the infinite lattice,
and 
\begin{equation}\label{infmats}
\bM(\be) = \{ \om_n=\frac{\pi}{\be}(2n+1):\; n\in \bZ\}
\end{equation}
the set of Matsubara frequencies in the limit $\Nta \to \infty$. 
For a bounded function $F_m: \Gainf^m \to \bC$, let
\begin{equation}
\abs{F_m}_0 = \sup\limits_{\ul{\Xp} \in \Gainf^m}
\abs{F_m (\ul{\Xp})}
\end{equation}
Up to now, the dependence of $\Irm(t)$ on $(\Nta,L)$ was not denoted 
explicitly. I now put it in a superscript and write 
$\Irmnl$. Let 
$\lli_{(\Nta,L) \to\infty} = \lim_{L\to\infty} \lim_{\Nta\to\infty}$.
\begin{lemma} Let $(\Irmnl (0))_{m,r,\Nta,L}$ 
be a family of bounded functions such that
$\Irmnl (0) = 0 $ if $m > 2r+2$,
$\lim_{(\Nta,L) \to\infty} \Irmnl (0) = \Irm (0)$
exists and is a bounded function on $\Gainf^m$,
and $\abs{\Irmnl (0)} \le K_{mr}^{(0)}$. 
Let $(\Irmnl(t))_{m,r,\Nta,L}$ be the solution to \Ref{equinteFou}. 
Then, for all $m$ and $r$, $\Irmnl(t) = 0 $ if $m>2r+2$,
and 
\begin{equation}\label{ih2}
\frac{\del}{\del t} \Irmnl (t) =0 \quad
\mbox{ for all } t>\log\frac{\be}{2}
,\end{equation}
there are bounded functions $\Irm(t):\Gainf^m \to \bC$ such that
\begin{equation}\label{ih3}
\lli_{(\Nta,L) \to\infty} \Irmnl (t) = \Irm (t)
.\end{equation}
Let $P_{mr}$ be the polynomials defined recursively as
\begin{equation}\label{polymr}
P_{mr} (x) = 6^{1-r} |\Irm(0)| + \epz^{-1} 
\int d\ka_{mr} \zil\;\zil!\; x^{\zil-1} 
P_{m_1r_1} (x) P_{m_2r_2}(x) 
\end{equation}
(in particular, the coefficients of $P_{mr}$ are independent of $\be$),
then for all $\Nta,L,\be,t$
\begin{equation}\label{ih4}
\abs{\Irmnl(t)}_0 \le (\epz\be)^{r-1}  
P_{mr} \left( 4\Vo \log\sfrac{\be\epz}{2}\right)
.\end{equation}
\end{lemma}

\begin{proof} Induction in $r$, with the statement of the Lemma
as the inductive hypothesis. Let $r=1$. Since $r_1\ge 1$ and $r_2 \ge 1$,
the right hand side of the equation is zero, so 
$I_{m1}^{(\Nta,L)} (t,\ul{\Xp}) = I_{m1}^{(\Nta,L)} (0,\ul{\Xp})$
for all $t$. Thus the statement follows from the hypotheses on 
$\Irmnl (0)$. 

Let $r \ge 2$, and the statement hold for all $r'<r$. 
Let $m > 2r+2$. Then $\Irmnl (0) = 0$. The five-tuple 
$(m_1,r_1,m_2,r_2,\zil)$ contributes to the right hand side
only if $r_1+r_2=r$, $m_1+m_2=m+2\zil$, $\zil \ge 1$, and 
by the inductive hypothesis, only if $m_1\le 2r_1+2$
and $m_2 \le 2r_2+2$. Thus $\Irmnl (t)$ can be nonzero only if
$m=m_1+m_2-2\zil \le m_1+m_2 -2 \le 2r_1+2 + 2r_2+2 -2 = 2r+2$.
The integral appearing on the right hand side of
\Ref{equinteFou} is a Riemann sum approximation to an integral 
over the $(t,\Nta,L)$-independent region $s \in [0,\infty)$, 
$k_j \in \bM(\be) \times \cB$. By the inductive hypothesis,
the factors $I_{m_k r_k} ^{(\Nta,L)}$ have a limit  \Ref{ih3} satisfying
\Ref{ih4}. Since for all $t$, $\CWb{t}$ is a bounded $C^2$ function,
the same holds for the propagators (boundedness holds because 
$\beta < \infty$). Thus the integrand converges pointwise, 
and it suffices to show that it is bounded by an integrable
function to get \Ref{ih3} and \Ref{ih4} by an application of the 
dominated convergence theorem. Let $\al = 4\Vo\log \sfrac{\be\epz}{2}$,
and let $g$ be the function on 
$[0,\infty) \times (\bM(\be) \times \cB)^{\zil -1}$ given by
\begin{eqnarray}
g(s,\ul{k}) &=& 
\frac{2}{\epz \E^{-s}} 
\True{s \le \log \sfrac{\be\epz}{2}}
\int d\ka_{mr}\; \zil\;\zil! P_{m_1r_1}(\al) \be^{r_1-1}
P_{m_2r_2}(\al) \be^{r_2-1}
\nonu\\
&&\pli_{j=2}^\zil 
\ili_0^s ds_j \True{|\om_j| \le \sfrac{\pi}{2}\epz \E^{-s_j}}
\True{|E(\Sk_j| \le 2 \epz \E^{-s_j}}
.\end{eqnarray}
The integrand on the right hand side of \Ref{equinteFou} is 
bounded by $g$ by Proposition \ref{elepro}, Lemma \ref{LemA1},
and the inductive hypothesis \Ref{ih4}. Because 
$g$ vanishes identically for $s > \log \frac{\be\epz}{2}$,
it is integrable. 
By \Ref{cwtdbe} and \Ref{cwtint},  
\begin{equation}
\ili_0^t ds \int g(\ul{k}) dk_2\ldots dk_\zil \le
\be^{r-1} \int d\ka_{mr} \zil\;\zil!\; \al^{\zil-1}
P_{m_1r_1}(\al) P_{m_2r_2}(\al) 
.\end{equation}
Thus \Ref{ih3} holds, and \Ref{ih4} holds with 
$P_{mr}$ given by \Ref{polymr} (the factor $6^{1-r}$ comes from
the assumption that $\be\epz \ge 6$).
Since the right hand side of \Ref{equinteFou} 
vanishes for $t > \log\sfrac{\be\epz}{2}$,
\Ref{ih2} holds.
\end{proof}

\begin{remark} $\Irmnl (0)$ is not the initial interaction because
at $t=0$, the propagator $C_0 \ne 0$. The $\Irmnl (0)$ are obtained
from the original interaction by Wick ordering with respect to $C$
and integrating over all fields with covariance $C_0$. 
The existence of the limit $(\Nta,L)\to \infty$ of the $\Irmnl (0)$
is not obvious because in the limit $\Nta \to \infty$, the 
absolute value of the propagator is not summable.
In perturbation theory, this is no serious problem, 
and the $\Irmnl (0)$  are controlled  
by a similar inductive proof as the above. An alternative proof
is in Appendix D of \cite{FST2}. 
\end{remark}

\subsection{The many-fermion system in the thermodynamic limit}
\label{sect55}
In this section, let the many-fermion system satisfy
the hypotheses stated in Section \ref{hyposuse} with $\kz > d$.

In the thermodynamic limit, 
the spatial part $\Sp$ of momentum becomes continuous, 
$\Sp \in \cB$, and the set of Matsubara frequencies becomes
$\bM(\be)$, given in \Ref{infmats}. It is convenient to take
$(p_0,\Sp) \in \bR \times \cB$ and put all the $\be$-dependence
into the integrand. To this end, I define the step function 
$\omb: \bR \to \bM(\be)$ by 
\begin{equation}
\omb(p_0) = \frac{\pi}{\be} (2n+1) 
\quad\mbox{if } p_0 \in (\frac{2\pi}{\be}n,\frac{2\pi}{\be}(n+1)]
.\end{equation}
For any continuous and integrable function $f$,
\begin{equation}\label{osu}
\ili_\bR \frac{dp_0}{2\pi} \; f(\omb(p_0)) = 
\frac{1}{\be}\sli_{\om\in\bM(\be)} f(\om)
.\end{equation}
Moreover, 
\begin{equation}\label{ombprop}
\sup\limits_{p_0 \in \bR} \abs{\omb(p_0)-p_0} = \frac{\pi}{\be}
\quad \mbox{and} \quad 
\inf\limits_{p_0 \in \bR} \abs{\omb(p_0)} = \frac{\pi}{\be}
,\end{equation}
$\abs{\omb(p_0)} \ge \frac{p_0}{2}$, and $\om_\be(-p_0)=
-\om_\be(p_0)$ holds Lebesgue-almost everywhere, so that 
in integrals like \Ref{osu}, $\omb$ can be treated as an 
antisymmetric function. With this, the propagator now reads
$\CWt{t} (p) = \cC_t (\omb(p_0),E(\Sp))$
with 
\begin{equation}\label{cCdef}
\cC_t(x,y) = \frac{1}{ix-y} \ch_1(\ept^{-2} (x^2+y^2))
\end{equation}
(and $\ept=\epz\;\E^{-t}$).

In infinite volume, the restriction on the spatial part 
$\Sp$ of momentum improves the bounds on the propagator
over the finite-volume ones given above.

\begin{lemma} \label{dtiprop}
$\CWt{t}$ is bounded, $C^\infty$ in $t$ and 
$C^\kz$ in $\Sp$. If $t > \log \frac{\be\epz}{\pi}$, then 
$\CWt{t} (p) = 0 $ for all $p \in \bR \times \cB$. 
For all multiindices $\al \in \bN_0^d$ with $|\al| \le \kz$, 
there is a constant $B_\al > 0$ such that 
\begin{eqnarray}\label{eist}
\abs{D^\al \dot\CWt{t} (p) } &\le& B_\al \ept^{-1-|\al|}
\True{\abs{i\omb(p_0) - E(\Sp)} \le \ept} \nonu\\
&\le & B_\al \ept^{-1-|\al|}
\True{\abs{\omb(p_0)} \le \ept}
\True{\abs{E(\Sp)} \le \ept}
.\end{eqnarray}
$B_0=4$. For $t \le \log \frac{\be\epz}{\pi}$, 
\begin{equation}\label{zwst}
\ili_\bR \sfrac{dp_0}{2\pi} \abs{D^\al \dot{\CWt{t}} (p)} \le
B_\al \ept^{-|\al|} \True{\abs{E(\Sp)} \le \ept}
,\end{equation}
\begin{equation}\label{drst}
\ili_{\bR\times \cB} \sfrac{d^{d+1}p}{(2\pi)^{d+1}}
\abs{D^\al \dot{\CWt{t}} (p) } \le 2 J_1 B_\al \ept^{1-|\al|}
,\end{equation}
and 
\begin{equation}\label{vist}
\ili_{\bR\times \cB} \sfrac{d^{d+1}p}{(2\pi)^{d+1}}
\abs{\CWt{t} (p) } \le 8 J_1 \ept
.\end{equation}
\end{lemma}

\begin{proof}
For $\CWt{t}(p)$ to be nonzero, $\abs{\omb(p_0)}\le \ept$
must hold. By \Ref{ombprop}, $\CWt{t} = 0 $ for
$t > \log \frac{\be\epz}{\pi}$. Since
$\dot\CWt{t} (p) = -2 \ept^{-2} (i\omb(p_0) + E(\Sp)) \; 
\ch_1'(\ept^{-2} (\omb(p_0)^2+E(\Sp)^2))$
and $\norm{\ch_1'}_\infty \le 2$, 
$|\dot\CWt{t} (p)| \le \sfrac{4}{\ept}
\True{\abs{i\omb(p_0) - E(\Sp)} \le \ept}$.
Thus  $B_0 = 4$. 
Derivatives with respect to $\Sp$ can act on $E(\Sp)$ or on 
$\ch_1'$, in which case they produce factors bounded by 
$\ept^{-1} \abs{\nabla E(\Sp)}$, so there is $B_\al $ such 
that \Ref{eist} holds. Inserting \Ref{eist} 
into the integral in \Ref{zwst} gives
\begin{equation}
\ili_\bR \frac{dp_0}{2\pi} \abs{D^\al \dot{\CWt{t}} (p)} \le
B_\al \ept^{-1-|\al|} \True{\abs{E(\Sp)} \le \ept}
M_\be
\end{equation}
with 
\begin{equation}\label{e150}
M_\be = 
\sfrac{1}{\be}\abs{\{ n \in \bZ: \abs{2n+1} \le \frac{\be\ept}{\pi}\}}
\le \frac{2}{\pi}\ept
,\end{equation}
which proves \Ref{zwst}. This gives for the integral in \Ref{drst}
\begin{equation}
\ili_{\bR\times \cB} \frac{d^{d+1}p}{(2\pi)^{d+1}}
\abs{D^\al \dot{\CWt{t}} (p) } \le B_\al \ept^{-|\al|}
\ili_{\cB} \frac{d^{d}\Sp}{(2\pi)^{d}}
\True{\abs{E(\Sp)} \le \ept}
\end{equation}
For $t \ge 0$, $\ept \le \epz$, so by a change of coordinates in 
the integral, 
\begin{equation}\label{e152}
\ili_{\cB} \frac{d^{d}\Sp}{(2\pi)^{d}}
\True{\abs{E(\Sp)} \le \ept}
\le \ili_{-\ept}^{\ept} d\rh\; \int d\th |J(\rh,\th)|
= 2 J_1 \ept
\end{equation}
which implies \Ref{drst}. \Ref{vist} follows by integration over $t$.
\end{proof}

\subsection{A bound in position space}

In this section, I prove a bound for the unrenormalized Green 
functions that motivates the statement that the unrenormalized 
expansion converges for $|\la|\be^d < $ const (if the solution
to \Ref{grecu} is exponentially bounded, it implies this convergence).
\begin{lemma}\label{tdlemma}
For many-fermion systems with $E \in C^\kz$, where $\kz > d$, 
there is $\De_2 >0$ such that for all $t\ge 0$ and all $\be$
\begin{equation}
\Norm{\dot \CW{t}} \le \De_2 \E^{td}
.\end{equation}
\end{lemma}

\begin{proof}
By definition of $\Norm{\cdot}$, 
$\Norm{\dot\CW{t}} \le 4 \int_{-\be/ 2}^{\be /2} dx_0 \;
\int_\SLa d\Sx\; |\bdtd (x_0,\Sx)|$
with 
\begin{equation}
\bdtd (x_0,\Sx) = \ili_\bR \sfrac{dk_0}{2\pi}\;
\ili_\cB \sfrac{d^d\Sk}{(2\pi)^d}
\E^{\I x_0 \omb (k_0) + \I \Sk\cdot \Sx} \dot\cC_t (\omb(k_0),E(\Sk))
.\end{equation}
By \Ref{cCdef}, 
$\dot \cC_t (\omb(k_0),E(\Sk)) = -\sfrac{2}{\ept^2}
(\I \omb (k_0) + E(\Sk)) \ch_1'(\sfrac{\omb(k_0)^2 + E(\Sk)^2}{\ept^2})$.
\hfill\break
{\em Claim:} There is a constant $N$, depending on $\kz$ and $E$,
such that for all $\be$
\begin{equation}
\label{wish}
|\bdtd (x_0,\Sx)| \le N\; \ept\; \frac{1}{(1+\ept|\Sx|)^\kz\;
(1+\ept\min\{|x_0|,\sfrac{\be}{2}-|x_0|\})^2}
.\end{equation}
The lemma follows from \Ref{wish} by 
\begin{equation}
\int d^d\Sx\; \frac{1}{(1+\ept|\Sx|)^\kz}= \ept^{-d}
\int d^d\xi (1+|\xi|)^{-\kz}
\end{equation}
and 
\begin{equation}
\ili_{-\be\over 2}^{\be\over2} dx_0 \;
\frac{1}{(1+\ept\min\{|x_0|,\sfrac{\be}{2}-|x_0|\})^2}
\le 4 \ept^{-1} \ili_0^\infty \frac{du}{(1+u)^2}
.\end{equation}
\Ref{wish} is proven by the standard integration by parts 
method. Since the $k_0$-dependence is via the step function 
$\omb$, one has to use summation by parts in the form
\begin{equation}
\bdtd (x_0,\Sx)\, ( 1- \E^{{2\pi\I x_0\over\be}})^q =
\ili_\bR \sfrac{dk_0}{2\pi}\;
\ili_\cB \sfrac{d^d\Sk}{(2\pi)^d}
\E^{\I x_0 \omb (k_0) + \I \Sk\cdot \Sx}
(\De_{2\pi\over\be}^q \dot\cC_t) (\omb(k_0),E(\Sk))
\end{equation}
where $(\De_a f)(k_0) = f(k_0) - f(k_0-a)$. Taking $q\le 2$ and using
a Taylor expansion of $\dot \cC_t (x,y)$ to order $q$ in $x$,
one sees that $\De_{2\pi\over\be}^q \dot\cC_t$ is still 
$C^\kz$ in $\Sk$ and that for any multiindex $\al$ with $|\al|\le \kz$,
\begin{equation}
\abs{(\sfrac{\del}{\del\Sk})^\al \De_{2\pi\over\be}^q \dot\cC_t\;
(\omb(k_0),E(\Sk))} \le N_1 \sfrac{1}{\be^2\ept^{1+|\al|+q}} \; 
\True{|i\omb(k_0)-E(\Sk)| \le \ept}
.\end{equation}
By \Ref{e150} and \Ref{e152}, this implies 
\begin{equation}
\abs{\Sx^\al \left( 1- \E^{{2\pi\over\be}\I x_0}\right)^2
\bdtd (x) } \le 4J_1 N_1 \frac{1}{\be^2}\;\ept^{-1-|\al|-q}
.\end{equation}
For $|\xi| \le \pi$, $|\sin\xi| \ge 
\sfrac{2}{\pi}\min\{|\xi|,\pi-|\xi|\}$,
so 
\begin{equation}
\abs{1-\E^{{2\pi\over\be}\I x_0}} \ge \abs{\sin{2\pi\over\be} x_0} \ge
\frac{4}{\be}\min\{|x_0|,\sfrac{\be}{2}-|x_0|\}
,\end{equation}
and thus 
\begin{equation}
\ept^{|\al|+q} \abs{\Sx^\al} (\min\{|x_0|,\sfrac{\be}{2}-|x_0|\})^q
\;\abs{\bdtd(x)} \le N_2 \ept
.\end{equation}
This implies \Ref{wish}.
\end{proof}

\begin{remark} This proof changes in finite volume: 
one needs the requirement that no point on the Fermi surface $\FS$
is a point of the momentum space lattice determined by 
the sidelength $L$ to 
get a bound that is uniform in $L$. This can be achieved by choosing
the chemical potential $\mu$ appropriately.
\end{remark}

\begin{theorem} \label{mfsge2}
For the many-fermion model in $d \ge 1$, and with the 
initial condition $\Norm{G_{mr}(0)} = \ga_{mr}^{(0)}$,
\begin{equation}\label{ihbeta}
\Norm{G_{mr}(t)} \le \ga_{mr} (\epz\be)^{(r-1)d}
\end{equation}
with the $\ga_{mr}$ given by \Ref{untrecu}, where
$A=\sfrac{8}{\epz d} \De_2$ and $B=8 J_1 \epz$.
\end{theorem}

\begin{proof} 
By \Ref{vist}, Corollary \ref{mfcor}, 
and Lemma \ref{gramdet}, $|\det\cDt{\zil-1}|\le (8J_1\ept)^{\zil-1}$,
so $\De_1\le 8J_1\epz$. 
By Lemma \ref{tdlemma}, $\Norm{\dot\CW{t}} \le \De_2\E^{td}$. 
The proof is by induction in $r$, with the inductive hypothesis
\begin{equation}\label{ihtd}
\Norm{G_{mr}(t)} \le \ga_{mr} \E^{td(r-1)}
.\end{equation}
The statement is
trivial for $r=1$. Let $r\ge 2$, and \Ref{ihtd} hold for
all $ r'< r$. By Lemma \ref{unidet} and \Ref{cwtint}, and since $r_1+r_2=r$,
\begin{equation}
\Norm{\QQrm{t}} \le \De_2 \E^{td(r-1)}\; \int d\ka_{mr} \zil^2 \;
\De_1^{\zil-1}\;
\ga_{m_1r_1}\ga_{m_2r_2} 
.\end{equation}
Since $r\ge 2$, $r-1 \ge \sfrac{r}{2}$, so 
$\int_0^{t} dt \; \E^{sd(r-1)} \le \frac{2}{dr}\E^{td(r-1)}$,
and
\Ref{ihtd} follows by integration and by
$r \ge \sfrac{m}{4}$. Because $\CW{t} = 0$ 
if $t > \log (\be\epz)$, \Ref{ihbeta} holds.
\end{proof}

\begin{remark}
The bounds here are rather crude because $\ept\le \epz$ was used.
However, these are bounds for the full Green functions, 
not for truncations. Improving them requires renormalization
and, for $d \ge 2$, the sector technique of \cite{FMRT}.
\end{remark}

\begin{proposition} 
The hypotheses of Theorem \ref{opoco}
are satisfied for the many-fermion model in $d=1$
in the thermodynamic limit.
\end{proposition}

\begin{proof} \Ref{einnorm} follows from \Ref{drst}. 
\Ref{sumnorm} follows from Lemma \ref{tdlemma}.
\end{proof}

\begin{proposition} 
Let $d \ge 2$. If the sum $\int d\ka_{mr}$ is truncated to 
$m_1 > 2(d+1)$, and $m_2 > 2 (d+1)$, then the $G_{mr}$ defined
by this truncation satisfy for all $m > 2d+2$
\begin{equation}
\Norm{G_{mr} (t) } \le \ga_{mr} \E^{t({m \over 2} - (d+1))}
\end{equation}
If these bounds were sharp, all $m$-point functions up to
$m= 2d$ would be relevant in the RG sense, and the 
$(2d+2)$-point function would be marginal. This is not really
the case; see \cite{FMRT}. It is, however, remarkable that such
a simple power counting ansatz goes through at all in $d \ge 2$.
\end{proposition}

\section{Regularity of the selfenergy}\label{sect6}
In this section, I study the regularity question 
for the skeleton self-energy, which is very nontrivial even 
in perturbation theory. I show that all notions introduced in 
\cite{FST1,FST2,FST3} arise in a natural way in the RGE and 
prove that the skeleton selfenergy is twice differentiable.
This verifies the regularity criterion $(1)$ for Fermi liquids
in perturbation theory. The proof given here is a considerable
simplification of the ones in \cite{FST1,FST2}.

The main technique in doing the regularity proofs, and in showing that
the generalized ladder graphs give the only contribution 
to the four-point function that is not bounded uniformly in $\be$, 
is the volume improvement technique 
invented in \cite{FST1}. Here, I show that the overlapping loops,
and all related concepts developed in \cite{FST1}--\cite{FST3}, appear
naturally in the Wick-ordered continuous RGE.
In particular, 
overlapping loops always appear in the skeleton two-point function, and
the only nonoverlapping part of the skeleton four-point function is 
$m_1=m_2=4$, which is precisely the ladder part of the four-point function.
Moreover, the double overlaps  of \cite{FST3} arise
in a natural way  when the integral equation \Ref{equinteFou}
is iterated. The main reason why all these effects are seen so 
easily is Wick ordering. The RGE without 
Wick ordering has too little structure to make these effects
explicit in a convenient way. For instance, the volume improvement
effect from overlapping loops is a two-loop effect, whereas the 
non-Wick-ordered RGE is a one-loop equation. 

\subsection{Volume-improved bounds}

The indicator functions restrict the spatial support of the 
propagator to regions
\begin{equation}
\cR (\ept) = \{ \Sp \in \cB : \abs{E(\Sp)} \le \ept\}
.\end{equation}
The volume  of the intersection of $\cR$ and its translates 
occurs in the bounds for the integrals in the RGE.
Good bounds for these volumes are the key to the analysis
of these systems. 

\begin{lemma} \label{twoloopvol}
Let $\kz \ge 2$, $\Spi(\rh,\th)$ be as defined in Section \ref{hyposuse},
and
\begin{equation}\label{cWdef}
\cW(\veps) = \sup\limits_{\Sq \in \cB}
\max\limits_{v_i \in \{ \pm 1\}} 
\ili_{S^{d-1}} d\th_1 \ili_{S^{d-1}} d\th_2 \,
\True{\abs{E(v_1 \Spi(0,\th_1) + v_1 \Spi (0,\th_2) + \Sq )} \leq \veps}
.\end{equation}
There is a constant $Q_V\ge 1$ such that for all $0 < \veps \le \epz$
\begin{equation}
\cW(\veps) \leq Q_V \; \veps \; \cases{1+|\log \veps | & if $d=2$\cr
1 & if $d \ge 3$.}
\end{equation}
\end{lemma}
\begin{proof} This is 
Theorem 1.2 of \cite{FST2}. The constant $Q_V$ depends on the curvature
of the Fermi surface, hence on $\epz$.
\end{proof}

I now prepare for the regularity proofs by using the volume improvement 
of Lemma \ref{twoloopvol} to bound the right hand side of the RGE.
This is possible because the graph drawn in Figure \ref{fig2} is 
overlapping according to the graph classification of \cite{FST1},
if $\zil \ge 3$.
\begin{equation}
D^\al \Qrmt{\ul{\Xp}} = \int d\ka_{mr} (m_1,r_1,m_2,r_2,\zil)
\zil\;\zil! \cX_{\al,\zil} (t \mid \ul{\Xp})
\end{equation}
where, after a change of variables $k_j \to (-1)^{l_j}k_j$,
\begin{eqnarray}
&&\cX_{\al,\zil} (t \mid \ul{\Xp})
= \sli_{\si\in \{-1,1\}^\zil}
\sli_{l \in \{1,2\}^\zil}
\sli_{\ul{\al}} \sfrac{\al!}{\al_0!\al_1!\al_2!}
\ili_{\bR\times\cB} dk_2\ldots \ili_{\bR\times\cB} dk_\zil 
 D^{\al_0} \dot \CWt{t} (k_1) \nonu\\
&&\qquad\pli_{j=2}^\zil \CWt{t}(k_j) \;
D^{\al_1} \Irmto{\ul{\Xp}^{(1)},\ul{\Xk}}\;
D^{\al_2} \Irmtt{\sim \ul{\Xk},\ul{\Xp}^{(2)}}
\end{eqnarray}
with $\Xk_j = (k_j,\si_j,l_j)$, the sum over $\ul{\al}$ running over
all triples $(\al_0,\al_1,\al_2)$  with $\al_0+\al_1+\al_2=\al$, and 
$k_1 = - \sum_{j=2}^\zil (-1)^{l_j} k_j + p_1 + \ldots + p_{m_1-\zil+1}$.
$\;\cX_{\al,\zil} $ depends also on $(m_1,r_1,m_2,r_2,m,r)$.
Application of $\abs{\,\cdot\,}_0$ gives
\begin{equation}
\abs{\cX_{\al,\zil}(t)}_0 \le 4^i
\sli_{\ul{\al}} \sfrac{\al!}{\al_0!\al_1!\al_2!}
\; \abs{D^{\al_1} I_{m_1r_1} (t)}_0
\; \abs{D^{\al_2} I_{m_2r_2} (t)}_0
\; \cY_{\al_0,\zil}(t)
\end{equation}
with 
\begin{equation}
\cY_{\al,\zil}(t) = \sup\limits_{Q \in \bM(\be) \times \cB}
\max\limits_{v_i\in \{-1,1\}} \;
\int \pli_{j=2}^\zil |\CWt{t}(k_j)| dk_j 
|D^\al  \dot \CWt{t} ( \sli_{j=2}^\zil v_j k_j +Q)|
\end{equation}

\begin{lemma}\label{Ylemma}
$ \cY_{\al,\zil}(t)=0$ for $t > \log\frac{\be\epz}{\pi}$, and 
\begin{equation}\label{ieq1bou}
 \cY_{\al,\zil}(t) \le (8J_1)^{\zil-1} B_\al\; \ept^{\zil-2-|\al|}
.\end{equation}
If $\zil \ge 3$, 
\begin{equation}\label{ige3bou}
 \cY_{\al,\zil}(t) \le (8J_1)^{\zil-1} B_\al \Ko
\half(1+t) \;\ept^{\zil-1-|\al|}
\end{equation}
with $\Ko$ given in \Ref{Kodef}.
\end{lemma}

\begin{proof} By \Ref{eist}, 
$|D^\al \dot{\CWt{t}}(p) | \le B_\al \ept^{-1-|\al|}$, so
\begin{equation}
 \cY_{\al,\zil}(t) \le B_\al \ept^{-1-|\al|} 
\left( \;\ili_{\bR\times\cB} \abs{\CWt{t} (k)} dk\right)^{\zil-1}
,\end{equation}
thus \Ref{ieq1bou} holds by \Ref{drst}. For $\zil \ge 3$,
\begin{equation}
 \cY_{\al,\zil}(t) \le \left(\;\ili_{\bR\times\cB} 
|\CWt{t} (k)| dk\right)^{\zil-3} 
\;\sup\limits_{Q \in \bM(\be) \times \cB}
\max\limits_{v_1,v_2 \in \{-1,1\}} \; Y_\al(t,Q,v_1,v_2)
\end{equation}
with
\begin{equation}
Y_\al(t,Q,v_1,v_2) = \int dk_1\int dk_2 
|\CWt{t}(k_1)| \; 
|\CWt{t}(k_2)| \; 
|D^\al \dot\CWt{t} (v_1k_1+v_2k_2+Q)|
.\end{equation}
In $Y_\al(t)$, I use $\dot \CWt{t}=0$ for $t > \log \frac{\be\epz}{\pi}$, to write $\CWt{t} (k_j) = - \int_t^{\log(\be\epz)} dt_j \CWt{t_j} (k_j)$.
By \Ref{eist}, 
$|D^\al \dot\CWt{t} (p) | \le B_\al \ept^{-1-|\al|}
\True{\abs{E(\Sp)} \le \ept}$. Inserting this and doing the 
integrals over $(k_1)_0$ and $(k_2)_0$, I get by \Ref{zwst}
\begin{equation}
Y_\al(t,Q,v_1,v_2) \le 8^2 B_\al \ept^{-1-|\al|} 
\ili_t^{\log(\be\epz)}dt_1
\ili_t^{\log(\be\epz)}dt_2\;
\cV(t,t_1,t_2)
\end{equation}
with 
\begin{equation}
\cV(t,t_1,t_2) = 
\ili_{\cR(\epz \E^{-t_1})} d\Sk_1\;
\ili_{\cR(\epz \E^{-t_2})} d\Sk_2\;
\True{\abs{E(v_1\Sk_1+v_2\Sk_2+\SQ)} \le \epz \E^{-t}}
\end{equation}
With the coordinates $(\rh,\th)$ defined in \Ref{stgz},
\begin{eqnarray}
\cV(t,t_1,t_2) &=& 
\ili_{-\epz \E^{-t_1}}^{\epz \E^{-t_1}} d\rh_1
\ili_{-\epz \E^{-t_2}}^{\epz \E^{-t_2}} d\rh_2
\int d\th_1\int d\th_2 \; J(\rh_1,\th_1) J(\rh_2,\th_2)
\nonu\\
&&\True{\abs{E(v_1\Spi(\rh_1,\th_1) + v_2 \Spi(\rh_2,\th_2) + \SQ)}
\le \epz \E^{-t}}
.\end{eqnarray}
Since $\abs{\rh_j} \le \epz \E^{-t_j} \le \epz \E^{-t}$, 
\Ref{stgz} implies 
$|E(v_1\Spi(\rh_1,\th_1) + v_2 \Spi(\rh_2,\th_2) + \SQ)$ $-
E(v_1\Spi(0,\th_1) + v_2 \Spi(0,\th_2) + \SQ)|$ $\le
\sfrac{4}{g_0} \abs{E}_1 \epz \E^{-t}$, with $|E|_1=|\nabla E|_0$.
Using $J(\rh_j,\th_j) \le J_0$ and doing the $\rh$-integrals, I get
\begin{equation}
\cV(t,t_1,t_2) \le (2\epz J_0)^2 \E^{-t_1-t_2} 
\cW \left( (1+ \sfrac{4|E|_1}{g_0}) \epz \E^{-t}\right)
\end{equation}
with $\cW$ given by \Ref{cWdef}. The integrals over $t_1$ and $t_2$ give
\begin{equation}
Y_\al(t,Q,v_1,v_2) \le 4^4 B_\al J_0^2 \; 
\ept^{1-|\al|} \cW \left( (1+ \sfrac{4|E|_1}{g_0}) \epz \E^{-t}\right)
.\end{equation}
This implies \Ref{ige3bou} with 
\begin{equation}\label{Kodef}
\Ko = 2 (\frac{J_0}{J_1})^2 Q_V (1+ \frac{4|E|_1}{g_0})
\left(1+\log (1+ \sfrac{4|E|_1}{g_0}) + \abs{\log\epz}\right)
.\end{equation}
\end{proof}
This gives the following 
\begin{lemma} \label{Xlemma}
For all $\zil$, 
\begin{eqnarray}\label{Xalli}
\abs{\cX_{\al,\zil}(t)}_0 &\le& 4^i (8J_1)^{\zil-1}
\sli_{\ul{\al}} \sfrac{\al!}{\al_0!\al_1!\al_2!}
\; \abs{D^{\al_1} I_{m_1r_1} (t)}_0
\; \abs{D^{\al_2} I_{m_2r_2} (t)}_0
\nonu\\
&& B_{\al_0} \ept^{\zil-2-|\al_0|}.
\end{eqnarray}
For $\zil \ge 3$
\begin{eqnarray}\label{Xige3}
\abs{\cX_{\al,\zil}(t)}_0 &\le& 4^i (8J_1)^{\zil-1}
\sli_{\ul{\al}} \sfrac{\al!}{\al_0!\al_1!\al_2!}
\; \abs{D^{\al_1} I_{m_1r_1} (t)}_0
\; \abs{D^{\al_2} I_{m_2r_2} (t)}_0 \nonu\\
&& B_{\al_0} \Ko \ept^{\zil-1-|\al_0|}
\cases{\half (1+t) & if $d=2$ \cr 1 & if $d\ge 3$.}
\end{eqnarray}
\end{lemma}
\Ref{Xige3} is the implementation of the volume improvement bound
from overlapping loops in the continuous RGE setting. 
In the next section I use it to prove regularity properties of the 
selfenergy.

\subsection{The ladder four-point function and the skeleton selfenergy}
\label{nRPAssuse}
\begin{definition} \label{d55}
The RGE for the 
skeleton functions $\IS_{mr}(t)$is given by \Ref{momcom}, but with
$\int d\ka_{mr}$ replaced by $\int d\kaS_{mr}$, where in 
the latter the sums over $m_1$ and $m_2$ start at four instead 
of two. The function $\IS_{2r} (t)$ is the {\em skeleton selfenergy}
of the model.
\end{definition}

This truncation of the sum prevents any two-legged insertions
from occurring in the graphical expansion. Thus only graphs
without two-legged proper subgraphs contribute to the skeleton 
functions. Since all effective vertices occurring in our model 
have an even number of legs,
the skeleton selfenergy $\Si_r = \IS_{2,r}$
is indeed given by the sum over all two-particle-irreducible
graphs (this is proven in Proposition 2.7 of \cite{FST3})
and the skeleton four-point function is the one-particle-irreducible
part of the four-point function (this is proven in Remark 2.23
of \cite{FST1}). 

\begin{figure}
\epsfxsize=5in
\centerline{\epsffile{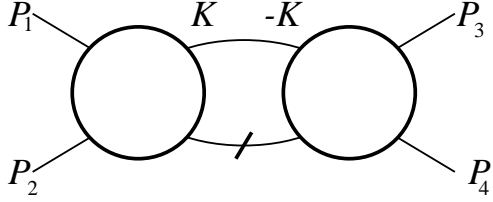}}
\caption{The graph corresponding to $Q_{4,r,2}$}
\label{fig3}
\end{figure}

To give a precise meaning to the statement that the 
ladder resummation takes into account the most singular contributions
to the four-point function,
I split the skeleton four-point function into two pieces, as
follows. In $Q_{4r}$, $m_1+m_2=4+2\zil$, so 
$\zil \ge \half (m_1+m_2-4) \ge 2$ since the skeleton 
condition has removed $m_1=2$ and $m_2=2$. Let  $Q_{4,r,2}$ be the 
$\zil=2$ term in this sum; it corresponds to the `bubble' graph drawn
in Figure \ref{fig3}. 

More explicitly, $Q_{4,r,2}$ is given by 
\begin{eqnarray}\label{Q4r2def}
Q_{4,r,2} (t\mid \Xp_1,&&\mkern-30mu\ldots,\Xp_4) = \half \bA_4 
\Big\lbrack 144 \sli_{r_1+r_2=r} \sli_{i_1,\si_1,i_2,\si_2}
\int dk\; \CWb{t}(\Xk)\;  \dot{\CWb{t}} (\Xk') \nonu\\
&&\IS_{4,r_1} (t\mid \Xp_1,\Xp_2,\Xk',\Xk)\;
\IS_{4,r_2} (t\mid \sim \Xk,\sim \Xk',\Xp_3,\Xp_4)
\Big\rbrack
\end{eqnarray}
with $\Xk=(k,i_1,\si_1)$ and $\Xk'=(p_1+p_2-k,i_2,\si_2)$.
Let 
\begin{equation}
Q_{4,r,\ge 3} (t\mid\ul{\Xp}) = 
Q_{4,r}(t\mid\ul{\Xp}) - Q_{4,r,2} (t\mid \ul{\Xp})
\end{equation}
be the contribution from all terms where at least $\zil \ge 3$ 
lines connect the two vertices in $Q_{4,r}$.
Correspondingly, let 
\begin{equation}
\IS_{4,r} (t) = B_{r} (t) + U_{r} (t)
\end{equation}
where $B_{4,r} (t)$ and $U_{4,r} (t)$ are defined by
\begin{equation}\label{bubcoup}
\frac{\del}{\del t}B_{r} (t) = \half \bA_4 Q_{4,r,2} (t), \quad
\frac{\del}{\del t}U_{r} (t) = \half \bA_4 Q_{4,r,\ge 3} (t)
\end{equation}
with the initial condition $B_{r} (0) + U_{r} (0) = \IS_{4,r}(0)$
(where it is understood that one of the two summands on the left
hand side is set to zero, see below). Note that because 
$\IS_{4,r}$ occurs on the right hand side of both equations
in \Ref{bubcoup}, \Ref{bubcoup} is a coupled system of 
differential equations. 

\begin{definition} \label{ISNUNRDef}
The function $\BL_{r}$ obtained from the skeleton RGE by the 
truncation 
\begin{equation}
\frac{\del}{\del t}U_{r} (t) = 0 , \quad U_{r}(0)=0
\end{equation}
is the ladder skeleton four-point function. The function 
$\UNR_r$ obtained by the truncation 
\begin{equation}
\frac{\del}{\del t}B_{r} (t) = 0 , \quad B_{r}(0)=0
\end{equation}
is the \nonRPA\  skeleton four-point function.
The functions $\ISN_{m,r}$, obtained with this truncation,
(i.e., where $m_1=4$ and $m_2=4$ are left out in the sum 
$\int d\kaS_{4,r}$) are the \nonRPA\  skeleton Green functions. 
\end{definition}
The motivation for the split is that in $Q_{4,r,\ge 3}$, the number
of internal lines of the graph in Figure \ref{fig2} is 
$\zil \ge 3$, so the volume improvement of Lemmas \ref{Ylemma}
and \ref{Xlemma} improves the power counting.

The constants in the following theorems are independent of $\be$.

\begin{theorem} \label{BLsatz}
There is a constant $L_2$ such that 
\begin{equation}
\abs{\BL_r(t)}_0 \le {L_2}^r \;(\half(1+t))^{r-1}
\le {L_2}^r\;\abs{\log(\be\epz)}^{r-1}
\end{equation}
The series $\sli_r \BL_r (t)$ converges uniformly in $t$ and $\ul{\Xp}$
if $\abs{\la \log\be\epz} < {L_2}^{-1}$.  
\end{theorem}

\begin{proof}
This follows immediately by induction from \Ref{Q4r2def} by use of
\Ref{Xalli} with $\zil=2$ and $\al=0$. 
\end{proof}

Using one-loop volume bounds one can show that both the 
particle-particle and the particle-hole ladder are
bounded uniformly in $t$ and $\be$ for $\SQ \ne 0$, so that
the only singularity in the four-point function can arise 
at zero momentum (for a discussion of this, see, e.g., \cite{S}). 
In the particle-particle ladder, this
singularity is really there, and it implies that Fermi liquid 
behaviour occurs only above a critical temperature:
the $\log \be$ in Definition \ref{FLDef} is the logarithm 
occurring in Theorem \ref{BLsatz}.

The next theorem states that the \nonRPA\  skeleton four-point function 
is bounded and that consequently, the \nonRPA\  skeleton 
selfenergy is $C^1$ uniformly in $\be$. This shows that 
indeed, only the ladder four-point function produces a nonuniformity
in $\be$ (this motivates the alternative criterion for
Fermi liquid behaviour in Section \ref{thepurpose}).
The second derivative, however, is only bounded by a power
of $\log \be$ in $d=2$; this motivates why at zero temperature,
the self-energy is only required to be $C^{2-\de}$ for some $\de >0$.

\begin{theorem}\label{skelnonRPA}
For all $r \ge 1$,  the \nonRPA\  skeleton functions 
$\ISN_{2,r} (t)$ converge for $t\to \infty$ to a $C^2$ function. 
There are constants $\Kdrr$ and $\Kvr$, independent of $\be$, 
such that
\begin{equation}
\abs{D^\al \ISN_{2,r} (t)}_0 \le \Kdrr 
\cases{1 & if $|\al| \le 1$ \cr
(\log\be\epz)^2 & if $|\al|=2$ and $d=2$ \cr
\log\be\epz & if $|\al|=2$ and $d \ge 3$.}
\end{equation}
and
\begin{equation}
\abs{D^\al \ISN_{4,r} (t)}_0 \le \Kvr 
\cases{1 & if $|\al| =0$ \cr
\log\be\epz & if $|\al|=1$ \cr
\be\epz & if $|\al|=2$.}
\end{equation}
\end{theorem}
If the ladder four-point function is left in the RGE, its logarithmic
growth (Theorem \ref{BLsatz}) shows up in all other Green functions,
and in the selfenergy:  

\begin{theorem}\label{skelsatz}
The skeleton functions $\IS_{2,r}$ and $\IS_{4,r}$ converge for
$t \to \infty$ and satisfy 
\begin{equation}
\abs{D^\al \IS_{m,r} (t)}_0 \le \Kfmr (\log\be\epz)^r
\end{equation}
for $m=2$, $|\al|\le 2$, and $m=4$, $|\al|=0$. 
\end{theorem}
Theorems \ref{skelnonRPA} and \ref{skelsatz} are proven in the 
next section.
\begin{remark} 
To prove bounds with a good $\be$ behaviour for $m \ge 6$
requires the use of different norms, where part of the 
momenta are integrated \cite{FT}. It is easy to see that
for $m \ge 6$, the connected $m$-point functions have singularities
and thus are not uniformly bounded functions of momentum;
for instance the second order six-point function
is given by $C(\Xp)$, which is $O(\be)$ if $\Sp$ is on the 
Fermi surface and if $\om=\pi/\be$.
\end{remark}

\begin{remark} 
A bound $\const^r$ for the constants in Theorem \ref{skelsatz} 
would suffice to show the first requirement of the Fermi liquid
behaviour defined in Definition \ref{FLDef}, namely the convergence of the 
perturbation expansion for the skeleton functions for
$\abs{\la\log\be} < \const$. The proofs given in this section do not imply 
this bound because in the momentum space equation, the factorial 
remains. It may, however, be possible to give such bounds in $d=2$
by combining the sector technique of \cite{FMRT} with the 
determinant bound. 
\end{remark}

\subsection{The regularity proofs}

At positive temperature, the frequencies are still discrete,
whereas the spatial part of momentum is a continuous variable. 
Whenever derivatives with respect to $p_0$ are written below, they 
are understood as a difference operation 
$\sfrac{\be}{2\pi} (f(p_0+\sfrac{2\pi}{\be})-f(p_0))$.
The RGE actually defines the Green functions for (almost all)
real values of $p_0$, not just the discrete set of 
Matsubara frequencies, so the effect of such a difference can 
be bounded by Taylor expansion. This changes at most constants,
so I shall not write this out explicitly in the proofs.

The following theorem implies Theorem \ref{skelnonRPA} about the 
\nonRPA\  skeleton selfenergy and four-point function. Note that
all bounds in this Theorem are independent of $\be$.

\begin{theorem} \label{vorletzt}
Let $\al$ be a multiindex with $|\al| \le 2$. 
For all $m,r$, let $I_{mr}(0)$ be $C^2$ in $(\Sp_1,\ldots,\Sp_m)$, 
and assume that there are $\Kzmr \ge 0$ such that 
$\abs{D^\al I_{mr}(0)} \le \Kzmr$, with $\Kzmr =0$ for $m > 2r+2$,
and $K_{m1}^{(0)} = \de_{m4} v$, where $v > 0$. 
Let $\ISN_{mr} (t)$ be the Green functions generated by the 
\nonRPA\  skeleton RGE, as given in Definition \ref{ISNUNRDef},
with initial values $I_{mr}(0)$. Let $\kK_{m1} = K_{m1}^{(0)}$ and for
$r \ge 2$, let
\begin{equation}\label{kKmrdef}
\kK_{mr} = \Kzmr + \frac{1}{m}\Mo \int d\tilde\ka_{mr}
\zil\;\zil! (32J_1)^{\zil-1} \kK_{m_1r_1}\kK_{m_2r_2}
\end{equation}
where $\Mo = 240 B\Ko$, with $B=\max_\al B_\al$. 
Then $\kK_{mr} =0$ if $m > 2r+2$, and for all $t\ge 0$ and all $m,r$
\begin{equation}\label{iha}
\abs{D^\al \QSN_{mr} (t)}_0 \le 2 \kK_{mr} 
\cases{ \ept^{2-{m\over 2} - |\al|} & if $m\ge 6$\cr
\ept^{1-|\al|} \frac{1+t}{2} & if $m = 4$ \cr
\ept^{2-|\al|} \left(\frac{1+t}{2}\right)^{\de_{d,2}} & if $m = 2$.} 
\end{equation}
Moreover, for $m\ge 6$,
\begin{equation}\label{ihb6}
\abs{D^\al \ISN_{mr} (t)}_0 \le \kK_{mr}\ept^{2-{m\over 2} - |\al| }
,\end{equation}
for $m=4$
\begin{equation}\label{ihb4}
\abs{D^\al \ISN_{mr} (t)}_0 \le \kK_{4r} 
\cases{ 1 & if $\al =0 $ \cr
\left(\frac{1+t}{2}\right)^2 & if $|\al|=1$ \cr
\ept^{-1} \frac{1+t}{2} & if $|\al|=2$,}
\end{equation}
and for $m=2$ 
\begin{equation}\label{ihb2}
\abs{D^\al \ISN_{mr} (t)}_0 \le \kK_{2r} 
\cases{ 1 & if $|\al| \le 1$ \cr
\left(\frac{1+t}{2}\right)^2 & if $|\al|=2$ and $d=2$\cr
\frac{1+t}{2} & if $|\al|=2$ and $d=3$.}
\end{equation}
\end{theorem}

\begin{proof}
Induction in $r$, with the statement of the Theorem as the inductive 
hypothesis. $r=1$ is trivial because $\QSN_{m1}=0$ and because the 
statement holds for $I_{m1}(0)$. Let $r \ge 2$ and the statement hold
for all $r' < r$. The inductive hypothesis applies to both factors 
$\ISN_{m_kr_k}$ in $\QSN_{mr}$. For $m_k=4$, it implies that 
\begin{equation}
\abs{\ISN_{m_kr_k}(t)}_0 \le \kK_{m_kr_k} \ept^{-|\al|} =
\kK_{m_kr_k} \epz^{-|\al|} \E^{t(|\al|+{m_k\over 2}-2)}
\end{equation}
for all $t \ge 0$. Recall \Ref{Xalli} and \Ref{Xige3}, and 
\begin{eqnarray}
\abs{D^\al \bA_m \QSN_{mr} (t)}_0 & = & 
\abs{\bA_m D^\al \QSN_{mr} (t)}_0 \le \abs{D^\al \QSN_{mr} (t)}_0 
\nonu\\
& \le & \int d\kaSN_{mr} \zil\; \zil! \abs{\cX_{\al,\zil}(t)}_0
.\end{eqnarray}

Let $m \ge 6$. By \Ref{Xalli} and $m_1+m_2=m+2\zil$, the $t$-dependent
factors in $\cX_{\al,\zil}(t)$ are 
\begin{equation}
\E^{t({m_1\over 2} -2 + |\al_1|)} \;
\E^{t({m_2\over 2}-2+|\al_2|)} 
\E^{t(2-\zil+|\al_0|)} = 
\E^{t({m\over 2} -2 + |\al|)}
\end{equation}
Since $\frac{m}{2}-2+|\al| \ge \frac{m}{2}-2 \ge 1$, integrating the 
RGE gives
\begin{eqnarray}
\abs{D^\al\ISN_{mr} (t)}_0 & \le & \abs{D^\al\ISN_{mr} (0)}_0 
+ \half\ili_0^t ds\; \abs{D^\al \QSN_{mr} (s)}_0
\nonu\\
& \le & \Kzmr + \frac{1}{m-4}\nu_{mr} \ept^{2-{m\over 2} - |\al|}
\end{eqnarray}
where 
\begin{equation}
\nu_{mr} =  4 B \int d\kaS \zil\; \zil! (32J_1)^{\zil-1}
\sli_{\ul{\al}} \sfrac{\al!}{\al_0!\al_1!\al_2!}
\kK_{m_1r_1}\kK_{m_2r_2}
.\end{equation}
For $m \ge 6$, $m-4 \ge \frac{m}{3}$. Moreover, 
$\sum \sfrac{\al!}{\al_0!\al_1!\al_2!} \le
3^{|\al|}$, so \Ref{iha} and \Ref{ihb6} follow.

Let $m=4$. One of $m_1$ and $m_2$ must be at least six because
the ladder part is left out in $\ISN_{mr}$. Thus
$\zil = \half(m_1+m_2-4) \ge 3$. By \Ref{Xige3}, there is an 
extra small factor $\ept=\epz\E^{-t}$, so 
\begin{equation}
\abs{D^\al \QSN_{4,r} (t)}_0 \le \frac{1+t}{2}\ept^{1-|\al|}
\; \Ko \nu_{mr}
,\end{equation}
which proves \Ref{iha}. \Ref{ihb4} follows by integration. 

Let $m=2$. The case $m_1=m_2=2$ is excluded since this is the skeleton 
RG. Thus $\zil =\half (m_1+m_2-2) \ge 3$, and \Ref{Xige3} implies
\begin{equation}
\abs{D^\al \QSN_{2,r} (t)}_0 \le 
\left(\frac{1+t}{2}\right)^{\de_{d,2}}\;
\ept^{1-|\al|}\; \Ko \nu_{2r}
.\end{equation}
Thus \Ref{iha} holds, and \Ref{ihb2} follows by integration.
\end{proof}

\begin{proof}[of Theorem \ref{skelnonRPA}]
Let $|\al| \le 1$. By \Ref{ihb2}, $|D^\al \ISN_{2r}(t)| \le \kK_{2,r}$.
By \Ref{iha}, $|D^\al \sfrac{\del}{\del t}\ISN_{2r}(t)| \le 
\kK_{2,r} \epz \E^{-t} \; \sfrac{1+t}{2} \to 0$ as $t \to \infty$. 
Thus the limit 
$t\to \infty$ of $\ISN_{2,r}(t)$ exists and is a $C^1$ function 
of $(\Sp_1,\Sp_2)$. All constants are uniform in $\be$. 
The second derivative of $\QSN_{2,r}$ is $O(t)$ in $d=2$ 
and $O(1)$ in $d=3$. Since $\QSN_{2,r} = 0$ for $t > \log \be\epz$, 
the integral for $\ISN_{2,r}$ over $t$ runs only up to $\log \be\epz$,
which gives the stated dependence on $\log \be\epz$.
\end{proof}
\begin{remark} Note that the bounds for $m=2$ and $|\al| =2$ in Theorem
\ref{vorletzt} do not imply convergence. This is the source of the logarithmic 
behaviour discussed in Section \ref{thepurpose} and in 
\cite{FST2}.
\end{remark}

The following theorem implies Theorem \ref{skelsatz}. 
Here the bounds have an explicit $\be$-dependence. 
One could avoid this $\be$-dependence by including 
polynomials in $t$, but this would also increase the 
combinatorial coefficients by factorials. 

\begin{theorem}
Let $\al$ be a multiindex with $|\al| \le 2$. 
For all $m,r$, let $I_{mr}(0)$ be $C^2$ in $(\Sp_1,\ldots,\Sp_m)$, 
and assume that there are $\Kzmr \ge 0$ such that 
$\abs{D^\al I_{mr}(0)} \le \Kzmr$, with $\Kzmr =0$ for $m > 2r+2$,
and $K_{m1}^{(0)} = \de_{m4} v$, where $v > 0$. 
Let $\IS_{mr} (t)$ be the Green functions generated by the 
skeleton RGE, as given in Definition \ref{d55},
with initial values $I_{mr}(0)$. Let $\kK_{m1} = K_{m1}^{(0)}$ and for
$r \ge 2$, let $\kK_{mr}$ be given by \Ref{kKmrdef}.
Then for $m \ge 4$
\begin{equation}
\abs{D^\al \QS_{mr}(t)}_0 \le 2\kK_{mr} \; (\log\be\epz)^{r-2}
\ept^{2-{m\over 2}-|\al|}
\end{equation}
and 
\begin{equation}
\abs{D^\al \IS_{mr}(t)}_0 \le \kK_{mr} \; (\log\be\epz)^{r-1}
\ept^{2-{m\over 2}-|\al|}
.\end{equation}
For $m=2$, 
\begin{equation}
\abs{D^\al \QS_{2r}(t)}_0 \le 2\kK_{2r} \; (\log\be\epz)^{r-2+\de_{d,2}}
\ept^{2-|\al|}
\end{equation}
and 
\begin{equation}
\abs{D^\al \IS_{2r}(t)}_0 \le \kK_{2r} \; (\log\be\epz)^{r-2}
\cases{ 1 & if $|\al| \le 1$ \cr (\log\be\epz)^{1+\de_{d,2}}
& if $|\al|=2$.}
\end{equation}
\end{theorem}

\begin{proof} The proof is by induction in $r$, with the statement of the 
theorem as the inductive hypothesis. It is similar to the proof of 
Theorem \ref{vorletzt}, with only a few changes. Note that $m_1=2$ 
and $m_2=2$ never appear on the right hand side of the RGE because
of the skeleton truncation in Definition \ref{d55}. For $m \ge 4$, 
use \Ref{Xalli}; this gives 
\begin{equation}\label{47stro}
\abs{D^\al \QS_{mr}(t)}_0 \le 9 \nu_{mr} \; (\log \be\epz)^{r-2}\;
\ept^{2-{m\over 2} -|\al|}
.\end{equation}
For $m \ge 6$, the scale integral is as in the proof of 
Theorem \ref{vorletzt}. For $m=4$, the scale integral is now
$\int_0^t ds\; = t \le \log \be\epz$. This produces the powers
of $\log \be\epz$ upon iteration. 
For $m \ge 2$, use \Ref{Xige3}; this gives 
$\Ko \ept^{2-|\al|} (\half(1+t))^{\de_{d,2}}$ instead of 
$\ept^{2-{m\over 2} -|\al|}$ in \Ref{47stro}. 
The theorem now follows by integration over $t$, recalling that 
the upper integration limit is at most $\log \be\epz$.
\end{proof}

\begin{proof} [ of Theorem \ref{skelsatz}] Convergence of the 
selfenergy follows for $|\al| \le 1$
as in the proof of Theorem \ref{skelnonRPA}.
For $m=4$, the function is bounded uniformly in $t$. 
For $|\al|=2$, convergence at $\be > 0$ holds 
because $\sfrac{\del}{\del t} \IS_{4r} (t) = 0$ for $t> \log \be\epz$.
\end{proof}

\section{Conclusion}
\label{sect7}

The determinant bound for the continuous Wick-ordered RGE 
removes a factorial in the recursion for the 
Green functions. If the model has a propagator with a pointlike
singularity, power counting bounds that include this combinatorial
improvement can be proven rather easily. The improvement may lead 
to convergence, but I have not proven this here. If it does, 
then Theorem \ref{opoco} implies analyticity of the Green functions
where all relevant and marginal couplings are left out 
in a region independent of the energy scale, and Theorem
\ref{mfsge2} implies analyticity of the full
Green functions for $|\la|\be^d < \const$
in many-fermion systems for all $d\ge 1$. 
Natural models to which Theorem \ref{opoco} applies 
are the Gross-Neveu model in two dimensions and the many-fermion 
system in one dimension. In both cases, I have only given 
bounds where the marginal and relevant terms where left out,
to give a simple application of the determinant bound derived
in Section \ref{sect4}. In both cases, analyses of the full models,
including the coupling flows,
have been done previously (\cite{GK,FMRS} and \cite{BG,BGPS,BM}). 

Many-fermion models in $d \ge 2$ are the most realistic physical systems 
where the interaction is regular enough for the analysis done here
to apply directly (i.e., without the introduction of boson fields).
I have defined a criterion for Fermi liquid behaviour for these
models. A proof that such behaviour occurs requires more detailed
bounds and a combination of the method with the sector method of \cite{FMRT}.
This may be feasible by an extension of the analysis done here. 
The Jellium dispersion relation
$E(\Sk) = \Sk^2/2 -\mu$ is the case where the proofs
are easiest because $\Si\vert_\FS$ is constant. The proof
that such a system is a Fermi liquid is possible by a combination
of the techniques of \cite{FMRT} and \cite{FST1}; 
is may also be within the reach of the continuous RGE method 
developed here. Perturbative bounds that implement the  
overlapping loop method of \cite{FST1} in a simple way 
were given in Section \ref{sect6}. 

The verification of regularity property $(2)$ for 
nonspherical Fermi surfaces is not as simple, but the 
perturbative analysis done here
can be extended rather easily to include the double 
overlaps used in \cite{FST3}, because the graph classification of \cite{FST3}
arises in a natural way when the integral equation for the 
effective action is iterated. Multiple overlaps can also be exploited
in the RGE; this may be necessary for the many-fermion systems in 
$d \ge 3$. The split of the four-point function 
into the ladder and \nonRPA\  part done in Section \ref{nRPAssuse} 
singles out the only singular contributions to the four-point 
function and the least regular 
contributions to the self-energy in a simple way. The treatment of 
these ladder contributions was done in \cite{FST2}, where bounds uniform in 
the temperature were shown for the second derivative in perturbation 
theory.  

The results in Section \ref{nRPAssuse} provide another proof 
that only the ladder flow corresponding to Figure \ref{fig3} leads to 
singularities and hence instabilities. It should be noted that
this statement depends on the assumptions stated in Section 
\ref{hyposuse}, in particular on the choice of $\epz$, 
in the following specific way. 
The curvature of the Fermi surface sets a 
natural scale which appears via the constants in the volume bounds. 
Above this scale,
the geometry of the Fermi surface provides no justification of
restricting to the ladder flow. 
This is of some relevance in the 
Hubbard model, where two scale regimes arise in a natural way: if 
$\ept > \tilde\mu$, 
where $\tilde\mu=\mu-{\rm t} d$, $t$ the hopping parameter,
is defined such that $\tilde\mu=0$ is 
half-filling, then the curvature of the Fermi surface is
effectively so small that one can replace the Fermi surface 
by a square.
More technically speaking, the constant $Q_V$, which depends on the
curvature of the Fermi surface, diverges for $\tilde\mu \to 0$.
Thus, for small $\tilde \mu$, $\ept$ has to be very small for
$Q_V \ept < 1$ to hold. If $Q_V \ept >1$, the improved volume estimate 
does not lead to a gain over ordinary power counting. 
Only below this energy scale, the curvature 
effects of the volume improvement bound, Lemma \ref{twoloopvol},
imply domination of the ladder part of the four-point 
function. To get to scale $\tilde\mu$, one has to calculate the 
effective action. Needless to say, the effective four-point interaction
at scale $\tilde \mu$ may look very different from the original interaction,
and RPA calculations suggest that the antiferromagnetic correlations
produced by the almost square Fermi surface lead to an 
attractive nearest-neighbour-interaction \cite{Vignale}. 
However, one should keep in mind that for the reasons just mentioned,
it is an ad hoc approximation to keep only the RPA part of the 
four-point function above scale $\tilde\mu$, and that a correct treatment
must either give a different justification of the ladder approximation,
or replace it by a better controlled approximation. 

Above, I have not discussed how one gets from the skeleton Green 
functions to the exact Green functions. The key to this 
is to take a Wick ordering covariance
which already contains part of the self-energy. That is, the Gaussian
measure changes in a nontrivial way with $t$. This can be done such 
that all two-legged insertions appear with the proper renormalization 
subtractions, and it gives a simple procedure for a rigorous 
skeleton expansion. Moreover, it makes clear why it is
so important to establish regularity of the self-energy: once the 
self-energy appears in the propagator, its regularity is needed to 
show Lemma \ref{dtiprop} (or its sector analogue), 
on which in turn, all other bounds depend.
The condition $\kz > d$ also enters in this lemma and seems 
indispensable from the point of view of the method developed here.
Details about this modified Wick ordering technique will appear later.

There is a basic duality in the technique applied to these fermionic
models. The determinant bound has to be done in position space, 
but the regularity bounds use geometric details that are most easily seen 
in momentum space. The continuous RGE shows very nicely that
those terms that require the very detailed regularity
analysis are very simple from the combinatorial point of view and
vice versa.

\ifspr\else
\newpage
\fi

\begin{appendix}

\section{Fourier transformation} 
\label{fourapp}
Recall that $\ps$ is defined on the doubled time direction 
$\bT_2$ and obeys the antiperiodicity with respect to 
translations by $\be$, \Ref{psanti}, and that 
$\Nta $ was chosen to be even. $\bT_2$ is the set 
$\bT_2 = \Teps \bZ / 2 \be \bZ = \bT \cup (\bT+\be)$
where 
\begin{equation}
\bT = \{ \ta \in \bT_2: \ta=\Teps k, \; 
k\in \{-\frac{\Nta}{2}, \ldots, \frac{\Nta}{2}-1\}\}
.\end{equation}
The dual to $\bT_2$ is 
$\bT_2^* = \frac{\pi}{\be}\bZ/2 \Nta\bZ =
\{ \om=\frac{\pi}{\be}k:\; k\in \{-\Nta, \ldots, \Nta-1\}\}$.
The Fourier transform on $\bT_2$ is 
\begin{equation}
\tilde f(\om) = \Teps \sli_{\ta \in \bT_2} \E^{-\I \om\ta} f(\ta),
\quad
f(\ta) = \frac{1}{2\be} \sli_{\om \in \bT_2} \E^{\I \om\ta} \tilde f(\om).
\end{equation}
If $f(\ta-\be) = -f(\ta)$, then $\tilde f(\om) =0$ if $\frac{\om\be}{\pi}$
is even. In that case, with $\hat f(\om) = \half \tilde f(\om)$, 
\begin{equation}
f(\ta) = \frac{1}{\be} \sli_{\om \in \Mats{\Nta}} \E^{\I \om\ta} \hat f(\om)
\end{equation}
with $\Mats{\Nta}$ given by \Ref{Matsdef}.  
The orthogonality relations are
\begin{equation}
\ili_{\bT} d\ta \; \E^{\I (\om_n\pm\om_m)\ta} =
\be \de_{mn},
\quad
\frac{1}{\be}\sli_{\om\in\Mats{\Nta}} \E^{\I \om\ta} =
\frac{1}{\Teps}
(\de_{\ta 0} -\de_{\ta\be})
.\end{equation}

\begin{lemma} \label{LemA1}
Let $\be > 0$, $\Ez \ge 0$, and $\Nta \ge 2\be(\Ez+E_{\rm max})$,
where $E_{\rm max}$ is defined in 
\Ref{EMAX}, and let $\Hat{\om}$ be defined as in \Ref{Hatomdef}.
Then for all $\Sk \in \cB$ and all $\om \in \Mats{\Nta}$,
$|\I\Hat{\om} - E(\Sk)| \le \Ez$ implies $|\om| \le \sfrac{\pi}{2}\Ez$
and $|E(\Sk)| \le 2 \Ez$, and 
\begin{equation}
\sfrac{1}{\be} \sli_{\om\in\Mats{\Nta}} 
\True{|\I\Hat{\om} - E(\Sk)| \le \Ez} \le 
\Ez \True{ |E(\Sk)| \le 2\Ez}.
\end{equation} 
\end{lemma}

\begin{proof}
By \Ref{Hatomdef}, $\mbox{Im }\Hat{\om} = \sfrac{1}{\Teps}(1-\cos(\om\Teps))$ and 
$\mbox{ Re }\Hat{\om} =  \sfrac{1}{\Teps} \sin(\om\Teps)$.
The condition
$|\I\Hat{\om} - E(\Sk)| \le \Ez$ implies $|\mbox{Re }\Hat{\om}|\le\Ez$
and $|\mbox{Im }\Hat{\om}+E(\Sk)| \le \Ez$.
Thus $|\mbox{Im }\Hat{\om}| \le \Ez+E_{\rm max}$. 
Since $1-\cos x \ge \sfrac{2}{\pi^2} x^2$, 
$(\Ez+E_{\rm max})\Teps \ge \Teps |\mbox{Im }\Hat{\om}| =
1-\cos(\om\Teps) \ge \sfrac{2}{\pi^2} (\om\Teps)^2$,
so $|\om\Teps| \le \pi (\be(\Ez+E_{\rm max})(2\Nta)^{-1})^{1/2}
\le \sfrac{\pi}{2}$. Since $\sfrac{\sin x}{x}$ is decreasing on
$[0,\sfrac{\pi}{2}]$, $\Ez \ge |\mbox{Re }\Hat{\om}| \ge 
|\om| |\sfrac{\sin (\om\Teps)}{\om\Teps} | \ge \sfrac{2}{\pi} |\om|$.
So $|\om| \le \sfrac{\pi}{2} \Ez$. Since $1-\cos x \le \sfrac{1}{2} x^2$,
\begin{equation}
\abs{\mbox{Im }\Hat{\om}} \le \sfrac{1}{2\Teps} (\om\Teps)^2 \le
\frac{\pi^2}{8} \be \Ez^2 \frac{1}{\Nta} \le \Ez
.\end{equation}
Thus $ |\mbox{Im }\Hat{\om}+E(\Sk)| \le \Ez$ implies
$|E(\Sk)| \le 2 \Ez$. In terms of indicator functions, this means that
$\True{|\I\Hat{\om} - E(\Sk)| \le \Ez} \le 
\True{|\om| \le \sfrac{\pi}{2}\Ez}
\True{ |E(\Sk)| \le 2\Ez}$.
The summation over $\Mats{\Nta}$ gives
\begin{equation}
\sfrac{1}{\be} \sli_{\om\in\Mats{\Nta}} \True{|\om| \le \sfrac{\pi}{2}\Ez}
= \sfrac{1}{\be} \sli_{n=-{\Nta\over2}}^{\Nta \over 2} 
\True{|2n+1| \le \sfrac{\be \Ez}{2}} \le \Ez
.\end{equation}
Note that the last inequality holds in particular 
if $\sfrac{\be\Ez }{2} < 1$, because then the sum is empty,
hence the result zero, because $2n+1$ is always odd.
\end{proof}


\section{Wick ordering} \label{appWick}
Recall the conventions fixed at the beginning of 
Section \ref{sect3}.
\begin{definition} 
Let $\cW_\Ga(\et ,\ps ) = \E^{(\et ,\ps )_\Ga - 
\frac{1}{2} (\et ,\, C\, \et )_\Ga}$,
and let $\cA_\Ga'$ be the Grassmann algebra generated by 
$(\ps (x))_{x \in \Ga}$. Wick ordering is the ${\Bbb C}$--linear map
$\Omega_C: \cA_\Ga' \to \cA_\Ga'$ 
that takes the following values on the monomials: $\WO{C}(1)=1$,
and for $n \ge 1$ and $\Xx_1,\ldots ,\Xx_n \in \Ga$,
\begin{equation}
\WO{C}\big(\ps (\Xx_1) \ldots \ps (\Xx_n)\big) =
\left\lbrack \pli_{k=1}^n 
\frac{\de}{\de \et (\Xx_k)} \; \cW_\Ga(\et ,\ps )\right\rbrack_{\et =0}
\end{equation}
\end{definition}

\begin{theorem} \label{allWick}
$\cW_\Ga(\et ,\ps ) = \WO{C}\left(\E^{(\et ,\ps )_\Ga}\right)$.
Let $\al_1, \ldots, \al_n$ be Grassmann variables,
and let $\al(\Xx) = \sum_{k=1}^n \al_k \; \de_\Ga (\Xx,\Xx_k)$.
Then 
\begin{equation}\label{nuetzli}
\WO{C}\big(\ps (\Xx_1) \ldots \ps (\Xx_n)\big) = 
\left\lbrack \left(\pli_{k=1}^n 
\frac{\del}{\del\al_k} \right) 
 \WO{C}\left(\E^{(\al ,\ps )_\Ga}\right)
\right\rbrack_{\al=0}
\end{equation}
\end{theorem}

\begin{proof} By Taylor expansion in $\et $, and by definition of $\WO{C}$,
\begin{eqnarray}
\cW_\Ga (\et ,\ps ) & = & \sli_{n \ge 0} \frac{1}{n!}\; 
\sli_{\Xx_1,\ldots,\Xx_n \in \Ga} 
\left\lbrack \pli_{k=1}^n \et (\Xx_k) \frac{\del}{\del \al (\Xx_k)} \; 
\cW_\Ga (\al ,\ps )\right \rbrack_{\al =0} \nonu \\
& = & \sli_{n \ge 0} \frac{1}{n!}\;
\int d\Xx_1 \ldots d\Xx_n\; 
\left\lbrack \pli_{k=1}^n \et (\Xx_k) \frac{\de}{\de \al (\Xx_k)} \; 
\cW_\Ga (\al ,\ps )\right \rbrack_{\al =0} \nonu \\
& = & \sli_{n \ge 0} \frac{1}{n!}\;
\int d\Xx_1 \ldots d\Xx_n\; 
\WO{C} \left(\pli_{k=1}^n \et (\Xx_k) \ps (\Xx_k)\right) \nonu\\
& = & \sli_{n \ge 0} \frac{1}{n!}\;
\WO{C} \left({(\et ,\ps )_\Ga}^n\right)
=\WO{C}\left( \E^{(\et ,\ps )_\Ga}\right)
.\end{eqnarray}
\Ref{nuetzli} follows directly from the definition of $\WO{C}$.
\end{proof}

The next theorem contains the alternative formula \Ref{WiLa} for the Wick
ordered monomials used in Section \ref{sect2}.
\begin{theorem}\label{ehLC}
Let $\Lp{C_t}$ be defined by \Ref{LpCtdef}. 
Then 
\begin{equation}\label{altrep}
\WO{C_t}\big(\ps (\Xx_1) \ldots \ps (\Xx_n)\big) =
\E^{-\Lp{C_t}} \;\ps (\Xx_1) \ldots \ps (\Xx_n)
.\end{equation}
In particular, if $C_t$ depends differentiably on $t$, then
\begin{equation}\label{Lapder}
\frac{\del}{\del t} \WO{C_t}\big(\ps (\Xx_1) \ldots \ps (\Xx_n)\big) = -
\frac{\del \Lp{C_t}}{\del t} \WO{C}\big(\ps (\Xx_1) \ldots \ps (\Xx_n)\big)
\end{equation}
\end{theorem}
\begin{proof} For any formal power series $f(z)=\sum f_k z^k$,
\begin{equation}
f(\Lp{C_t} ) \; \E^{(\et ,\ps )_\Ga} = f\left(\frac{1}{2} 
(\et \;,C_t\; \et )_\Ga\right)
\; \E^{(\et ,\ps )_\Ga}
,\end{equation}
so $\E^{-\Lp{C_t}} \; \E^{(\et ,\ps )_\Ga} = 
\E^{-\frac{1}{2} (\et ,C_t\et )_\Ga + (\et ,\ps )_\Ga}$,
from which \Ref{altrep} follows by definition of $\WO{C_t}$, because 
derivatives with respect to $\et $ commute with $\Lp{C_t}$. If $C_t$ depends
differentiably on $t$, \Ref{Lapder} follows by taking 
a derivative of \Ref{altrep} because $\Lp{C_t}$ commutes with 
$\frac{\del \Lp{C_t}}{\del t}$. 
\end{proof}


\section{Wick reordering} \label{ReWick}

By \Ref{expaWO}  and \Ref{toWm},
$\cQ_r(t,\ps) = \sfrac{1}{2} \sli_{r_1\ge 1,r_2\ge 1\atop r_1+r_2=r}
\sli_{m_1=1}^{\bar m(r_1)}
\sli_{m_2=1}^{\bar m(r_2)}
\cY_{r_1,m_1,r_2,m_2}$
with
\begin{equation}
\cY_{r_1,m_1,r_2,m_2} = \int d\ul{\Xx}\; 
\Grmto{\ul{\Xx}^{(1)}}  \; \Grmtt{\ul{\Xx}^{(2)}}\;
\PP(\ul{\Xx},\ps)
\end{equation}
where $\ul{\Xx}^{(1)} = (\Xx_1, \ldots, \Xx_{m_1})$,
$\ul{\Xx}^{(2)} = (\Xx_{m_1+1}, \ldots, \Xx_{m_1+m_2})$,
$\ul{\Xx} = (\Xx_1, \ldots,$ $ \Xx_{m_1+m_2})$,
$\PP(\ul{X},\ps) = \int d\Xx d\Xy \dot C_t (\Xx,\Xy) \; \om_{\Xx,\Xy}(\ps)$,
and
\begin{equation}
\om_{\Xx,\Xy}(\ps) = \left(\frac{\de}{\de \psi(\Xx)}
\WO{\CW{t}}\left( \pli_{k_1=1}^{m_1} \psi(\Xx_{k_1})\right)\right) \; 
\frac{\de}{\de \psi(\Xy)}
\WO{\CW{t}}\left( \pli_{k_2=1}^{m_2} \psi(\Xx_{m_1+k_2})\right) 
.\end{equation}
Only even $m_1$ and $m_2$ contribute to the sum for $\cQ_r(t,\ps)$ 
because $\cG_r(t,\psi)$ is an element of the even subalgebra for all $t$.
Let 
\begin{equation}
\et^{(1)} = \sli_{k=1}^{m_1} \et_k \; \de_\Ga (\Xx,\Xx_k), \quad
\et^{(2)} = \sli_{k=1}^{m_2} \et_{m_1+k} \; \de_\Ga (\Xx,\Xx_{m_1+k}), \quad
\end{equation}
and $\et = \et^{(1)}+\et^{(2)}$. 
With this, 
\begin{eqnarray}
\frac{\de}{\de \psi(\Xx)}
\WO{\CW{t}}&& \kern-26pt\left( \pli_{k_1=1}^{m_1} \psi(\Xx_{k_1})\right) =
\left\lbrack\frac{\de}{\de \psi(\Xx)}
\pli_{k=1}^{m_1} \frac{\del}{\del \et_k} 
\WO{\CW{t}}\left(\E^{(\et^{(1)},\ps)_\Ga}\right)\right\rbrack_{\et=0}
\nonu\\
&=& \vv^{m_1} \left\lbrack
\pli_{k=1}^{m_1} \frac{\del}{\del \et_k} 
\frac{\de}{\de \psi(\Xx)}
\E^{(\et^{(1)},\ps)_\Ga - \half (\et^{(1)},\CW{t}\et^{(1)} )_\Ga}
\right\rbrack_{\et=0}
\nonu \\
& = & \left\lbrack \pli_{k=1}^{m_1} \frac{\del}{\del \et_k} 
\vv\et^{(1)}(\Xx)
\E^{(\et^{(1)},\ps)_\Ga - \half (\et^{(1)},\CW{t}\et^{(1)} )_\Ga}
\right\rbrack_{\et=0}
\end{eqnarray}
since $\vv^{m_1} = 1$. Similarly, 
\begin{equation}
\WO{\CW{t}}\left( \pli_{k=1}^{m_2} \psi(\Xx_{k})\right) =
\left\lbrack \pli_{k=1}^{m_2} \frac{\del}{\del \et_{m_1+k}} 
\vv\et^{(2)}(\Xy)
\E^{(\et^{(2)},\ps)_\Ga - \half (\et^{(2)},\CW{t}\et^{(2)} )_\Ga}
\right\rbrack_{\et=0}
\end{equation}
Since $\et^{(1)}$ is independent of $\et_{m_1+1}, \ldots, \et_{m_1+m_2}$, 
the derivatives with respect to $\et^{(2)}$ can be commuted through 
so that 
\begin{equation}
\PP(\ul{\Xx},\ps) =
\left\lbrack \frac{\del^{m_1+m_2}}{\del\et_1 \ldots \del\et_{m_1+m_2}}
\; Z(\et,\ps) \right\rbrack_{\et=0}
\end{equation}
with 
\begin{equation}
Z(\et,\ps) = (\et^{(1)}, \;\dot C_t \;\et^{(2)}) \; 
\E^{(\et,\ps)_\Ga - \half (\et^{(1)},\; \CW{t} \;\et^{(1)})_\Ga 
-\half (\et^{(2)},\;\CW{t} \;\et^{(2)})_\Ga}
\end{equation}
Wick ordering of $ \E^{(\et,\ps)_\Ga }$, the antisymmetry of 
$\CW{t}$, and $\dot C_t = -\dot \CW{t}$, give 
\begin{equation}\label{Z3}
Z(\et,\ps)=
\left(-\; \sfrac{\del}{\del t}\; 
\E^{(\et^{(1)}, \;\CW{t}\; \et^{(2)})_\Ga} 
\right) \WO{\CW{t}} \left(\E^{(\et,\ps)_\Ga }\right)
.\end{equation}

\begin{lemma}\label{umdiff}
Let $A$ and $B$ be elements of the Grassmann algebra 
generated by $(\et(\Xx))_{\Xx\in \Ga}$, 
that is, $A = \sum_{I \subset \Ga} a_I \; \eta^I$ and 
$B = \sum_{I \subset \Ga} b_I \; \eta^I$,
where $a_I, b_I \in {\Bbb C}$, and let $F(x)$ be the formal power series
$F(x) = \sum_{r\ge 0} f_r x^r$. Then 
\begin{equation}
\left\lbrack 
A(\frac{\del}{\del\eta}) \; B(\eta) \; F\big( (\eta,\psi)\big)
\right\rbrack_{\eta=0}
 =
\left\lbrack 
A(\eta) \; B(\frac{\del_L}{\del\eta})\;  
F\big( (\frac{\del_L}{\del\eta},\psi)\big)
\right\rbrack_{\eta=0}
\end{equation}
where $\frac{\del_L}{\del\eta}$ is the derivative with respect to $\eta$,
acting to the left. 
\end{lemma}

\begin{proof} The proof is an easy exercise in Grassmann algebra and 
is left to the reader.
\end{proof}

Let $m_1+m_2=\mu$.  By Lemma \ref{umdiff}
\begin{equation}\label{schwupp}
\left\lbrack \frac{\del^{\mu}}{\del\et_1 \ldots \del\et_{\mu}}
\; Z(\et,\ps) \right\rbrack_{\et=0} =
\left\lbrack \et_1 \ldots \et_{\mu}
\; Z(\dlinks{\et},\ps) \right\rbrack_{\et=0}
\end{equation}
By definition of $\et^{(1)}$ and $\et^{(2)}$, 
\begin{equation}
\left( \dlinks{\et^{(1)}}, \CW{t} 
\dlinks{\et^{(2)}}\right) = 
\sli_{k_1=1}^{m_1} 
\sli_{k_2=m_1+1}^{\mu} 
\dlinks{\et_{k_1}}\CW{t} (\Xx_{k_1},\Xx_{k_2})
\dlinks{\et_{k_2}}
.\end{equation}
Every derivative can act only once on $\et_1 \ldots \et_\mu$
without giving zero because the $\et$'s are all different. So 
\begin{eqnarray}
\et_1\ldots\et_\mu &&\kern-20pt\exp\left(\dlinks{\et^{(1)}}, \CW{t} 
\dlinks{\et^{(2)}}\right)  
\nonu\\
&=&\sli_{L\subset M_1\times M_2} \et_1\ldots\et_\mu \pli_{(k_1,k_2)\in L} 
\left(\dlinks{\et_{k_1}}\CW{t} (\Xx_{k_1},\Xx_{k_2})
\dlinks{\et_{k_2}}\right)
\end{eqnarray}
where $M_1=\nat{m_1}$ and $M_2=\{m_1+1,\ldots,\mu\}$.
 $L=\emptyset$ contributes to the sum, but gives
the $t$-independent result $\et_1\ldots\et_\mu $, so the 
$t$-derivative removes this term in \Ref{Z3}. 
Let $\pi_1(k_1,k_2) = k_1$ and $\pi_2(k_1,k_2) = k_2$. 
For a term given by $L\ne \emptyset$
to be nonzero, $\pi_1\vert_L$ and $\pi_2\vert_L$ must be injective,
because otherwise a derivative would act twice. Thus the sum can
be restricted to the set 
\begin{equation}
\cL = \{ L \subset M_1\times M_2 : 
L \ne \emptyset, \mbox{ and } \pi_k\mid_L \mbox{ injective for }
k=1,2\}
\end{equation}
If $L \in \cL$, $\pi_1(L) = \pi_2(L)$. Thus 
\begin{equation}
\cL = \bigcup\limits_{\zil=1}^{\min\{m_1,m_2\}} 
\; \bigcup\limits_{B_i \subset M_i \atop |B_1|=|B_2|=\zil}
\cL(B_1,B_2)
\end{equation}
with 
\begin{equation}
\cL(B_1,B_2) = \{ L \in \cL: \pi_1(L)=B_1 \mbox{ and }\pi_2(L)=B_2\}
.\end{equation}
Let
\begin{eqnarray}
B_1=\{ b_1,\ldots,b_\zil\}, && 1 \le b_1 < \ldots b_\zil \le m_1,
\nonu\\
B_2=\{ b_{\zil+1},\ldots,b_{2\zil}\}, && 
m_1+1 \le b_{\zil+1} < \ldots b_{2\zil} \le m_1+m_2
,\end{eqnarray}
then for any $L \in \cL(B_1,B_2)$, there is a unique permutation
$\pi \in \Perm{\zil}$ such that 
\begin{equation}
L=\{ (b_k,b_{\zil+\pi(k)}): k \in \nat{\zil}\}
.\end{equation}
Thus the sum over $L$ splits into a sum over $\zil \ge 1$, a sum over
sequences $b=(b_1,\ldots,b_{2\zil})$ with 
\begin{equation}\label{beinsch}
1 \le b_1 < b_2 < \ldots < b_{\zil} \le m_1 < b_{\zil+1} <
\ldots < b_{2\zil} \le m_1+m_2,
\end{equation}
and a sum over permutations
$\pi \in \Perm{\zil}$. Therefore
\begin{eqnarray}
\et_1\ldots\et_\mu&&\kern-20pt \exp\left(\dlinks{\et^{(1)}}, \CW{t} \;
\dlinks{\et^{(2)}} \right) 
\nonu\\
&=& \sli_{\zil=1}^{\min\{m_1,m_2\}} \sli_{\pi\in \Perm{\zil}}
\sli_b \; \pli_{k=1}^\zil D(\Xx_{b_k},\Xx_{b_{\zil+\pi(k)}})\;
H(\zil,b,\pi)
\end{eqnarray}
with 
\begin{equation}
H(\zil,b,\pi)= \et_1\ldots\et_\mu \pli_{k=1}^\zil \left(
\dlinks{\et_{b_k}} \;\dlinks{\et_{b_{\zil+\pi(k)}}} \right)
.\end{equation}

The derivatives give (since $m_1 $ and $m_2$ are even and hence 
$\vv^{m_k}=1$)
\begin{equation}
H(\zil,b,\pi) = 
\vv^{\frac{\zil(\zil+1)}{2}-\sli_{k=1}^{2\zil} b_k} \; \Sign(\pi)
\;\pli_{k=1\atop k \not\in\{b_1,\ldots,b_{2\zil}\}}^{\mu}
\et_k \;
\end{equation}
The remaining derivatives acting in \Ref{schwupp}
come from the Wick ordered exponential in \Ref{Z3}.
They now act on $H$. By Lemma \ref{umdiff}, they give
\begin{eqnarray}
\left\lbrack \pli_{k=1\atop k \not\in\{b_1,\ldots,b_{2\zil}\}}^{\mu}
\et_k \; \WO{\CW{t}} \left( \E^{(\dlinks{\et},\ps)}\right)\right\rbrack_{\et=0}
&=& \left\lbrack 
\pli_{k=1\atop k \not\in\{b_1,\ldots,b_{2\zil}\}}^{\mu}
\frac{\del}{\del \et_k}\WO{\CW{t}} 
\left( \E^{(\et,\ps)}\right)\right\rbrack_{\et=0}
\nonu\\
&=& \WO{\CW{t}} 
\left( 
\pli_{k=1\atop k \not\in\{b_1,\ldots,b_{2\zil}\}}^{m_1+m_2}
\ps(\Xx_k)\right)
.\end{eqnarray}
The final step is to rename the integration variables to rewrite
the Wick ordered product in the form in which it appears in 
\Ref{beingQ}. Before this is done, it is necessary to permute
the arguments of the $\Grmt{\Xx^{(k)}}$ such that 
$\Xx_1,\ldots,\Xx_{b_\zil}$ appear as the first $\zil$
entries. Since $b_1 < \ldots < b_{\zil}$, one can first permute
$(\Xx_1,\ldots,\Xx_{b_1}) \to (\Xx_{b_1}, \Xx_1,\ldots,\Xx_{b_1-1})$
This takes $b_1-1$ transpositions and hence gives a factor 
$\vv^{b_1-1}$.  
The next permutation  
\begin{equation}
(\Xx_{b_1}, \Xx_1,\ldots,\Xx_{b_1-1},\Xx_{b_1+1}, 
\ldots \Xx_{b_2}) 
\to 
(\Xx_{b_1}, \Xx_{b_2}, \Xx_1,
\ldots ,\Xx_{b_2-1}) 
\end{equation}
gives a factor $\vv^{b_2-2}$ because $\Xx_{b_1}$ has already been moved,
etc. Thus
\begin{equation}
\Grmto{\Xx_1, \ldots, \Xx_{m_1}} = 
\vv^{\sli_{k=1}^\zil (b_k-k)} 
\Grmto{\Xx_{b_1}, \ldots, \Xx_{b_\zil},\tilde\Xx}
\end{equation}
with $\tilde\Xx = (\Xx_{\rh_1}, \ldots, \Xx_{\rh_{m_1-\zil}})$
where $\nat{m_1} \setminus \{b_1,\ldots,b_{\zil}\} = 
\{ \rh_1, \ldots, \rh_{m_1-\zil}\}$
and $\rh_1 < \ldots < \rh_{m_1-\zil}$. Similarly,
\begin{equation}
\Grmtt{\ul{\Xx_2}} = 
\vv^{\sli_{k=1}^\zil (b_{\zil+k}-k-m_1)} 
\Grmtt{\Xx_{b_{\zil+1}}, \ldots, \Xx_{b_{2\zil}},\overline{\Xx}}
.\end{equation}
Thus the $b$--dependent sign factor cancels, and upon renaming of the 
integration variables, $\Xv_k = \Xx_{b_k}$, $\Xw_k=\Xx_{b_{\zil+k}}$,
etc., 
and with $m=m_1+m_2-2\zil$, $\cY$ becomes
\begin{equation}
\cY_{r_1,m_1,r_2,m_2} = \sli_{\zil=1}^{\min\{m_1,m_2\}}
\int d\ul{\Xx} \;\WO{\CW{t}} \left( \pli_{k=1}^m \ps(\Xx_k)\right)\;
Y(\ul{\Xx})
\end{equation}
with 
\begin{eqnarray}
Y (\ul{\Xx}) &= &    
\vv^{\frac{\zil(\zil+1)}{2}} 
\int d\ul{\Xv} \; d\ul{\Xw}
\sli_b \Grmto{\ul{\Xv},\ul{\Xx_1}}\;\Grmtt{\ul{\Xw},\ul{\Xx_2}}
\nonu\\
&&\sli_{\pi\in \Perm{\zil}} \Sign(\pi) 
\left(- \sfrac{\del}{\del t} \pli_{k=1}^\zil \CW{t}(\Xv_k,\Xw_{\pi(k)})
\right)
\end{eqnarray}
and $\ul{\Xx}$, $\ul{\Xx_1}$, $\ul{\Xx_2}$, $\ul{\Xv}$, and $\ul{\Xw}$ 
given in Proposition \ref{Qmrtvors}. 
The summand does not depend on $b$ any more, so 
the sum over $b$, with the constraint \Ref{beinsch}, gives
\begin{equation}
\sli_{B_1\subset M_1, B_2\subset M_2\atop
|B_1|=|B_2|=\zil} = {m_1 \choose \zil} \; 
{m_2\choose \zil} = \ka_{m_1m_2\zil}
\end{equation}
The sum over permutations $\pi$ gives the determinant of $\cDt{\zil}$.
Finally, 
\begin{equation}
\Grmto{\Xv_1,\ldots, \Xv_\zil,\ul{\Xx_1}}=
\vv^{(m_1-\zil)\zil}
\Grmto{\ul{\Xx_1},\Xv_1,\ldots, \Xv_\zil}
\end{equation}
and 
\begin{equation}
\Grmtt{\Xw_1,\ldots, \Xw_\zil,\ul{\Xx_2}}=
\vv^{\frac{\zil(\zil-1)}{2}}
\Grmtt{\Xw_\zil,\ldots, \Xw_1,\ul{\Xx_2}}
\end{equation}
which cancels the sign factor, and thus proves \Ref{Qrmtdef}.

The graphical interpretation of the 
above derivation is that 
$\Grmto{\ul{X_1}}$ and $\Grmtt{\ul{X_2}}$ are vertices of 
a graph, with the set of legs of vertex $1$ given by 
$\et_1 \ldots \et_{m_1}$ and the set of legs of vertex $2$ given by 
$\et_{m_1+1} \ldots \et_{m_1+m_2}$. 
The operator $(\dlinks{\et^{(1)}}, \CW{t} \dlinks{\et^{(2)}})$
generates lines between these two vertices by removing factors of
$\et$ in the monomial $\et_1 \ldots \et_{m_1+m_2}$. The number of
these internal lines is $\zil$, and the remaining $m=m_1+m_2-2\zil$
factors $\et_k$ correspond to external legs of the graph. 
$\zil \ge 1$ must hold because a derivative with respect to $t$ 
is taken, so the graph is connected. The rearrangement using 
permutations is simply the counting of all those graphs that
have the same value. There are ${m_1 \choose \zil}$ ways of picking 
$\zil$ legs from the $m_1$ legs of vertex number $1$ and
${m_2 \choose \zil}$ ways of picking 
$\zil$ legs from the $m_2$ legs of vertex number $2$,
which gives $\ka_{m_1m_2\zil}$.
And there are $\zil !$ ways of pairing these legs to form 
internal lines. 
One could also have used the antisymmetry of 
$\GGrmt{t}$ to get this factor explicitly, i.e.\ to
permute to get
\begin{equation}\label{riddle}
\zil! \; \pli_{k=1}^\zil \CW{t}(\Xv_k,\Xw_k)
.\end{equation}
instead of the determinant. 
This is the result one would have got for bosons since there 
are no sign factors in that case. The order of the $W_k$
in $\Grmtt{W_\zil, \ldots , W_1,\ul{X_2}}$ is chosen 
reversed to cancel the $\zil$-dependent sign factor for 
$\pi =id$. The graph for $\pi=id$ is the planar graph
drawn in Figure \ref{fig2}. 
The sum over permutations $\pi\ne id$ corresponds to the sum 
over all graphs with the fixed two vertices and $\zil$ internal
lines. In particular, it contains the {\em nonplanar} graphs. 
If one restricts the sum
to planar graphs, only the shifts $\pi_k(j)=j+k$ mod $\zil$,
$k\in \natz{\zil-1}$, remain, and this is only the case if 
$m_1=\zil$ of $m_2=\zil$. 
The factor $\zil!$ makes the combinatorial 
difference between the exact theory and the `planarized' theory.

\end{appendix}

\ifspr

\end{document}